 \documentclass[journal]{IEEEtran}

\usepackage[latin1]{inputenc}
\usepackage[T1]{fontenc}
\usepackage[english]{babel}
\usepackage{amssymb,amsmath,amsfonts}
\usepackage{times}
\usepackage{graphicx}
\usepackage{color}
\usepackage{verbatim}
\usepackage{cite}

\usepackage{subfig}		
\usepackage{pgf}		
\usepackage{tikz}
\usepackage{pgfplots}	
\usetikzlibrary{decorations}
\usetikzlibrary{patterns}
\usetikzlibrary{positioning}
\usetikzlibrary{backgrounds}

\usepackage{stmaryrd}

\usepackage{rotating}

\usepackage{cancel}
\usepackage[normalem]{ulem}

\newtheorem{definition}	{Definition}
\newtheorem{theorem}	{Theorem}
\newtheorem{proposition}{Proposition}
\newtheorem{lemma}		{Lemma}
\newtheorem{corollary}	{Corollary}
\newtheorem{remark}		{Remark}

\newcommand{\cA}{{\mathcal A}}
\newcommand{\cC}{{\mathcal C}}
\newcommand{\cE}{{\mathcal E}}
\newcommand{\cG}{{\mathcal G}}
\newcommand{\cN}{{\mathcal N}}
\newcommand{\cR}{{\mathcal R}}
\newcommand{\cS}{{\mathcal S}}
\newcommand{\cT}{{\mathcal T}}
\newcommand{\cU}{{\mathcal U}}
\newcommand{\cV}{{\mathcal V}}
\newcommand{\cW}{{\mathcal W}}
\newcommand{\cX}{{\mathcal X}}
\newcommand{\cY}{{\mathcal Y}}

\newcommand{\bN}{{\mathbb N}}
\newcommand{\bR}{{\mathbb R}}

\newcommand{\bE}{{\mathbb E}}

\newcommand{\pr}[1]{\Pr\left\{#1\right\}}

\newcommand{\Var}[1]{\text{Var}\left[#1\right]}

\newcommand{\mkv}{-\!\!\!\!\minuso\!\!\!\!-}

\newcommand{\err}{{\mathsf E}}
\newcommand{\Nerr}{\cancel{\mathsf E}}

\newcommand{\typ}[1]{T_\delta^n(#1)}

\providecommand{\abs}[1]{\left|#1\right|}

\providecommand{\norm}[1]{\lVert#1\rVert}

\newcommand{\lessnoisy}[1]{\succeq_{\scriptscriptstyle #1}}

\newcommand{\ind}[1]{\mathbb{I}_{#1}}

\newcommand{\toas}[1]{\xrightarrow[#1]{}}

\title{Secure Multiterminal Source Coding\\ with Side Information at the Eavesdropper} 
\author{
	Joffrey Villard and Pablo Piantanida
\thanks{The work of J. Villard is supported by DGA (French Armement Procurement Agency). This research is partially supported by the FP7 Network of Excellence in Wireless COMmunications NEWCOM\#.
The material in this paper was presented in part at the 2010 48th Annual Allerton Conference on Communication, Control, and Computing, Monticello, IL, USA; and the 1st International ICST Workshop on Secure Wireless Networks, Cachan, France.}
\thanks{J. Villard and P. Piantanida are with the Department of Telecommunications, SUPELEC, 91192 Gif-sur-Yvette, France (e-mail: joffrey.villard@supelec.fr; pablo.piantanida@supelec.fr).}
}

\begin{document}
\maketitle

\begin{abstract}
The problem of secure multiterminal source coding with side information at the eavesdropper is investigated. 
This scenario consists of a main encoder (referred to as Alice) that wishes to compress a single source but simultaneously satisfying the desired requirements on the distortion level at a legitimate receiver (referred to as Bob) and the equivocation rate --average uncertainty-- at an eavesdropper (referred to as Eve). 
It is further assumed the presence of a (public) rate-limited link between Alice and Bob. 
In this setting, Eve perfectly observes the information bits sent by Alice to Bob and has also access to a correlated source which can be used as side information. 
A second encoder (referred to as Charlie) helps Bob in estimating Alice's source by sending a compressed version of its own correlated observation via a (private) rate-limited link, which is only observed by Bob. 
For instance, the problem at hands can be seen as the unification between the Berger-Tung and the secure source coding setups. 
Inner and outer bounds on the so called rates-distortion-equivocation region are derived. 
The inner region turns to be tight for two cases: (i)  uncoded side information at Bob and (ii) lossless reconstruction of both sources at Bob --secure distributed lossless compression--. 
Application examples to secure lossy source coding of Gaussian and binary sources in the presence of Gaussian and binary/ternary (resp.) side informations are also considered. 
Optimal coding schemes are characterized for some cases of interest where the statistical differences between the side information at the decoders and the presence of a non-zero distortion at Bob can be fully exploited to guarantee secrecy.
\end{abstract}

\section{Introduction}

Consider the classical problem of compressing a source at a sensor node (referred to as Alice) which must be estimated at a remote destination (referred to as Bob) within a certain distortion level. 
Assume also that a (public) rate-limited link is available between the two devices. 
In addition to this, the encoder wishes to leak the least possible amount of information about its source to an eavesdropper (referred to as Eve) \emph{e.g.}, an untrusted sensor, who perfectly observes the information bits sent by Alice and may have access to an observation correlated to the source. 
Another sensor (referred to as Charlie) will help Bob in estimating Alice's source by sending a compressed version of its own correlated observation on a (private) rate-limited link, which is only observed by Bob.
In this setting, the correlation between the observations can be useful not only to decrease the rate needed for the communications, but also to increase secrecy, which means the average uncertainty of Eve about Alice's source. 
From a theoretical viewpoint, the problem at hands is therefore very rich and still quite open, as it contains, as subproblems, the long-standing information-theoretic problem of distributed lossy source coding, as well as recent ones \emph{e.g.}, source coding with security constraints.

Slepian and Wolf~\cite{slepian1973noiseless} introduced the problem of distributed lossless compression \emph{i.e.}, when Bob wants to perfectly estimate both sources of Alice and Charlie. 
Wyner~\cite{wyner1975source} and Ahlswede and K\"orner~\cite{ahlswede1975source} characterized the achievable region when only one source is to be estimated \emph{i.e.}, source coding with coded (or partial) side information. 
Generalization of the Slepian-Wolf setup to arbitrary distortion levels on both sources was introduced by Berger~\cite{berger1977multiterminal}, who provided inner and outer bounds on the achievable region which do not match in general. 
When Bob is intended to estimate only one source, Berger \emph{et al.}~\cite{berger1979upper} provided a new inner bound which was further proved in~\cite{jana2008partial} to be equivalent to the one of~\cite{berger1977multiterminal}, and strictly sub-optimal~\cite{wagner2009lossy}. 
Several results of optimality were proved in case of uncoded side information~\cite{wyner1976rate}, lossless reconstruction of at least one source~\cite{berger1989multiterminal}, and in some special cases, including Gaussian sources with quadratic distortion measure~\cite{oohama1997gaussian,wagner2008rate}. 
Over the years, these topics have been the focus of intense study and some remarkable progress has been made in theoretical and practical aspects, including general frameworks for lossless compression with multiple terminals~\cite{han1980unified,csiszar1980towards}, lossy source coding with uncertain side information at the decoder~\cite{heegard1985rate,kaspi1994rate}, lossy compression with partially separated encoders~\cite{kaspi1982rate-distortion} or with many decoders~\cite{tian2007multistage,tian2008side,timo2010rate}, some results of optimality for Gaussian sources in various contexts~\cite{tavildar2010gaussian,rahman2010rate}, as well as the design of nested codes for distributed compression \emph{e.g.}, using parity-check~\cite{zamir2002nested}, lattice~\cite{zamir2002nested,liveris2002compression,servetto2007lattice}, or algebraic trellis~\cite{pradhan2003distributed} codes. 
Nevertheless, in spite of these efforts, the simplest scenario of distributed lossy compression first introduced in~\cite{berger1977multiterminal} still remains open.

On the other hand, extensive research has been done on secure communication. 
The traditional focus was on cryptography, based on computational complexity where security only depends on the intractability assumption of some hard problems (\emph{e.g.}, factoring large integers). 
As a matter of fact, the security requirements were only taken into account in the upper layers of the OSI model (\emph{e.g.}, the application layer), assuming that reliable communication/compression schemes were already available.
Shannon in~\cite{shannon1949communication} introduced the information-theoretic notion of secrecy, where security is measured through the equivocation rate --the remaining uncertainty about the message-- at the eavesdropper. 
This information-theoretic approach of secrecy allows to consider security issues at the physical layer, and ensures unconditionally (regardless of the eavesdropper's computing power and time) secure schemes, since it only relies on the statistical properties of the system. 
Adopting this approach in a channel coding perspective, Wyner introduced the wiretap channel in~\cite{wyner1975wire} and showed that it is possible to send information at a positive rate with perfect secrecy as long as the channel of the eavesdropper is a degraded version of the legitimate user's one. 
Csisz\`ar and K\"orner~\cite{csiszar1978broadcast} extended this result to the setting of general broadcast channels with any arbitrary equivocation rate. 
Since then, several extensions have been proposed \emph{e.g.}, for fading channels~\cite{liang2008secure}, arbitrary \emph{i.e.}, not necessarily stationary memoryless, channels~\cite{bloch2008secrecy}, channels with state information at the encoder~\cite{chen2008wiretap}, cooperative relay broadcast channels~\cite{ekrem2011secrecy} (see also~\cite{it2008special,liang2009information,liu2010securing} for a review of recent results), as well as practical coding schemes for secure communication \emph{e.g.}, nested codes for (Gaussian and binary) type-II wiretap channels~\cite{liu2007secure}, LDPC~\cite{klinc2009ldpc,muramatsu2009construction} and lattice~\cite{oggier2011lattice} codes for the Gaussian wiretap channel, polar codes for binary symmetric channels~\cite{mahdavifar2010achieving}, and construction of secure codes using sparse matrices~\cite{thangaraj2007applications} or ordinary channel codes~\cite{hayashi2010construction}.
So far, very few work has been reported on source coding problems with security constraints, while early work~\cite{ahlswede1993common,maurer1993secret} showed that the presence of correlation between the different observations may guarantee some secrecy. 

Researchers have employed two approaches in the literature of secure source coding. 
In fact, it is assumed either that there already exists a secure rate-limited link between Alice and Bob, which allows the system to use secret keys, or at least the decoders have access to some side information about the source. 
In the scenario of secret key sharing, both lossless and lossy compression have been studied in various contexts~\cite{yamamoto1983source,yamamoto1988rate,yamamoto1994coding,yamamoto1997rate,merhav2006shannon,merhav2008shannon}.
Classical lossy source coding followed by encryption using the secret key was proved to be optimal when the receivers have no side information~\cite{yamamoto1997rate}.
For the second scenario, recent work~\cite{prabhakaran2007secure} considered the case of lossless source coding with (uncoded) side information at both decoders under the assumption of no rate constraint in the communication between Alice and Bob. 
In such a case, the usual Slepian-Wolf scheme is proved to be insufficient. 
Lossless source coding with coded side information, resp. distributed lossless compression, has been studied in~\cite{gunduz2008secure,tandon2009securea}, resp.~\cite{gunduz2008lossless}. 
In their ``one-sided helper'' scenario, the authors of~\cite{tandon2009securea} characterized the achievable region when only one source is to be perfectly estimated and Eve does not have side information. 
In particular, they proved that the achievable scheme of Wyner~\cite{wyner1975source} and Ahlswede and K\"orner~\cite{ahlswede1975source} achieves the whole region. 
Inner and outer bounds on the achievable region for secure distributed lossless compression have been proposed in~\cite{gunduz2008lossless}.
Secure \emph{lossy} source coding with side information at the decoders received less attention. 
As a matter of fact, if the (uncoded) side informations at the decoders are degraded then the achievable region can be derived as a special case of~\cite{merhav2008shannon} where Wyner-Ziv coding~\cite{wyner1976rate} is optimal.

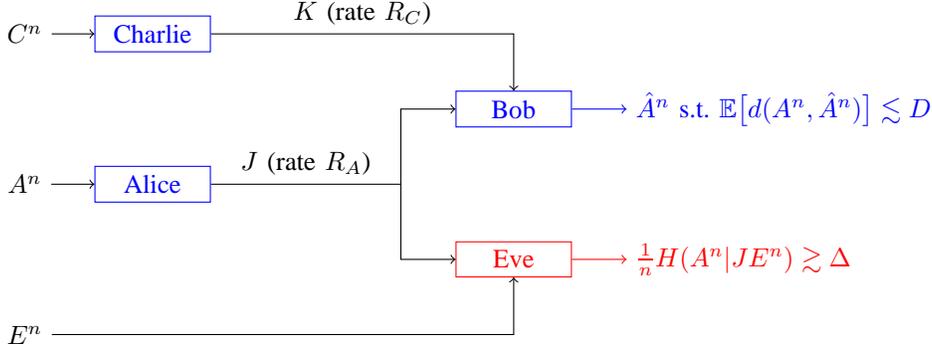
\begin{figure*}[!ht]
\centering
\begin{tikzpicture}
	\useasboundingbox (0,-2) rectangle (12,3);

	\tikzstyle{sensor}=[rectangle,draw,text centered,text width=1.3cm]
	\tikzstyle{gentil}=[blue]	
	\tikzstyle{mechant}=[red]

	\node	(A) 		at (0,0) 					{$A^n$};
	\node	(alice)		at (1.7,0)	[sensor,gentil] {Alice};
	\node	(C) 		at (0,2) 					{$C^n$};	
	\node	(charlie) 	at (1.7,2)	[sensor,gentil] {Charlie};
	\node	(bob) 		at (6.5,1)	[sensor,gentil]	{Bob};
	\node	(hatA) 		at (8,1)	[right,gentil]	{$\hat A^n$ s.t. $\bE\big[ d(A^n,\hat A^n) \big] \lesssim D$};
	\node	(eve) 		at (6.5,-1)	[sensor,mechant]{Eve};
	\node	(E) 		at (0,-2)					{$E^n$};
	\node	(D) 		at (8,-1)	[right,mechant]	{$\frac1n H(A^n|J E^n)\gtrsim\Delta$};
		
	\draw[->]		(A)			to (alice);
	\draw[->]		(alice)		to node[auto]{$J$ (rate $R_A$)} (5,0) to (5,1) to (bob);
	\draw[->]		(5,0)		to (5,-1)to (eve);
	
	\draw[->]		(C)			to (charlie);
	\draw[->]		(charlie)	to node[auto]{$K$ (rate $R_C$)} (6.5,2) to (bob);
	
	\draw[->]		(E)			to (6.5,-2) to (eve);
	
	\draw[->,blue]	(bob)		to (hatA);
	\draw[->,red]	(eve)		to (D);
\end{tikzpicture}
\caption{Secure lossy source coding with coded side information.}
\label{fig:schema}
\end{figure*}

In this paper, we investigate the general problem of secure lossy source coding of memoryless sources with coded side information at the legitimate receiver in the presence of an eavesdropper, who in addition to observe the information bits can also have access to correlated side information, as depicted in Fig.~\ref{fig:schema}.  
It is assumed that all links between encoders and decoders are noiseless so that they cannot provide any advantage to increase secrecy. 
This setting can be seen as the extension of the Berger \emph{et al.} problem~\cite{berger1979upper} to the one with security constraints. 
We provide inner and outer bounds on the achievable region, referred to as the \emph{rates-distortion-equivocation region}. It should be noted that the central difficulty lies in the evaluation of the equivocation at Eve and that these bounds do not match in general because of a long Markov chain condition. From the proposed inner region, we derive two novel results of optimality for the cases of: (i) uncoded side information, generalizing the results in~\cite{prabhakaran2007secure,gunduz2008secure} to any arbitrary distortion level, and (ii) lossless reconstruction of both sources at the legitimate receiver --distributed lossless compression--, refining~\cite{gunduz2008lossless}. 
When dealing with the lossy case in the presence of uncoded side information, it should be mentioned here that if one side information (either at Bob or Eve) is less noisy than the other, then Wyner-Ziv coding is sufficient. 
Similarly, for the distributed lossless compression setting it is shown that if the side information at Eve is less noisy than the observation of Charlie, then Slepian-Wolf coding achieves the whole region. 
As an application example, we consider the case of secure lossy source coding of a Gaussian source with Gaussian side informations, extending \cite{oohama1997gaussian} to the scenario with security constraints. 
We also consider the case of secure lossy source coding of a binary source, where the (uncoded) side information at Bob (resp. Eve) is the output of a binary erasure channel (resp. a binary symmetric channel) with the source as the input. 
This model is of theoretical interest since neither Bob nor Eve can always be a lessnoisy decoder. 

The rest of this paper is organized as follows. 
Section~\ref{sec:lossy} states definitions along with the main results on secure lossy source coding with coded side information at the legitimate receiver. 
Section~\ref{sec:uncoded} (resp. Section~\ref{sec:lossless}) provides an optimal characterization of the achievable region for the case of uncoded side information at Bob (resp. distributed lossless compression). 
The detailed proofs are relegated to the Appendices as well as a reminder on some useful notions and results. 
Section~\ref{sec:examples} presents application examples to Gaussian and binary sources. 
Finally, Section~\ref{sec:summary} summarizes the paper and provides discussions.

\subsection*{Notation}
For any sequence~$(x_i)_{i\in\bN^*}$, notation $x_k^n$
stands for the collection $(x_k,x_{k+1},\dots, x_n)$.
$x_1^n$ is simply denoted by $x^n$.
Let $\cT$ be an arbitrary finite set.
The cardinality of $\cT$ is denoted by $\norm{\cT}$. 
For any subset $\cS\subset\cT$, notation $\ind{\cS}$ stands for the indicator function of $\cS$ in $\cT$ \emph{i.e.}, for each $t\in\cT$, $\ind{\cS}(t)=1$ if $t\in\cS$, and $\ind{\cS}(t)=0$ otherwise.
Entropy is denoted by $H(\cdot)$, and mutual information by $I(\cdot;\cdot)$.
We denote typical and conditional typical sets by $\typ{X}$ and $\typ{Y|x^n}$, respectively (see Appendix~\ref{sec:typical} for details).
Let $X$, $Y$ and $Z$ be three random variables on some alphabets with probability distribution~$p$. If $p(x|y,z)=p(x|y)$ for each $x,y,z$, then $X$, $Y$ and $Z$ form a Markov chain, which is denoted by $X\mkv Y\mkv Z$.
Random variable $Y$ is said to be \emph{less noisy} than $Z$ w.r.t. $X$ if $I(U;Y) \geq I(U;Z)$ for each random variable $U$ such that $U\mkv X\mkv (Y,Z)$ form a Markov chain.
This relation is denoted by $Y\lessnoisy{X}Z$.
For each $x\in\bR$, notation $[x]_+$ stands for $\max(0;x)$.
Logarithms are taken in base $2$ and denoted by $\log(\cdot)$.
For each $a,b\in[0,1]$, $a\star b = a(1-b) + (1-a)b$.

\section{Secure Lossy Source Coding with Coded Side Information}
\label{sec:lossy}

\subsection{Definitions}

In this section, we give a more rigorous formulation of the context depicted in Fig.~\ref{fig:schema}. 
Let $\cA$, $\cC$ and $\cE$ be three finite sets. Alice, Charlie and Eve observe sequences of random variables $(A_i)_{i\in\bN^*}$, $(C_i)_{i\in\bN^*}$ and $(E_i)_{i\in\bN^*}$ respectively, which take values on $\cA$, $\cC$ and $\cE$, resp. For each $i\in\bN^*$, random variables $A_i$, $C_i$ and $E_i$ are distributed according to the joint distribution $p(a,c,e)$ on $\cA\times\cC\times\cE$. Moreover, they are independent across time $i$.

Let $d : \cA\times\cA \to [0\,;d_\text{max}]$ be a finite distortion measure \emph{i.e.}, such that $0\leq d_\text{max} < \infty$. We also denote by $d$ the component-wise mean distortion on $\cA^n\times\cA^n$ \emph{i.e.}, for each $a^n,b^n\in\cA^n$, $d(a^n,b^n) \triangleq \frac1n\,\sum_{i=1}^n d(a_i,b_i)$.\vspace{1mm}

\begin{definition}
An $(n,R_A,R_C)$-code for source coding in this setup is defined by
\begin{itemize}
\item An encoding function at Alice	denoted by	$f_A : \cA^n \to \{1,\dots,2^{nR_A}\}$,
\item An encoding function at Charlie	  denoted by $f_C : \cC^n \to \{1,\dots,2^{nR_C}\}$,
\item A decoding function at Bob 	denoted by	$g : \{1,\dots,2^{nR_A}\}\times\{1,\dots,2^{nR_C}\} \to \cA^n$.
\end{itemize}
\end{definition}

\begin{definition}
\label{def:achievable}
A tuple $(R_A,R_C,D,\Delta)\in\bR_+^4$ is said to be \emph{achievable} if,
for any $\varepsilon>0$, there exists an $(n,R_A+\varepsilon,R_C+\varepsilon)$-code $(f_A,f_C,g)$ such that:
\begin{IEEEeqnarray*}{rCl}
\bE\big[ d\big(A^n,g(f_A(A^n),f_C(C^n))\big) \big]	&\leq& D+\varepsilon 		\ ,\\
\dfrac1n\,H(A^n|f_A(A^n),E^n) 						&\geq& \Delta-\varepsilon 	\ .
\end{IEEEeqnarray*}
The set of all such achievable tuples is denoted by $\cR^*$
and is referred to as the \emph{rates-distortion-equivocation region}.
\end{definition}

\begin{remark}
Region $\cR^*$ is closed and convex.
\end{remark}

\begin{remark}
Quantities $\bE\big[ d(A^n,g(f_A(A^n),f_C(C^n)) \big]$ and $\frac1n\,H(A^n|f_A(A^n),E^n)$ in Definition~\ref{def:achievable} only depend on the marginal distributions $p(a,c)$ and $p(a,e)$, respectively.
The same holds for region $\cR^*$.
\end{remark}

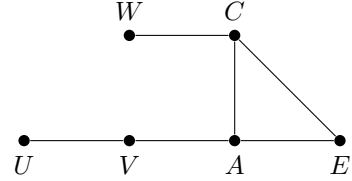
\begin{figure}[ht]
\centering
\begin{tikzpicture}[scale=1.4]	
	\tikzstyle{point}=[circle,fill,inner sep=0pt,minimum size=1.5mm]
	
	\node[point,label=below:$U$]	(U) at (0,0)	{};
	\node[point,label=below:$V$]	(V)	at (1,0)	{};
	\node[point,label=below:$A$]	(A) at (2,0) 	{};
	\node[point,label=below:$E$]	(E) at (3,0)	{};
	\node[point,label=above:$C$]	(C) at (2,1)	{};
	\node[point,label=above:$W$]	(W) at (1,1)	{};
		
	\draw	(U)	to (V);
	\draw	(V)	to (A);
	\draw	(A)	to (E);
	\draw	(E)	to (C);
	\draw	(A)	to (C);
	\draw	(C)	to (W);
\end{tikzpicture}
\caption{Inner bound--Graphical representation of probability distribution $p(uvwace)$.}
\label{fig:prob_inner_bound}
\end{figure}

\subsection{Inner and Outer Bounds on the Rates-Distortion-Equivocation Region}
\label{sec:lossy:regions}

The following theorem gives an inner bound on region $\cR^*$ \emph{i.e.}, it defines region $\cR_\text{in}\subset\cR^*$. 

\begin{theorem}
\label{th:inner_region}
A tuple $(R_A,R_C,D,\Delta)\in\bR_+^4$ is achievable if there exist random variables $U$, $V$, $W$ on some finite sets $\cU$, $\cV$, $\cW$, respectively, s.t. the joint distribution writes $p(uvwace)=p(u|v)p(v|a)p(w|c)p(ace)$, and a function $\hat A : \cV\times\cW \to \cA$, that verify the following inequalities:
\begin{IEEEeqnarray}{rCl}
R_A			&\geq&	I(V;A|W) 						\ ,\label{eq:inner:Ra}		\\
R_C			&\geq&	I(W;C|V) 						\ ,							\\
R_A+R_C		&\geq&	I(VW;AC)						\ ,\label{eq:inner:RaRc}	\\
D 			&\geq&	\bE\big[d(A,\hat A(V,W))\big] 	\ ,							\\
\Delta		&\leq&	H(A|VW) + I(A;W|U) - I(A;E|U) 	\ ,\label{eq:inner:Delta}	\\	
\Delta-R_C	&\leq&	H(A|V)-I(A;E|U)-I(W;C|V)		\ .\label{eq:inner:DRc}
\end{IEEEeqnarray}
Region $\cR_\text{in}$ is defined as the convex hull of the set of all such tuples.
\end{theorem}
\begin{table*}[!ht]
\centering
\caption{Corner points.}
\label{tab:corner_points}
\begin{IEEEeqnarraybox}[\IEEEeqnarraystrutmode]{x/t/V/c/v/c/v/c/x}
&Corner point		&& (I)							&& (II)					
					&& (III)						&						\\
\IEEEeqnarrayrulerow														\\
&Communication order&& W,\, U,\, V					&&	U,\, W,\, V
					&& U,\, V,\, W					&						\\
\IEEEeqnarrayrulerow														\\
&$R_A$				&& I(V;A|W)						&& I(U;A) + I(V;A|UW)
					&& I(V;A)						&						\\
&$R_C$				&& I(W;C)						&& I(W;C|U)
					&& I(W;C|V)						&						\\
&$D$					&& \bE\big[d(A,\hat A(V,W))\big]&& \bE\big[d(A,\hat A(V,W))\big]
					&& \bE\big[d(A,\hat A(V,W))\big]&							\\
&$\Delta$			&& H(A|UE) - I(V;A|UW)			&& H(A|UE) - I(V;A|UW)
					&& H(A|UE) - I(V;A|U)			&
\end{IEEEeqnarraybox}
\end{table*}

The proof of Theorem~\ref{th:inner_region} is based on superposition coding  and random binning at both encoders Alice and Charlie, and joint decoding at Bob. In the proposed scheme, layer $V$ (on the top of $U$) encodes source $A$ at Alice while layer $W$ encodes source $C$ at Charlie. A careful analysis  of this scheme along with standard properties of typical sequences  enables to characterize the equivocation rate at Eve. The detailed proof is relegated to Appendix~\ref{sec:inner_region}. The above inner region can also be achieved using a time-sharing combination of three complementary families of codes. Since this approach may yield better intuition, its proof is sketched below. 

Inequalities~\eqref{eq:inner:Ra}--\eqref{eq:inner:RaRc} are identical to the ones of Berger and Tung~\cite{berger1977multiterminal}. They ensure perfect reconstruction of both variables $V$ and $W$ at Bob, who can hence compute estimate $\hat A(V,W)$ of $A$.
The sum-rate constraint~\eqref{eq:inner:RaRc} captures the trade-off between rates $R_A$ and $R_C$. The information must be transmitted by one or the other encoder.

Let us now give some intuition on Equations~\eqref{eq:inner:Delta} and~\eqref{eq:inner:DRc}.
The first term $H(A|VW)$ corresponds to the equivocation rate at Bob.
Alice thus exploits the admissible distortion at Bob to increase the equivocation rate at Eve. 
Moreover, for given variables $V$ and $W$, which determine the rates and the distortion level at Bob, the auxiliary variable $U$ can be tuned to make Bob \emph{more capable} than Eve \emph{i.e.}, maximize $I(A;W|U) - I(A;E|U)$.
This quantity represents the gain (or the loss) at Eve in terms of equivocation rate.
At the same time, Equation~\eqref{eq:inner:DRc} imposes a trade-off between the equivocation rate at Eve $\Delta$ and the rate of Charlie $R_C$, which captures the fact that $\Delta$ cannot be too large if $R_C$ is not.
If the secrecy requirement is harsh, more information must be sent through the private link (between Charlie and Bob).
We will refer to quantity $\Delta-R_C$ as the \emph{public-link secrecy rate}.

Note that Equation~\eqref{eq:inner:Delta} also writes
\[
\Delta	\leq	H(A|UE) - I(V;A|UW)	\ .
\] 
Variable $U$ is thus considered as a \emph{common message} \emph{i.e.}, as if Eve could decode it.
As a matter of fact, in case of uncoded side information at Bob (resp. distributed lossless compression), Proposition~\ref{prop:uncoded_bis} (resp. \ref{prop:lossless_bis}) shows that it is optimal to encode $U$ so that Eve can reliably estimate it.
The remaining information rate of Alice (on the public link) \emph{i.e.}, $I(V;A|UW)$, is directly subtracted from the equivocation rate, meaning that it is treated as ``raw'' bits of $A$.

\begin{figure}[th]
\centering
\begin{tikzpicture}[line join=round,scale=.8]
\tikzstyle{surf} = [draw=none,style=nearly opaque];
\tikzstyle{axe} = [arrows=->,line width=.4pt];
\draw[axe](-2.981,.711)--(-8.572,2.044);
\draw[axe](0,0)--(-2.981,.711);
\filldraw[fill=white,draw=none](-3.578,.489)--(-2.981,.711)--(-2.981,1.622)--(-3.578,1.4)--cycle;
\draw[draw=none](-3.578,.489)--(-2.981,.711)--(-2.981,1.622);
\draw[draw=black!30](-2.981,.711)--(-2.981,1.622);
\draw[draw=black!30](-2.981,.711)--(-3.578,.489);
\filldraw[fill=white,draw=none](-3.578,.489)--(-3.578,1.4)--(-4.323,1.578)--(-4.323,.667)--cycle;
\draw[draw=none](-4.323,1.578)--(-4.323,.667)--(-3.578,.489);
\draw[draw=black!30](-3.578,.489)--(-4.323,.667);
\draw[draw=black!30](-4.323,.667)--(-4.323,1.578);
\filldraw[fill=white,draw=none](-3.578,.489)--(-.596,-.222)--(-.596,.689)--(-3.578,1.4)--cycle;
\draw[draw=none](-3.578,.489)--(-.596,-.222)--(-.596,.689);
\filldraw[fill=white,draw=none](-3.578,.489)--(-3.578,1.4)--(-7.752,-.156)--(-7.752,-1.067)--cycle;
\draw[draw=none](-7.752,-.156)--(-7.752,-1.067)--(-3.578,.489);
\draw[draw=black!30](-3.578,.489)--(-5.963,-.4);
\draw[draw=black!30](-.596,-.222)--(-3.578,.489);
\draw[draw=black!30](-3.578,.489)--(-3.578,1.4);
\draw[draw=black!30](-3.578,.489)--(-3.578,1.4);
\draw[draw=black!30](-2.832,.311)--(-2.832,1.222);
\draw[draw=black!30](-4.174,.267)--(-4.174,1.178);
\draw[draw=black!30](-2.087,.133)--(-2.087,1.044);
\draw[draw=black!30](-4.77,.044)--(-4.77,.956);
\draw[draw=black!30](-1.342,-.044)--(-1.342,.867);
\draw[draw=black!30](-5.367,-.178)--(-5.367,.733);
\draw[draw=black!30](-5.963,-.4)--(-5.963,.511);
\draw[draw=black!30](-5.963,-.4)--(-7.752,-1.067);
\draw[draw=black!30](-6.559,-.622)--(-6.559,.289);
\draw[draw=black!30](-7.155,-.844)--(-7.155,.067);
\filldraw[fill=white,draw=none](0,.911)--(-3.727,1.8)--(-8.497,.022)--(-4.77,-.867)--cycle;
\draw[draw=black!30](-2.981,1.622)--(-3.727,1.8);
\draw[draw=black!30](-3.727,1.8)--(-4.323,1.578);
\filldraw[fill=white,draw=none](-3.578,1.4)--(-3.578,5.956)--(-4.323,6.133)--(-4.323,1.578)--cycle;
\draw[draw=none](-3.578,5.956)--(-4.323,6.133)--(-4.323,1.578);
\draw[draw=black!30](-4.323,1.578)--(-8.497,.022);
\draw[draw=black!30](-4.323,1.578)--(-4.323,4.311);
\draw[draw=black!30](-3.578,1.4)--(-4.323,1.578);
\draw[draw=black!30](-3.578,1.4)--(-4.323,1.578);
\draw[draw=black!30](-3.578,2.311)--(-4.323,2.489);
\draw[draw=black!30](-3.578,3.222)--(-4.323,3.4);
\draw[draw=black!30](-4.174,1.178)--(-4.919,1.356);
\draw[draw=black!30](-3.578,4.133)--(-4.323,4.311);
\draw[draw=black!30](-4.323,4.311)--(-4.323,6.133);
\draw[draw=black!30](-3.578,5.044)--(-4.323,5.222);
\draw[draw=black!30](-4.77,.956)--(-5.516,1.133);
\draw[draw=black!30](-5.367,.733)--(-6.112,.911);
\draw[draw=black!30](-5.963,.511)--(-6.708,.689);
\draw[draw=black!30](-6.559,.289)--(-7.304,.467);
\draw[draw=black!30](-7.155,.067)--(-7.901,.244);
\filldraw[fill=white,draw=none](-2.981,1.622)--(-2.981,6.178)--(-7.752,4.4)--(-7.752,-.156)--cycle;
\draw[draw=none](-2.981,1.622)--(-2.981,6.178)--(-7.752,4.4)--(-7.752,-.156);
\draw[draw=black!30](0,.911)--(-2.981,1.622);
\draw[draw=black!30](-2.981,1.622)--(-2.981,6.178);
\draw[draw=black!30](-2.981,1.622)--(-3.578,1.4);
\draw[draw=black!30](-2.981,1.622)--(-3.578,1.4);
\draw[draw=black!30](-2.981,2.533)--(-3.578,2.311);
\draw[draw=black!30](-2.981,3.444)--(-3.578,3.222);
\draw[draw=black!30](-2.236,1.444)--(-2.832,1.222);
\draw[draw=black!30](-2.981,4.356)--(-3.578,4.133);
\draw[draw=black!30](-2.981,5.267)--(-3.578,5.044);
\draw[draw=black!30](-1.491,1.267)--(-2.087,1.044);
\draw[draw=black!30](-.745,1.089)--(-1.342,.867);
\filldraw[fill=white,draw=none](-3.578,1.4)--(-.596,.689)--(-.596,5.244)--(-3.578,5.956)--cycle;
\draw[draw=none](-.596,.689)--(-.596,5.244)--(-3.578,5.956);
\draw[draw=black!30](-3.578,1.4)--(-5.367,.733);
\draw[draw=black!30](-3.578,1.4)--(-5.367,.733);
\draw[draw=black!30](-.596,.689)--(-3.578,1.4);
\draw[draw=black!30](-.596,.689)--(-3.578,1.4);
\draw[draw=black!30](-3.578,1.4)--(-3.578,4.133);
\draw[draw=black!30](-3.578,1.4)--(-3.578,4.133);
\draw[draw=black!30](-3.578,2.311)--(-4.77,1.867);
\draw[draw=black!30](-.596,1.6)--(-3.578,2.311);
\draw[draw=black!30](-3.578,3.222)--(-4.174,3);
\draw[draw=black!30](-2.832,1.222)--(-4.621,.556);
\draw[draw=black!30](-.596,2.511)--(-3.578,3.222);
\draw[draw=black!30](-2.832,1.222)--(-2.832,3.956);
\draw[draw=black!30](-2.683,.822)--(-4.174,1.178);
\draw[draw=black!30](-4.174,1.178)--(-4.174,3);
\draw[draw=black!30](-2.981,6.178)--(-3.578,5.956);
\draw[draw=black!30](-3.578,4.133)--(-7.752,2.578);
\draw[draw=black!30](-.596,3.422)--(-3.578,4.133);
\draw[draw=black!30](-3.578,4.133)--(-3.578,5.956);
\draw[draw=black!30](-2.087,1.044)--(-3.876,.378);
\draw[draw=black!30](-2.087,1.044)--(-2.087,3.778);
\draw[draw=black!30](-3.28,.6)--(-4.77,.956);
\draw[dashed](-3.876,.378)--(-2.981,.711)--(-4.472,2.433);
\draw[draw=black!30](-1.342,.867)--(-3.13,.2);
\draw[draw=black!30](-1.342,.867)--(-1.342,3.6);
\draw[draw=black!30](-1.193,.467)--(-2.683,.822);
\draw[draw=black!30](-1.789,.244)--(-3.28,.6);
\filldraw[fill=green!60,surf](-3.876,.378)--(-4.472,2.433)--(-3.578,4.133)--(-.596,3.422)--(-2.385,.022)--(-3.876,.378)--cycle;
\draw[draw=black!30](-3.578,5.956)--(-4.323,6.133);
\draw[draw=black!30](-.596,4.333)--(-3.578,5.044);
\draw[draw=black!30](-1.342,3.6)--(-1.342,5.422);
\filldraw[fill=black!60,surf](-3.578,4.133)--(-3.578,5.956)--(-.596,5.244)--(-.596,3.422)--(-3.578,4.133)--cycle;
\draw[draw=black!30](-3.578,5.044)--(-7.752,3.489);
\draw[draw=black!30](-4.174,3)--(-4.174,5.733);
\draw[draw=black!30](-4.174,3)--(-7.752,1.667);
\draw[draw=black!30](-4.77,2.322)--(-4.77,5.511);
\draw[draw=black!30](-5.367,2.1)--(-5.367,5.289);
\draw[draw=black!30](-5.963,1.878)--(-5.963,5.067);
\draw[draw=black!30](-6.559,1.656)--(-6.559,4.844);
\draw[draw=black!30](-7.155,1.433)--(-7.155,4.622);
\filldraw[fill=black!60,surf](-3.578,4.133)--(-3.578,5.956)--(-7.752,4.4)--(-7.752,1.211)--(-4.472,2.433)--(-3.578,4.133)--cycle;
\draw[draw=black!30](-3.578,4.133)--(-3.578,5.956);
\draw(-3.578,4.133)--(-.596,3.422);
\draw(-3.578,5.956)--(-3.578,4.133)--(-4.472,2.433)--(-3.876,.378)--(-2.385,.022);
\draw[axe](0,0)--(0,5.467);
\draw[axe](0,0)--(-5.367,-2);
\draw[draw=black!30](-3.578,5.956)--(-7.752,4.4);
\draw[draw=black!30](-.596,5.244)--(-3.578,5.956);
\draw[draw=black!30](-4.77,.956)--(-4.77,2.322);
\draw[draw=black!30](-2.832,3.956)--(-2.832,5.778);
\draw[dashed](-3.876,.378)--(-5.367,.733)--(-4.472,2.433);
\draw[draw=black!30](-4.77,1.867)--(-7.752,.756);
\draw[draw=black!30](-3.876,.378)--(-5.367,.733);
\draw[draw=black!30](-5.367,.733)--(-5.367,2.1);
\draw[draw=black!30](-5.367,.733)--(-7.752,-.156);
\draw[draw=black!30](-5.367,.733)--(-7.752,-.156);
\draw[draw=black!30](-4.621,.556)--(-7.006,-.333);
\draw[draw=black!30](-4.472,.156)--(-5.963,.511);
\draw[draw=black!30](-5.963,.511)--(-5.963,1.878);
\draw[draw=black!30](-5.068,-.067)--(-6.559,.289);
\draw[draw=black!30](-6.559,.289)--(-6.559,1.656);
\draw[draw=black!30](-5.665,-.289)--(-7.155,.067);
\draw[draw=black!30](-7.155,.067)--(-7.155,1.433);
\filldraw[fill=blue!60,surf](-4.472,2.433)--(-7.752,1.211)--(-6.261,-.511)--(-3.876,.378)--(-4.472,2.433)--cycle;
\draw(-4.472,2.433)--(-7.752,1.211);
\draw[draw=black!30](0,.911)--(-.596,.689);
\draw[draw=black!30](-.596,-.222)--(-.596,.689);
\draw[draw=black!30](-2.087,3.778)--(-2.087,5.6);
\draw[draw=black!30](-.596,.689)--(-4.77,-.867);
\draw[draw=black!30](-.596,.689)--(-.596,5.244);
\draw(-.596,5.244)--(-.596,3.422)--(-2.385,.022)--(-4.77,-.867);
\draw[draw=black!30](-2.385,.022)--(-3.876,.378);
\filldraw[fill=black!60,surf](-3.876,.378)--(-6.261,-.511)--(-4.77,-.867)--(-2.385,.022)--(-3.876,.378)--cycle;
\draw[draw=black!30](-3.876,.378)--(-6.261,-.511);
\draw(-3.876,.378)--(-6.261,-.511);
\draw[draw=black!30](-3.13,.2)--(-5.516,-.689);
\draw[draw=black!30](-2.981,-.2)--(-4.472,.156);
\draw[draw=black!30](-7.752,-.156)--(-8.497,.022);
\draw[draw=black!30](-3.578,-.422)--(-5.068,-.067);
\draw[draw=black!30](-7.752,-1.067)--(-7.752,-.156);
\draw[draw=black!30](-4.77,-.867)--(-7.752,-.156);
\draw[draw=black!30](-7.752,-.156)--(-7.752,4.4);
\draw[draw=black!30](-4.174,-.644)--(-5.665,-.289);
\path 	(-3.578,4.133) node[right] {$(I)$}
					(-4.472,2.433) node[right] {$(II)$}
					(-3.876,.378) node[right] {$(III)$};
\path 	(-5.367,-2) node[left]	{$R_A$}
					(0,5.467) node[above]	{$R_C$}
					(-8.572,2.044) node[below]	{$\Delta$};
\end{tikzpicture}
\caption{Achievable tuples $(R_A,R_C,\Delta)$ for some fixed distortion level $D$.}
\label{fig:region_3D}
\end{figure}
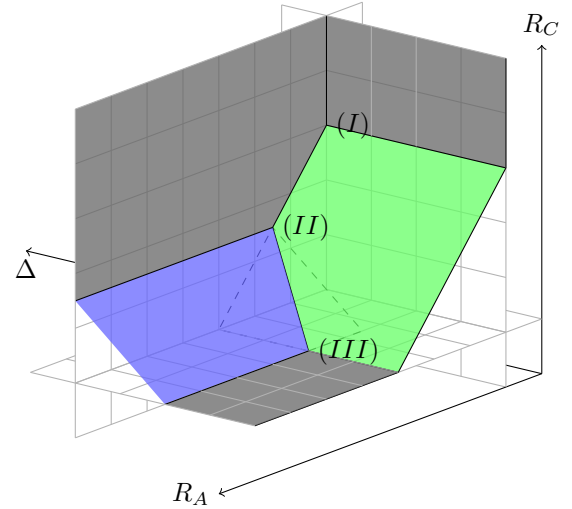

\begin{figure*}[!ht]
\centering
\begin{tabular}{p{4cm}p{3cm}p{4cm}}
\begin{tikzpicture}[scale=1.4]	
	\tikzstyle{point}=[circle,fill,inner sep=0pt,minimum size=1.5mm]
	
	\node[point,label=below:$U$]	(U)	at (0,0)	{};
	\node[point,label=below:$V$]	(V)	at (1,0)	{};
	\node[point,label=below:$A$]	(A) at (2,0) 	{};
	\node[point,label=below:$E$]	(E) at (3,0)	{};
	\node[point,label=above:$C$]	(C) at (2,1)	{};
		
	\draw	(U)	to (V);
	\draw	(V)	to (A);
	\draw	(A)	to (E);
	\draw	(E)	to (C);
	\draw	(A)	to (C);
\end{tikzpicture}
&&
\begin{tikzpicture}[scale=1.4]	
	\tikzstyle{point}=[circle,fill,inner sep=0pt,minimum size=1.5mm]
	
	\node[point,label=below:$A$]	(A) at (2,0) 	{};
	\node[point,label=below:$E$]	(E) at (3,0)	{};
	\node[point,label=above:$C$]	(C) at (2,1)	{};
	\node[point,label=above:$W$]	(W) at (1,1)	{};
		
	\draw	(A)	to (E);
	\draw	(E)	to (C);
	\draw	(A)	to (C);
	\draw	(C)	to (W);
\end{tikzpicture}
\end{tabular}

\caption{Outer bound--Graphical representation of probability distributions $p(uvace)$ and $p(wace)$.}
\label{fig:prob_outer_bound}
\end{figure*}
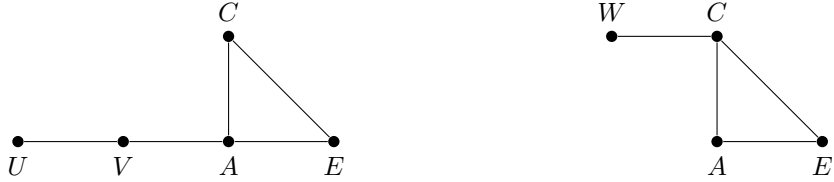

\emph{Sketch of proof of Theorem~\ref{th:inner_region} (Time-sharing combination technique): }
We first construct three codes achieving corner points $(I)$, $(II)$ and $(III)$ illustrated in Fig.~\ref{fig:region_3D},~\ref{fig:region_RaRc} and~\ref{fig:region_RcD}. 
Each corner point is achieved using a three-step communication scheme which aim is to reliably deliver variables $(U,V)$ and $W$, descriptions of $A$ at Alice and $C$ at Charlie, respectively, to Bob.
Note that $V$ is on the top of $U$ (\emph{superposition coding}).
At each step, the information previously received (and decoded) is used as side-information at Bob. \emph{Random binning} \emph{a la} Wyner-Ziv~\cite{wyner1976rate} is performed to take advantage of this side information.
These schemes correspond to all possible combinations of the set $\{U,V,W\}$, provided that $U$ is decoded prior to $V$, as summarized in row~\#2 of Table~\ref{tab:corner_points}.
For each scheme, the equivocation rate at Eve can be characterized following the argument of Appendix~\ref{sec:inner_equivocation}.
After Fourier-Motzkin elimination and classical manipulation, we can prove that the three proposed schemes can achieve corner points $(I)$, $(II)$ and $(III)$, which coordinates are given in Table~\ref{tab:corner_points}.

Points $(I)$ and $(II)$ correspond to identical distortion and equivocation rate levels, say $D$ and $\Delta$ (see Fig.~\ref{fig:region_RcD}). By a time-sharing combination of these schemes, each point on segment $(I)$--$(II)$ is also achievable and presents distortion $D$ and equivocation rate $\Delta$. This segment can be easily described since the quantity $R_A+R_C$ is identical for both points $(I)$ and $(II)$ (see Fig.~\ref{fig:region_RaRc}).

Points $(II)$ and $(III)$ correspond to identical distortion level, say $D$. By a time-sharing combination of these schemes, each point on segment $(II)$--$(III)$ is also achievable and presents distortion $D$. This segment can be easily described since quantities $R_A+R_C$ and $\Delta-R_C$ are identical for both points $(II)$ and $(III)$ (see Fig.~\ref{fig:region_RaRc} and~\ref{fig:region_RcD}, respectively).

Segments $(I)$--$(II)$ and $(II)$--$(III)$ define regions which union is delimited by six hyperplanes given by the equations of Theorem~\ref{th:inner_region}.
\endproof

\begin{remark}
The simple union of the regions given by the equations of Theorem~\ref{th:inner_region} is not convex. In fact, a time-sharing variable $T$ cannot be included in auxiliary variables $U$, $V$ and $W$. This would break the long Markov chain $U\mkv V\mkv A\mkv C\mkv W$ which is essential in our coding scheme.
\end{remark}

\begin{remark}
Projections of points~$(I)$ and~$(III)$ on the plane $\Delta=0$ \emph{i.e.}, when there is no secrecy constraint, are those obtained using Berger-Tung coding~\cite{berger1977multiterminal}. In this case, point $(II)$ is useless since it is achievable by a time-sharing combination of points~$(I)$ and~$(III)$, as shown by Fig.~\ref{fig:region_RaRc}. In the general case, the proposed scheme can however improve the security of the transmission, as shown in Fig.~\ref{fig:region_RcD}.
\end{remark}

\begin{remark}
When there is no security requirement, Jana and Blahut~\cite{jana2008partial} recently proved the equivalence of the inner bounds of~\cite{berger1977multiterminal} and~\cite{berger1979upper}, meaning that point~$(I)$ alone yields the same region that points~$(I)$ and $(III)$ (after the convex hull operation).
A similar result in our secure setting does not seem obvious.
\end{remark}

The following proposition gives upper bounds on the cardinalities of alphabets $\cU$, $\cV$ and $\cW$. The proof, which is given in Appendix~\ref{sec:card}, relies on Fenchel-Eggleston-Carath\'eodory's theorem and follow standard cardinality bounding argument (see~\cite[Appendix C]{elgamal2010lecture}).


\begin{proposition}
\label{prop:card}
In the inner region~$\cR_\text{in}$ given by Theorem~\ref{th:inner_region}, it suffices to consider sets~$\cU$,~$\cV$ and~$\cW$ such that 
$\norm{\cU}\leq \norm{\cA}+5$,
$\norm{\cV}\leq (\norm{\cA}+5)(\norm{\cA}+3)$ and
$\norm{\cW}\leq\norm{\cC}+3$.
\end{proposition}

The following theorem gives an outer bound on region $\cR^*$ \emph{i.e.}, it defines region $\cR_\text{out}\supset\cR^*$. The proof is given in Appendix~\ref{sec:outer_region}.

\begin{theorem}
\label{th:outer_region}
Region $\cR^*$ is included in $\cR_\text{out}$, defined as the closure of the set of all tuples $(R_A,R_C,D,\Delta)\in\bR_+^4$ such that there exist random variables $U$, $V$, $W$ on some finite sets $\cU$, $\cV$, $\cW$, respectively, and a function $\hat A : \cV\times\cW \to \cA$
satisfying $p(wace)=p(w|c)p(ace)$, $p(uvace)=p(u|v)p(v|a)p(ace)$, and
\begin{IEEEeqnarray*}{rCl}
R_A			&\geq&	I(V;A|W) 						\ ,\\
R_C			&\geq&	I(W;C|V) 						\ ,\\
R_A+R_C		&\geq&	I(VW;AC)						\ ,\\
D 			&\geq&	\bE\big[d(A,\hat A(V,W))\big] 	\ ,\\
\Delta 		&\leq&	H(A|VW) + I(A;W|U) - I(A;E|U) 	\ ,\\	
\Delta-R_C	&\leq&	H(A|V) - I(A;E|U) - I(W;C|V)	\ .
\end{IEEEeqnarray*}
\end{theorem}

\begin{figure}[ht]
\centering
\begin{minipage}{7.2cm}
\begin{tikzpicture}[line join=round,scale=.6]
\tikzstyle{surf} = [draw=none,style=nearly opaque];
\tikzstyle{axe} = [arrows=->,line width=.4pt];
\draw[draw=black!30](-3,0)--(-3,6);
\draw[draw=black!30](-2,0)--(-2,6);
\draw[draw=black!30](0,3)--(-8,3);
\draw[draw=black!30](0,2)--(-8,2);
\filldraw[fill=green!60,surf](-4,1)--(-2.5,2.5)--(-1,4)--(-1,4)--(-4,1)--(-4,1)--cycle;
\draw[draw=black!30](0,5)--(-8,5);
\filldraw[fill=black!60,surf](-1,4)--(-1,6)--(-1,6)--(-1,4)--(-1,4)--cycle;
\draw[draw=black!30](-7,0)--(-7,6);
\draw[draw=black!30](-6,0)--(-6,6);
\draw[draw=black!30](-5,0)--(-5,6);
\draw[draw=black!30](-4,0)--(-4,6);
\filldraw[fill=blue!60,surf](-2.5,2.5)--(-8,2.5)--(-8,1)--(-4,1)--(-2.5,2.5)--cycle;
\filldraw[fill=black!60,surf](-1,4)--(-1,6)--(-8,6)--(-8,2.5)--(-2.5,2.5)--(-1,4)--cycle;
\draw[draw=black!30](0,0)--(-8,0);
\draw[draw=black!30](0,1)--(-8,1);
\draw[draw=black!30](-8,0)--(-8,6);
\draw[draw=black!30](-1,0)--(-1,6);
\draw[draw=black!30](0,0)--(0,6);
\draw[draw=black!30](0,6)--(-8,6);
\draw[draw=black!30](0,4)--(-8,4);
\draw(-1,4)--(-1,4);
\draw(-2.5,2.5)--(-8,2.5);
\draw(-1,6)--(-1,4)--(-2.5,2.5)--(-4,1)--(-4,1);
\filldraw[fill=black!60,surf](-4,1)--(-8,1)--(-8,1)--(-4,1)--(-4,1)--cycle;
\draw(-4,1)--(-8,1);
\draw[axe](0,0)--(-8.2,0);
\draw[axe](0,0)--(0,6.2);
\draw(-1,6)--(-1,4)--(-4,1)--(-8,1);
\path 	(-1,4) node[right] {$(I)$}
					(-2.5,2.5) node[right] {$(II)$}
					(-4,1) node[right] {$(III)$};
\path 	(-8.2,0) node[left]	{$R_A$}
					(0,6.2) node[above]	{$R_C$};
\end{tikzpicture}
	\caption{Projection on the plane $\Delta=0$.}
	\label{fig:region_RaRc}
\end{minipage}
\hfill
\begin{minipage}{7.2cm}
\begin{tikzpicture}[line join=round,scale=.6]
\tikzstyle{surf} = [draw=none,style=nearly opaque];
\tikzstyle{axe} = [arrows=->,line width=.4pt];
\draw[draw=black!30](0,0)--(0,6);
\draw[draw=black!30](-1,0)--(-1,6);
\draw[draw=black!30](0,6)--(-5,6);
\draw[draw=black!30](0,5)--(-5,5);
\draw[draw=black!30](0,4)--(-5,4);
\draw[draw=black!30](0,3)--(-5,3);
\draw[draw=black!30](0,2)--(-5,2);
\draw[draw=black!30](0,1)--(-5,1);
\draw[draw=black!30](0,0)--(-5,0);
\draw[draw=black!30](-5,0)--(-5,6);
\draw[draw=black!30](-4,0)--(-4,6);
\draw[draw=black!30](-3,0)--(-3,6);
\draw[draw=black!30](-2,0)--(-2,6);
\draw[axe](0,0)--(-5.2,0);
\draw[axe](0,0)--(0,6.2);
\filldraw[fill=green!60,surf](-2,1)--(-4,2.5)--(-4,4)--(0,4)--(0,1)--(-2,1)--cycle;
\filldraw[fill=black!60,surf](-4,4)--(-4,6)--(0,6)--(0,4)--(-4,4)--cycle;
\filldraw[fill=black!60,surf](-4,4)--(-4,6)--(-4,6)--(-4,2.5)--(-4,2.5)--(-4,4)--cycle;
\draw(-4,6)--(-4,4)--(-4,2.5)--(-2,1)--(0,1);
\draw(-4,4)--(0,4);
\draw(0,6)--(0,4)--(0,1)--(0,1);
\filldraw[fill=blue!60,surf](-4,2.5)--(-4,2.5)--(-2,1)--(-2,1)--(-4,2.5)--cycle;
\draw(-4,2.5)--(-4,2.5);
\filldraw[fill=black!60,surf](-2,1)--(-2,1)--(0,1)--(0,1)--(-2,1)--cycle;
\draw(-2,1)--(-2,1);
\path 	(-4,4) node[right] {$(I)$}
					(-4,2.5) node[right] {$(II)$}
					(-2,1) node[right] {$(III)$};
\path 	(0,6.2) node[above]	{$R_C$}
					(-5.2,0) node[left]	{$\Delta$};
\end{tikzpicture}
	\caption{Projection on the plane $R_A=0$.}
	\label{fig:region_RcD}
\end{minipage}
\end{figure}

As in the classical multiterminal source coding setup~\cite{berger1977multiterminal}, the outer region resembles the inner region except that it is convex without time-sharing and that Markov chain conditions $W\mkv C\mkv (A,E)$ and $U\mkv V\mkv A\mkv (C,E)$ are weaker than the long Markov chain of Theorem~\ref{th:inner_region} (compare Fig.~\ref{fig:prob_inner_bound} and~\ref{fig:prob_outer_bound}, and see Appendix~\ref{sec:graphical} for details on such graphical representations).

\subsection{Special Case: Lossless Reconstruction of $A$}

In case of lossless reconstruction of $A$ at Bob,\footnote{This case is included in the general setup choosing $d$ as the Kronecker delta and $D=0$.} if Eve has no side information ($E=\emptyset$), then point~$(I)$ yields the optimal performance choosing auxiliary variables $U=\emptyset$ and $V=A$ \emph{i.e.}, using Wyner-Ahlswede-K\"orner coding~\cite{wyner1975source,ahlswede1975source}, as stated by  Tandon \emph{et al.}~\cite[Theorem 1]{tandon2009securea}:
In this case, region $\cR^*$ writes as the closure of the set of all tuples $(R_A,R_C,D=0,\Delta)\in\bR_+^4$ such that there exists a random variable $W$ on some finite set $\cW$ s.t. $W\mkv C\mkv A$ form a Markov chain and
\begin{IEEEeqnarray*}{rCl}
R_A			&\geq&	H(A|W) 					\ ,\\
R_C			&\geq&	I(W;C) 					\ ,\\
\Delta		&\leq&	I(A;W) 	\ .
\end{IEEEeqnarray*}

\begin{figure*}[!ht]
\centering
\begin{tikzpicture}
	\useasboundingbox (0,-2) rectangle (12,3);

	\tikzstyle{sensor}=[rectangle,draw,text centered,text width=1.3cm]
	\tikzstyle{gentil}=[blue]	
	\tikzstyle{mechant}=[red]

	\node	(A) 		at (0,0) 					{$A^n$};
	\node	(alice)		at (1.7,0)	[sensor,gentil] {Alice};
	\node	(C) 		at (0,2) 					{$C^n$};	
	\node	(bob) 		at (6.5,1)	[sensor,gentil]	{Bob};
	\node	(hatA) 		at (8,1)	[right,gentil]	{$\hat A^n$ s.t. $\bE\big[ d(A^n,\hat A^n) \big] \lesssim D$};
	\node	(eve) 		at (6.5,-1)	[sensor,mechant]{Eve};
	\node	(E) 		at (0,-2)					{$E^n$};
	\node	(D) 		at (8,-1)	[right,mechant]	{$\frac1n H(A^n|J E^n)\gtrsim\Delta$};
		
	\draw[->]		(A)			to (alice);
	\draw[->]		(alice)		to node[auto]{$J$ (rate $R_A$)} (5,0) to (5,1) to (bob);
	\draw[->]		(5,0)		to (5,-1)to (eve);
	
	\draw[->]		(C)			to (6.5,2) to (bob);
	
	\draw[->]		(E)			to (6.5,-2) to (eve);
	
	\draw[->,blue]	(bob)		to (hatA);
	\draw[->,red]	(eve)		to (D);
\end{tikzpicture}
\caption{Secure lossy source coding with uncoded side information.}
\label{fig:uncoded:schema}
\end{figure*}
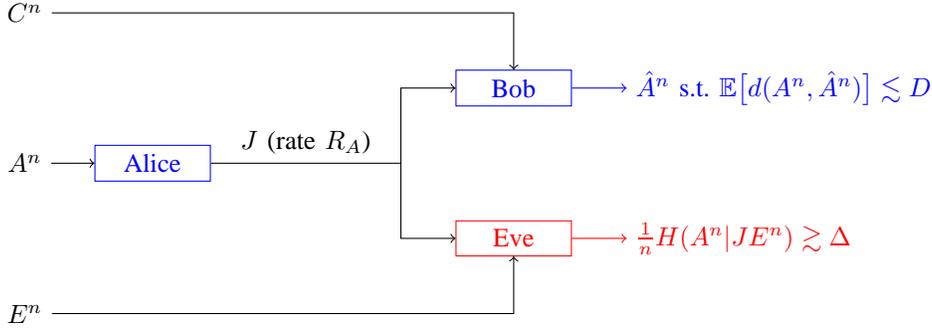

\subsection{Joint Estimation and Equivocation of Both Sources}

Definition~\ref{def:achievable} only involves the distortion level at Bob and the equivocation rate at Eve about Alice's source.
As a matter of fact, the proofs of Theorems~\ref{th:inner_region} and~\ref{th:outer_region} can be used to obtain inner and outer bounds on the achievable region when also considering a distortion constraint on Charlie's source at Bob.
This requires the following additional inequality in the definition of the achievability:
\[
\bE\big[ d_C\big(C^n,g_C(f_A(A^n),f_C(C^n))\big) \big] \leq D_C+\varepsilon \ ,
\]
for some distortion measure $d_C$ and decoding function $g_C$.
The resulting bounds will only differ from the ones of Theorems~\ref{th:inner_region} and~\ref{th:outer_region} by adding the following inequality:
\[
D_C \geq \bE\big[d_C(C,\hat C(V,W))\big] \ ,
\]
for some function $\hat C: \cV\times\cW \to \cC$.
For the sake of readability, we did not include this fifth dimension in the main definitions.
In Section~\ref{sec:lossless}, we remove the distortion, addressing the case of lossless reconstruction of both sources, and prove that region $\cR_\text{in}$ yields an optimal characterization of the corresponding achievable region. 

Furthermore, the \emph{joint} equivocation rate writes:
\begin{IEEEeqnarray*}{rCl}
\frac1n H(A^n C^n|f_A(A^n),E^n) &=& \frac1n H(A^n |f_A(A^n),E^n) \\
	&+& \frac1n H(C^n|A^n E^n) \ ,
\end{IEEEeqnarray*}
and the last term $\frac1n H(C^n|A^n E^n)$ is constant \emph{i.e.}, independent of the coding scheme.
Hence, the results involving $\frac1n H(A^n |f_A(A^n),E^n)$ directly apply to the joint equivocation rate.

\section{Secure Lossy Source Coding with Uncoded Side Information}
\label{sec:uncoded}

\subsection{Definitions}

In this section, we consider the special case depicted in Fig.~\ref{fig:uncoded:schema} where Bob has access to \emph{uncoded} side information \emph{i.e.}, Bob and Charlie are collocated.
We need the following new definitions:

\begin{definition}
An $(n,R_A)$-code for source coding in this setup is defined by
\begin{itemize}
\item An encoding function at Alice		$f : \cA^n \to \{1,\dots,2^{nR_A}\}$,
\item A decoding function at Bob		$g : \{1,\dots,2^{nR_A}\}\times\cC^n \to \cA^n$.
\end{itemize}
\end{definition}

\begin{definition}
A tuple $(R_A,D,\Delta)\in\bR_+^3$ is said to be \emph{achievable} if,
for any $\varepsilon>0$, there exists an $(n,R_A+\varepsilon)$-code $(f,g)$ such that:
\begin{IEEEeqnarray*}{rCl}
\bE\left[ d\big(A^n,g(f(A^n),C^n)\big) \right]	&\leq& D+\varepsilon		\ ,\\
\dfrac1n\,H(A^n|f(A^n),E^n) 					&\geq& \Delta-\varepsilon	\ .
\end{IEEEeqnarray*}
The set of all such achievable tuples is denoted by $\cR_\text{uncoded}^*$
and is referred to as the \emph{rate-distortion-equivocation region}.
\end{definition}

\subsection{Optimal Characterization}

\begin{figure}
\centering
\begin{tikzpicture}[line join=round,scale=.6]
\tikzstyle{surf} = [draw=none,style=nearly opaque];
\tikzstyle{axe} = [arrows=->,line width=.4pt];
\draw[axe](0,0)--(-8.2,0);
\draw[axe](0,0)--(0,5.2);
\draw[draw=black!30](-8,0)--(-8,5);
\draw[draw=black!30](-7,0)--(-7,5);
\draw[draw=black!30](-6,0)--(-6,5);
\draw[draw=black!30](-5,0)--(-5,5);
\draw[draw=black!30](-4,0)--(-4,5);
\draw[draw=black!30](-3,0)--(-3,5);
\draw[draw=black!30](-2,0)--(-2,5);
\draw[draw=black!30](-1,0)--(-1,5);
\draw[draw=black!30](0,0)--(0,5);
\draw[draw=black!30](0,5)--(-8,5);
\draw[draw=black!30](0,4)--(-8,4);
\draw[draw=black!30](0,3)--(-8,3);
\draw[draw=black!30](0,2)--(-8,2);
\draw[draw=black!30](0,1)--(-8,1);
\draw[draw=black!30](0,0)--(-8,0);
\filldraw[fill=green!60,surf](-4,2)--(-2.5,4)--(-1,4)--(-1,0)--(-4,0)--(-4,2)--cycle;
\filldraw[fill=black!60,surf](-4,2)--(-8,2)--(-8,0)--(-4,0)--(-4,2)--cycle;
\filldraw[fill=blue!60,surf](-2.5,4)--(-8,4)--(-8,2)--(-4,2)--(-2.5,4)--cycle;
\draw(-1,4)--(-1,4)--(-2.5,4)--(-4,2)--(-4,0);
\draw(-4,2)--(-8,2);
\draw(-1,0)--(-1,0)--(-4,0)--(-8,0);
\filldraw[fill=black!60,surf](-1,4)--(-1,4)--(-8,4)--(-8,4)--(-2.5,4)--(-1,4)--cycle;
\draw(-2.5,4)--(-8,4);
\filldraw[fill=black!60,surf](-1,4)--(-1,4)--(-1,0)--(-1,0)--(-1,4)--cycle;
\draw(-1,4)--(-1,0);
\path 	(-1,4) node[right] {$(I)$}
					(-2.5,4) node[right] {$(II)$}
					(-4,2) node[right] {$(III)$};
\path 	(-8.2,0) node[left]	{$R_A$}
					(0,5.2) node[above]	{$\Delta$};
\end{tikzpicture}
\caption{Projection on the plane $R_C=0$.}
\label{fig:region_RaD}
\end{figure}
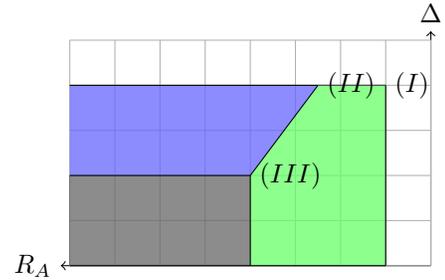

In the setup considered in this section, the following theorem provides a single-letter characterization of region $\cR_\text{uncoded}^*$. 
The achievability follows from the one of Theorem~\ref{th:inner_region}, choosing auxiliary variable $W=C$, and removing constraints on $R_C$ (letting $R_C$ tend to $\infty$) \emph{i.e.}, from the achievability of point~$(I)$ (see Fig.~\ref{fig:region_RaD}).
A new proof is needed for the converse part (see Appendix~\ref{sec:uncoded:converse}).

Note that if Eve is a legitimate decoder that wishes to estimate source $A$ within a certain distortion criterion (instead of an eavesdropper that other terminals must contend with), then \cite{timo2010rate} provides inner and outer bounds on the corresponding rate-distortion function (with two decoders and side-information).
Finding an optimal characterization of the achievable region in such a case is still an open problem.

\begin{theorem}
\label{th:uncoded}
Region $\cR_\text{uncoded}^*$ writes as the closure of the set of all tuples $(R_A,D,\Delta)\in\bR_+^3$ such that there exist random variables $U$, $V$ on some finite sets $\cU$, $\cV$, respectively, and a function $\hat A : \cV\times\cC \to \cA$ such that $U\mkv V\mkv A\mkv (C,E)$ form a Markov chain and
\begin{IEEEeqnarray}{rCl}
R_A		&\geq& I(V;A|C)							\ ,\label{eq:uncoded:Ra}	\\
D 		&\geq& \bE\big[d(A,\hat A(V,C))\big]	\ ,\label{eq:uncoded:D}	\\
\Delta	&\leq& H(A|VC) + I(A;C|U) - I(A;E|U)	\ .\label{eq:uncoded:Delta}
\end{IEEEeqnarray}
\end{theorem}

Comments similar to the ones of Section~\ref{sec:lossy:regions} about Theorem~\ref{th:inner_region} are also relevant here:
Equations~\eqref{eq:uncoded:Ra} and~\eqref{eq:uncoded:D} are classical in rate-distortion theory,
Alice can exploit the admissible distortion at Bob to increase the equivocation rate at Eve (see term $H(A|VC)$ in Equation~\eqref{eq:uncoded:Delta}), and
auxiliary variable~$U$ can be tuned to maximize $I(A;C|U)-I(A;E|U)$.

The following proposition gives upper bounds on the cardinalities of alphabets $\cU$ and $\cV$. The proof is similar to the one of Proposition~\ref{prop:card} (given in Appendix~\ref{sec:card}) and is therefore omitted.

\begin{proposition}
\label{prop:uncoded:card}
In the single-letter characterization of the rate-distortion-equivocation region~$\cR_\text{uncoded}^*$ given by Theorem~\ref{th:uncoded}, it suffices to consider sets~$\cU$ and~$\cV$ such that 
$\norm{\cU}\leq\norm{\cA}+2$ and 
$\norm{\cV}\leq(\norm{\cA}+2)(\norm{\cA}+1)$.
\end{proposition}

\subsection{Alternative Characterization}

The following proposition can be easily proved from Theorem~\ref{th:uncoded}.

\begin{proposition}
\label{prop:uncoded_bis}
Region $\cR_\text{uncoded}^*$ writes as the closure of the set of all tuples $(R_A,D,\Delta)\in\bR_+^3$ such that there exist random variables $U$, $V$ on some finite sets $\cU$, $\cV$, respectively, and a function $\hat A : \cV\times\cC \to \cA$ such that $U\mkv V\mkv A\mkv (C,E)$ form a Markov chain and
\begin{IEEEeqnarray}{rCl}
R_A		&\geq& \Big[ I(U;C) - I(U;E) \Big]_+ + I(V;A|C)	\ ,
								\label{eq:uncoded_bis:Ra}	\\
D 		&\geq& \bE\big[d(A,\hat A(V,C))\big]			\ ,	\\
\Delta	&\leq& H(A|VC) + I(A;C|U) - I(A;E|U)			\ .
								\label{eq:uncoded_bis:Delta}
\end{IEEEeqnarray}
\end{proposition}

\begin{IEEEproof}
Inequalities~\eqref{eq:uncoded_bis:Ra}--\eqref{eq:uncoded_bis:Delta} yield a smaller region than \eqref{eq:uncoded:Ra}--\eqref{eq:uncoded:Delta}. The achievability of the above proposition thus follows from the one of Theorem~\ref{th:uncoded}.

Notice that the r.h.s. of~\eqref{eq:uncoded:Delta} and~\eqref{eq:uncoded_bis:Delta} writes
\[
H(A|VC) + I(A;C) - I(A;E) - \Big[ I(U;C) - I(U;E) \Big]	\ .
\]
Maximizing this term w.r.t. $U$ thus boils down to minimizing $I(U;C) - I(U;E)$.
In the worst case, setting $U=\emptyset$ makes this term zero, meaning that the optimal choice $U^*$ always leads to $I(U^*;C) - I(U^*;E)\leq 0$, and makes Equations~\eqref{eq:uncoded:Ra} and~\eqref{eq:uncoded_bis:Ra} identical.
\end{IEEEproof}

Proposition~\ref{prop:uncoded_bis}, along with the above proof, indicates that the optimal choice of $U$ is a random variable $U^*$ that can be decoded by Eve.
Since minimizing quantity $I(U;C)-I(U;E)$ w.r.t. $U$ corresponds to looking for a part of $V$ which conveys more information about $E$ than $C$, this \emph{common message} should however give little information to Eve.

\subsection{Special Cases of Interest}

\subsubsection{Lossless secure source coding}

In case of lossless reconstruction of $A$ at Bob, the following corollary directly follows from Theorem~\ref{th:uncoded}.

\begin{corollary}
\label{coro:uncoded:lossless}
In case of lossless reconstruction of $A$ at Bob, region $\cR_\text{uncoded}^*$ reduces to the closure of the set of all tuples $(R_A,D=0,\Delta)\in\bR_+^3$ such that there exists a random variable $U$ on some finite set $\cU$,
such that $U\mkv A\mkv (C,E)$ form a Markov chain and
\begin{IEEEeqnarray*}{rCl}
R_A		&\geq& H(A|C) \ ,\\
\Delta	&\leq& I(A;C|U) - I(A;E|U) \ .
\end{IEEEeqnarray*}
\end{corollary}

\begin{remark}
In case of a noiseless public link of unlimited capacity \emph{i.e.}, $R_A\to\infty$, the authors of~\cite{prabhakaran2007secure} studied the so-called \emph{leakage rate}, defined as $\liminf\frac1n I(A^n; J E^n)$, which equals $H(A)-\Delta$. Their result ``When Bob remains silent''~\cite[Theorem~1]{prabhakaran2007secure} thus follows as a special case of Corollary~\ref{coro:uncoded:lossless}.
\end{remark}

\subsubsection{Bob has less noisy side information than Eve ($C\lessnoisy{A}E$)}

\begin{corollary}
\label{coro:uncoded:lessnoisy}
If Bob has less noisy side information than Eve, then region $\cR_\text{uncoded}^*$ reduces to the closure of the set of all tuples $(R_A,D,\Delta)\in\bR_+^3$ such that there exist a random variable $V$ on some finite set $\cV$, and a function $\hat A : \cV\times\cC \to \cA$ such that $V\mkv A\mkv (C,E)$ form a Markov chain and
\begin{IEEEeqnarray*}{rCl}
R_A		&\geq& I(V;A|C) \ ,\\
D 		&\geq& \bE\big[d(A,\hat A(V,C))\big] \ ,\\
\Delta	&\leq& H(A|VC) + I(A;C) - I(A;E) \ .
\end{IEEEeqnarray*}
\end{corollary}
In this case, random variable~$U$ of Theorem~\ref{th:uncoded} is set to a constant value, and hence Wyner-Ziv coding~\cite{wyner1976rate} achieves the whole region. Also note that, by the Markov condition, the upper bound on the equivocation rate can be written as $H(A|E)-I(V;A|C)$, emphasizing that the reduction of the equivocation at Eve is equivalent to the amount of information $I(V;A|C)$ transmitted by Alice.

\subsubsection{Eve has less noisy side information than Bob ($E\lessnoisy{A}C$)}

\begin{corollary}
If Eve has less noisy side information than Bob, then region $\cR_\text{uncoded}^*$ reduces to the closure of the set of all tuples $(R_A,D,\Delta)\in\bR_+^3$ such that there exist a random variable $V$ on some finite set $\cV$, and a function $\hat A : \cV\times\cC \to \cA$ such that $V\mkv A\mkv (C,E)$ form a Markov chain and
\begin{IEEEeqnarray*}{rCl}
R_A		&\geq& I(V;A|C) \ ,\\
D 		&\geq& \bE\big[d(A,\hat A(V,C))\big] \ ,\\
\Delta	&\leq& H(A|VE) \ .
\end{IEEEeqnarray*}
\end{corollary}

In this case, random variable~$U$ of Theorem~\ref{th:uncoded} is set to $V$, and hence Wyner-Ziv coding~\cite{wyner1976rate} achieves the whole region.
The equivocation rate at Eve corresponds to the case where Eve can reliably decode $V$.
Here, Alice can only exploit the available distortion at Bob to achieve a non-zero 
equivocation rate at Eve.

\section{Secure Distributed Lossless Compression}
\label{sec:lossless}

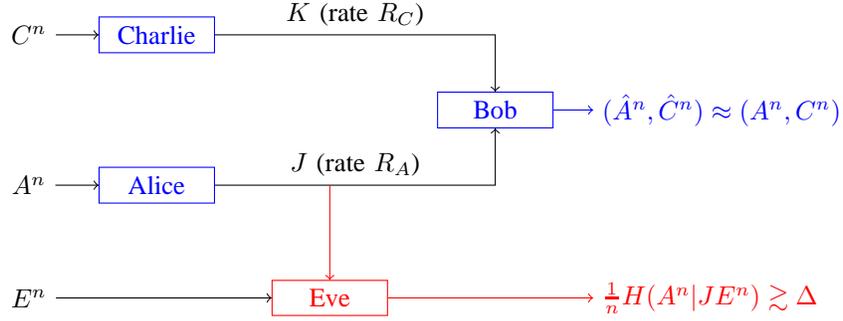
\begin{figure*}[!ht]
\centering
\begin{tikzpicture}
	\tikzstyle{sensor}=[rectangle,draw,text centered,text width=1.3cm]
	\tikzstyle{gentil}=[blue]	
	\tikzstyle{mechant}=[red]	
		
	\node	(A) 		at (0,0) 						{$A^n$};
	\node	(alice)		at (1.7,0)		[sensor,gentil] {Alice};
	\node	(C) 		at (0,2) 						{$C^n$};	
	\node	(charlie)	at (1.7,2)		[sensor,gentil] {Charlie};
	\node	(bob) 		at (6.2,1)		[sensor,gentil]	{Bob};
	\node	(hatA) 		at (7.5,1)		[right,gentil]	{$(\hat A^n,\hat C^n)\approx(A^n,C^n)$};
	\node	(eve) 		at (4,-1.5)		[sensor,mechant]{Eve};
	\node	(E) 		at (0,-1.5)						{$E^n$};
	\node	(D) 		at (7.5,-1.5)	[right,mechant]	{$\frac1n H(A^n|J E^n)\gtrsim\Delta$};
		
	\draw[->]			(A)			to (alice);
	\draw[->]			(alice)		to node[auto]{$J$ (rate $R_A$)} (6.2,0) to (bob);
	\draw[->,mechant]			(4,0)		to (eve);
	
	\draw[->]			(C)			to (charlie);
	\draw[->]			(charlie)	to node[auto]{$K$ (rate $R_C$)} (6.2,2) to (bob);
	
	\draw[->]	(E)			to (eve);
	
	\draw[->,gentil]	(bob)		to (hatA);
	\draw[->,mechant]	(eve)		to (D);
\end{tikzpicture}
\caption{Secure distributed lossless compression.}
\label{fig:lossless:schema}
\end{figure*}

\subsection{Definitions}

In this section, we consider the case where Bob wants to perfectly reconstruct both sources $A$ and $C$, from messages $J$ and $K$ \emph{i.e.}, \emph{distributed lossless compression}, as depicted in Fig.~\ref{fig:lossless:schema}.
We need the following new definitions:

\begin{definition}
An $(n,R_A,R_C)$-code for distributed compression in this setup is defined by
\begin{itemize}
\item An encoding function at Alice denoted by	$f_A : \cA^n \to \{1,\dots,2^{nR_A}\}$,
\item An encoding function at Charlie	 denoted by $f_C : \cC^n \to \{1,\dots,2^{nR_C}\}$,
\item A decoding function at Bob	denoted by	$g : \{1,\dots,2^{nR_A}\}\times\{1,\dots,2^{nR_C}\} \to \cA^n\times\cC^n$.
\end{itemize}
\end{definition}

\begin{definition}
\label{def:lossless:achievable}
A tuple $(R_A,R_C,\Delta)\in\bR_+^3$ is said to be \emph{achievable} if,
for any $\varepsilon>0$, there exists an $(n,R_A+\varepsilon,R_C+\varepsilon)$-code $(f_A,f_C,g)$ such that:
\begin{IEEEeqnarray*}{rCl}
\pr{ g(f_A(A^n),f_C(C^n)) \neq (A^n,C^n) }	&\leq& \varepsilon	 		\ ,\\
\dfrac1n\,H(A^n|f_A(A^n),E^n) 				&\geq& \Delta-\varepsilon 	\ .
\end{IEEEeqnarray*}
The set of all such achievable tuples is denoted by $\cR_\text{lossless}^*$
and is referred to as the \emph{compression-equivocation rates region}.
\end{definition}

\subsection{Optimal Characterization}

In the setup considered in this section, the following theorem provides a single-letter characterization of region $\cR_\text{lossless}^*$. The achievability follows from the one of Points~$(I)$ and $(II)$, choosing auxiliary variables $V=A$ and $W=C$ (see Section~\ref{sec:lossy:regions}).
A new proof is needed for the converse part (see Appendix~\ref{sec:lossless:converse}).

\begin{theorem}
\label{th:lossless}
Region $\cR_\text{lossless}^*$ writes as the closure of the set of all tuples $(R_A,R_C,\Delta)\in\bR_+^3$ such that there exists a random variable $U$ on some finite set $\cU$ verifying the Markov chain $U\mkv A\mkv(C,E)$, and the following inequalities:
\begin{IEEEeqnarray}{rCl}
R_A			&\geq&	H(A|C) 					\ ,\label{eq:lossless:Ra}	\\
R_C			&\geq&	H(C|U) 					\ ,							\\
R_A+R_C		&\geq&	H(AC)					\ ,\label{eq:lossless:RaRc}	\\
\Delta		&\leq&	I(A;C|U) - I(A;E|U) 	\ .\label{eq:lossless:Delta}
\end{IEEEeqnarray}
\end{theorem}

Inequalities~\eqref{eq:lossless:Ra}--\eqref{eq:lossless:RaRc} resemble the ones of Slepian and Wolf~\cite[Section~III]{slepian1973noiseless}. They ensure perfect reconstruction of both variables $A$ and $C$ at Bob.
Depending on the distribution of $(A,C,E)$, variable $U$ can be tuned to allow non-zero equivocation rate at Eve (see Equation~\eqref{eq:lossless:Delta}).
If the side information at Eve $E$ is \emph{less noisy} than $C$ \emph{i.e.}, $E\lessnoisy{A}C$, then setting $U=A$ is optimal, and hence Slepian-Wolf coding achieves the whole region (with $\Delta=0$).

In case of uncoded side information at Bob, Theorem~\ref{th:lossless} directly yields Corollary~\ref{coro:uncoded:lossless} letting $R_C$ tend to infinity.

\begin{remark}
As a matter of fact, the above result refines recent ones~\cite{gunduz2008lossless,liu2010securing} which only provide inner and outer bounds on $\cR_\text{lossless}^*$. 
It should be mentioned here that the outer bound of~\cite[Chapter~8]{liu2010securing}, \cite{gunduz2008lossless,gunduz2008secure} is incorrect.
We use \cite{gunduz2008lossless} as the main reference, but comments below also apply to \cite[Chapter~8]{liu2010securing} and~\cite{gunduz2008secure} as well.
In~\cite{gunduz2008lossless}, Equation~(5) writes $\Delta \geq [H(A|E)-R_A]_+$, meaning that points with $\Delta=0$ are not always included in the considered region, while zero equivocation rate is achievable by any coding scheme.
This inequality can thus not be proved in the converse part.
In fact, Equation~(29) is derived using $H(A^N|E^N,J)\leq\Delta$, while only the reverse inequality holds.
\end{remark}

\subsection{Alternative Characterization}

As in Section~\ref{sec:uncoded} for lossy source coding with uncoded side information, here we can also provide an alternative characterization of region $\cR_\text{lossless}^*$.
The achievability follows from the one of Theorem~\ref{th:lossless}.
A new proof is needed for the converse part (see Appendix~\ref{sec:lossless_bis:proof}).

\begin{proposition}
\label{prop:lossless_bis}
Region $\cR_\text{lossless}^*$ writes as the closure of the set of all tuples $(R_A,R_C,\Delta)\in\bR_+^3$ such that there exists a random variable $U$ on some finite set $\cU$ s.t. $U\mkv A\mkv (C,E)$ form a Markov chain and
\begin{IEEEeqnarray}{rCl}
R_A			&\geq&	\big[ I(U;C) - I(U;E) \big]_+ + H(A|C)	\ ,\\
R_C			&\geq&	H(C|U) 									\ ,\\
R_A+R_C		&\geq&	H(AC)									\ ,\\
\Delta		&\leq&	I(A;C|U) - I(A;E|U) 					\ .
\end{IEEEeqnarray}
\end{proposition}
 
This new single-letter characterization means that \emph{giving $U$ to Eve is also optimal}. The corresponding additional rate $[ I(U;C) - I(U;E)]_+$ does not lead to a lower equivocation at Eve.
This should be considered with reference to known results on the wiretap channel~\cite{csiszar1978broadcast,liang2009information}, where the so called \emph{common message} can be chosen so that Eve also decodes it, without changing the achievable region.

\section{Application Examples}
\label{sec:examples}

\subsection{Gaussian Sources with Coded Side Information}

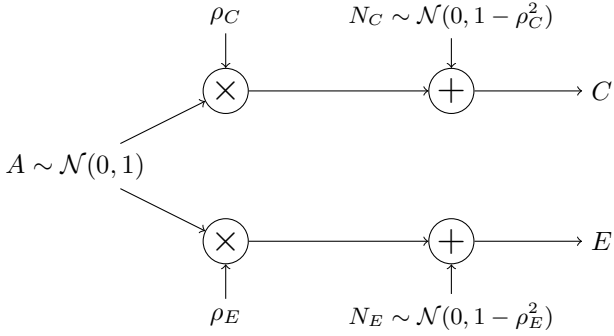
\begin{figure}
\centering
\begin{tikzpicture}
	\node	(A) 	at (0,0) 	{$A\sim\cN(0,1)$};
	\tikzstyle{oper}=[inner sep=0pt,minimum width=.6cm,draw,circle];
	
	\node		(rhoC)	at (2,2)	{$\rho_C$};
	\node[oper]	(multC)	at (2,1)	{\Large $\times$};
	
	\node		(NC)	at (5,2)	{\small $N_C\sim\cN(0,1-\rho_C^2)$};
	\node[oper]	(addC)	at (5,1)	{\Large $+$};
	
	\node		(C) 	at (7,1) 	{$C$};
		
	\node		(rhoE)	at (2,-2)	{$\rho_E$};
	\node[oper]	(multE)	at (2,-1)	{\Large $\times$};
	
	\node		(NE)	at (5,-2)	{\small $N_E\sim\cN(0,1-\rho_E^2)$};
	\node[oper]	(addE)	at (5,-1)	{\Large $+$};
	
	\node		(E) 	at (7,-1) 	{$E$};	
	
	\draw[->] (A) 		to (multC);	
	\draw[->] (rhoC)	to (multC);	
	\draw[->] (multC)	to (addC);	
	\draw[->] (NC) 		to (addC);	
	\draw[->] (addC) 	to (C);	
	
	\draw[->] (A) 		to (multE);	
	\draw[->] (rhoE)	to (multE);	
	\draw[->] (multE)	to (addE);	
	\draw[->] (NE) 		to (addE);	
	\draw[->] (addE) 	to (E);		
\end{tikzpicture}
\caption{A model for Gaussian sources.}
\label{fig:gaussian}
\end{figure}

Consider the source model depicted in Fig.~\ref{fig:gaussian} where the source at Alice is standard Gaussian, and observations at Charlie and Eve are the outputs of independent additive white Gaussian noise (AWGN) channels with input $A$, gains $\rho_C$,  $\rho_E$, and noise powers $(1-\rho_C^2)$, $(1-\rho_E^2)$, resp., for some $0<\rho_C,\rho_E<1$.

Although Theorem~\ref{th:inner_region} is stated and proved for finite alphabet sources, we take the liberty to use its statement, with the appropriate quadratic distortion measure \emph{i.e.}, the Euclidean distance on $\bR$ ($d(a,b) = (a-b)^2$, for each $a,b\in\bR$), as an achievable region also for Gaussian sources (using differential entropy $h(\cdot)$, and considering any equivocation rates $\Delta\in\bR$).
In this setup, the \emph{rates-distortion-equivocation region} is denoted by $\cR_\text{Gaussian}^*$.
Notice that the results should be generalizable to more general cases of continuous-alphabet sources.

Proposition~\ref{prop:gaussian:inner_region} below provides an inner bound on $\cR_\text{Gaussian}^*$ based on the achievability of point~$(I)$ (see Section~\ref{sec:lossy:regions}) with Gaussian auxiliary variables.
This choice is motivated by~\cite[Theorem~1]{oohama1997gaussian} where Oohama proved that it is optimal when only one source is to be estimated within a certain distortion level (with no security constraint).

\begin{proposition}
\label{prop:gaussian:inner_region}
In the Gaussian setup considered in this section, a tuple $(R_A,R_C,D,\Delta)\in\bR_+^2\times\bR_+^*\times\bR$ is achievable if:
\begin{IEEEeqnarray*}{rCl}
R_A		&\geq&	\frac12 \left[ \log\left( \frac{1-\rho_C^2+\rho_C^2\,2^{-2R_C}}{D} \right)\right]_+		\ ,\\
\Delta	&\leq&	\frac12\log\left( 2\pi e(1-\rho_E^2) \right)	\\
		&&\hspace{1.5cm}		- \frac12 \min\Bigg\{	\left[ \log\left( \frac{1-\rho_C^2+\rho_C^2\,2^{-2R_C}}{D} \right)\right]_+ ; \\
		&&\log\left( 1 + (1-\rho_E^2)\left[ \frac1D - \frac1{1-\rho_C^2+\rho_C^2\,2^{-2R_C}} \right]_+ \right)	\Bigg\} \ .
\end{IEEEeqnarray*}
\end{proposition}

Fig.~\ref{fig:gaussian:inner_region} shows a numerical evaluation of the above inner region setting $\rho_C=0.8$, $\rho_E=0.6$ and $D=0.1$.

\begin{IEEEproof}
\setcounter{subsubsection}{0}
Corner point~$(I)$ defines a region $\cR_{(I)}$ given by the following inequalities (see Table~\ref{tab:corner_points} in Section~\ref{sec:lossy:regions}):
\begin{IEEEeqnarray}{rCl}
R_A		&\geq&	I(V;A|W)						\ ,\label{eq:gaussian:RA}	\\
R_C		&\geq&	I(W;C)							\ ,\label{eq:gaussian:RC}	\\
D		&\geq&	\bE\big[d(A,\hat A(V,W))\big]	\ ,\label{eq:gaussian:D}	\\
\Delta	&\leq&	h(A|VW) + I(A;W|U) - I(A;E|U)	\ .\label{eq:gaussian:Delta}
\end{IEEEeqnarray}

For some fixed $R_C\geq0$ and $D>0$, auxiliary random variables $U$, $V$ and $W$ are chosen so that bounds on $R_A$ and $\Delta$ given by Proposition~\ref{prop:gaussian:inner_region} yields a point $(R_A,R_C,D,\Delta)$ in region $\cR_{(I)}$.
More precisely, function $\hat A$ is chosen as the minimum mean square error (MMSE) estimator of $A$ given $V$ and $W$, and auxiliary variables $V$ and $W$ are defined as the outputs of independent AWGN channels with respective inputs $A$ and $C$. 
The gains of these two channels are tuned to meet constraints~\eqref{eq:gaussian:RC} and~\eqref{eq:gaussian:D}, respectively.
Then, since variables $A$, $E$ and $W$ are Gaussian, either $W\lessnoisy{A}E$, or $E\lessnoisy{A}W$.
The upper bound~\eqref{eq:gaussian:Delta} is thus maximized setting $U=\emptyset$, or $U=V$.

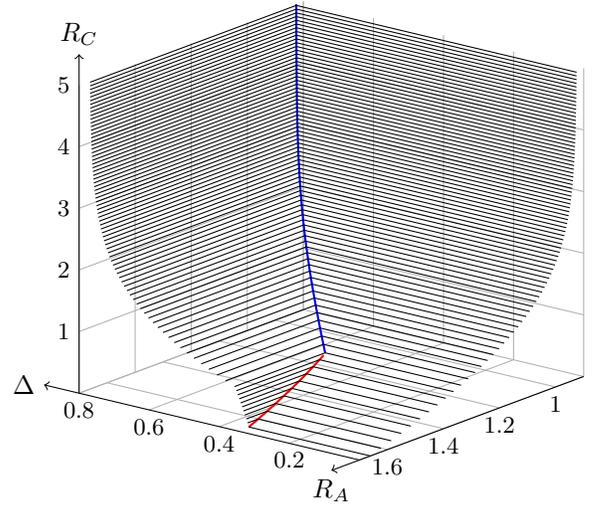
\begin{figure}
\centering
\begin{tikzpicture}[line join=round,]
\draw[draw=black!30](-7.081,-.361)--(-10.062,-1.472);
\draw[draw=black!30](-7.081,.459)--(-10.062,-.652);
\draw[draw=black!30](-7.081,1.279)--(-10.062,.168);
\draw[draw=black!30](-7.081,2.099)--(-10.062,.988);
\draw[draw=black!30](-7.081,2.919)--(-10.062,1.808);
\draw[draw=black!30](-7.081,-.361)--(-7.081,3.739);
\draw[draw=black!30](-7.826,-.639)--(-7.826,3.461);
\draw[draw=black!30](-8.572,-.917)--(-8.572,3.183);
\draw[draw=black!30](-9.317,-1.194)--(-9.317,2.906);
\draw[draw=black!30](-10.062,-1.472)--(-10.062,2.628);
\draw[draw=black!30](-7.081,-.361)--(-10.062,-1.472);
\draw[draw=black!30](-6.149,-.583)--(-9.131,-1.694);
\draw[draw=black!30](-5.217,-.806)--(-8.199,-1.917);
\draw[draw=black!30](-4.286,-1.028)--(-7.267,-2.139);
\draw[draw=black!30](-3.354,-1.25)--(-6.336,-2.361);
\draw[draw=black!30](-3.727,-1.389)--(-7.454,-.5);
\draw[draw=black!30](-4.472,-1.667)--(-8.199,-.778);
\draw[draw=black!30](-5.217,-1.944)--(-8.944,-1.056);
\draw[draw=black!30](-5.963,-2.222)--(-9.69,-1.333);
\draw[draw=black!30](-7.081,-.361)--(-7.081,3.739);
\draw[draw=black!30](-6.149,-.583)--(-6.149,3.517);
\draw[draw=black!30](-5.217,-.806)--(-5.217,3.294);
\draw[draw=black!30](-4.286,-1.028)--(-4.286,3.072);
\draw[draw=black!30](-3.354,-1.25)--(-3.354,2.85);
\draw[draw=black!30](-3.354,-1.25)--(-7.081,-.361);
\draw[draw=black!30](-3.354,-.43)--(-7.081,.459);
\draw[draw=black!30](-3.354,.39)--(-7.081,1.279);
\draw[draw=black!30](-3.354,1.21)--(-7.081,2.099);
\draw[draw=black!30](-3.354,2.03)--(-7.081,2.919);
\tikzstyle{axe} = [arrows=->,line width=.4pt];
\draw[draw=black](-3.53,1.046)--(-7.154,1.91);
\draw[draw=black](-9.814,.919)--(-7.154,1.91);
\draw[draw=blue,thick](-6.79,-.935)--(-6.799,-.897)--(-6.807,-.859)--(-6.815,-.822)--(-6.822,-.785)--(-6.83,-.748)--(-6.837,-.712)--(-6.845,-.676)--(-6.851,-.64)--(-6.859,-.605)--(-6.866,-.569)--(-6.873,-.535)--(-6.879,-.5)--(-6.886,-.466)--(-6.892,-.432)--(-6.898,-.399)--(-6.905,-.366)--(-6.911,-.333)--(-6.917,-.3)--(-6.923,-.268)--(-6.929,-.236)--(-6.934,-.205)--(-6.94,-.173)--(-6.945,-.142)--(-6.951,-.112)--(-6.956,-.081)--(-6.961,-.051)--(-6.966,-.021)--(-6.971,.009)--(-6.976,.038)--(-6.981,.067)--(-6.985,.096)--(-6.99,.125)--(-6.995,.153)--(-6.999,.181)--(-7.003,.209)--(-7.007,.237)--(-7.011,.264)--(-7.015,.291)--(-7.019,.318)--(-7.023,.345)--(-7.027,.371)--(-7.03,.398)--(-7.034,.424)--(-7.037,.449)--(-7.041,.475)--(-7.044,.5)--(-7.047,.526)--(-7.051,.551)--(-7.054,.575)--(-7.057,.6)--(-7.06,.624)--(-7.063,.649)--(-7.066,.673)--(-7.068,.696)--(-7.071,.72)--(-7.074,.744)--(-7.076,.767)--(-7.079,.79)--(-7.081,.813)--(-7.084,.836)--(-7.086,.859)--(-7.088,.881)--(-7.091,.904)--(-7.093,.926)--(-7.095,.948)--(-7.097,.97)--(-7.099,.992)--(-7.101,1.014)--(-7.103,1.035)--(-7.105,1.057)--(-7.107,1.078)--(-7.109,1.099)--(-7.11,1.121)--(-7.112,1.141)--(-7.114,1.162)--(-7.115,1.183)--(-7.117,1.204)--(-7.118,1.224)--(-7.12,1.245)--(-7.121,1.265)--(-7.123,1.285)--(-7.124,1.305)--(-7.126,1.325)--(-7.127,1.345)--(-7.128,1.365)--(-7.129,1.385)--(-7.131,1.405)--(-7.132,1.424)--(-7.133,1.444)--(-7.134,1.463)--(-7.135,1.483)--(-7.136,1.502)--(-7.137,1.521)--(-7.138,1.54)--(-7.139,1.559)--(-7.14,1.578)--(-7.141,1.597)--(-7.142,1.616)--(-7.143,1.635)--(-7.144,1.654)--(-7.145,1.672)--(-7.146,1.691)--(-7.147,1.71)--(-7.147,1.728)--(-7.148,1.746)--(-7.149,1.765)--(-7.149,1.783)--(-7.15,1.802)--(-7.151,1.82)--(-7.152,1.838)--(-7.152,1.856)--(-7.153,1.874)--(-7.153,1.892)--(-7.154,1.91)--(-7.155,1.928)--(-7.155,1.946)--(-7.156,1.964)--(-7.156,1.982)--(-7.157,2)--(-7.157,2.017)--(-7.158,2.035)--(-7.158,2.053)--(-7.159,2.071)--(-7.159,2.088)--(-7.16,2.106)--(-7.16,2.123)--(-7.161,2.141)--(-7.161,2.158)--(-7.161,2.176)--(-7.162,2.193)--(-7.162,2.211)--(-7.162,2.228)--(-7.163,2.245)--(-7.163,2.263)--(-7.164,2.28)--(-7.164,2.297)--(-7.164,2.315)--(-7.165,2.332)--(-7.165,2.349)--(-7.165,2.366)--(-7.165,2.383)--(-7.166,2.401)--(-7.166,2.418)--(-7.166,2.435)--(-7.167,2.452)--(-7.167,2.469)--(-7.167,2.486)--(-7.167,2.503)--(-7.167,2.52)--(-7.168,2.537)--(-7.168,2.554)--(-7.168,2.571)--(-7.168,2.588)--(-7.168,2.605)--(-7.169,2.622)--(-7.169,2.639)--(-7.169,2.656)--(-7.169,2.673)--(-7.169,2.69)--(-7.17,2.706)--(-7.17,2.723)--(-7.17,2.74)--(-7.17,2.757)--(-7.17,2.774)--(-7.17,2.791)--(-7.171,2.807)--(-7.171,2.824)--(-7.171,2.841)--(-7.171,2.858)--(-7.171,2.874)--(-7.171,2.891)--(-7.171,2.908)--(-7.171,2.925)--(-7.172,2.941)--(-7.172,2.958)--(-7.172,2.975)--(-7.172,2.991)--(-7.172,3.008)--(-7.172,3.025)--(-7.172,3.041)--(-7.172,3.058)--(-7.172,3.075)--(-7.172,3.091)--(-7.173,3.108)--(-7.173,3.125)--(-7.173,3.141)--(-7.173,3.158)--(-7.173,3.174)--(-7.173,3.191)--(-7.173,3.208)--(-7.173,3.224)--(-7.173,3.241)--(-7.173,3.257)--(-7.173,3.274)--(-7.173,3.291)--(-7.174,3.307)--(-7.174,3.324)--(-7.174,3.34)--(-7.174,3.357)--(-7.174,3.373)--(-7.174,3.39)--(-7.174,3.407)--(-7.174,3.423)--(-7.174,3.44)--(-7.174,3.456)--(-7.174,3.473)--(-7.174,3.489)--(-7.174,3.506)--(-7.174,3.522)--(-7.174,3.539)--(-7.174,3.555)--(-7.174,3.572)--(-7.174,3.588)--(-7.174,3.605)--(-7.174,3.621)--(-7.174,3.638)--(-7.175,3.654)--(-7.175,3.671)--(-7.175,3.687)--(-7.175,3.704);
\draw[draw=black](-3.538,.994)--(-7.152,1.856);
\draw[draw=black](-9.805,.868)--(-7.152,1.856);
\draw[draw=black](-9.822,.97)--(-7.156,1.964);
\draw[draw=black](-3.523,1.098)--(-7.156,1.964);
\draw[draw=black](-3.546,.942)--(-7.15,1.802);
\draw[draw=black](-9.795,.816)--(-7.15,1.802);
\draw[draw=black](-9.83,1.021)--(-7.157,2.017);
\draw[draw=black](-3.517,1.149)--(-7.157,2.017);
\draw[draw=black](-3.511,1.201)--(-7.159,2.071);
\draw[draw=black](-9.838,1.072)--(-7.159,2.071);
\draw[draw=black](-9.784,.764)--(-7.148,1.746);
\draw[draw=black](-3.554,.889)--(-7.148,1.746);
\draw[draw=black](-3.506,1.252)--(-7.16,2.123);
\draw[draw=black](-9.844,1.123)--(-7.16,2.123);
\draw[draw=black](-9.772,.712)--(-7.146,1.691);
\draw[draw=black](-3.564,.837)--(-7.146,1.691);
\draw[draw=black](-3.501,1.303)--(-7.161,2.176);
\draw[draw=black](-9.85,1.174)--(-7.161,2.176);
\draw[draw=black](-9.759,.66)--(-7.143,1.635);
\draw[draw=black](-3.574,.784)--(-7.143,1.635);
\draw[draw=black](-3.497,1.354)--(-7.162,2.228);
\draw[draw=black](-9.856,1.224)--(-7.162,2.228);
\draw[draw=black](-9.746,.607)--(-7.14,1.578);
\draw[draw=black](-3.585,.73)--(-7.14,1.578);
\draw[draw=black](-3.492,1.404)--(-7.164,2.28);
\draw[draw=black](-9.861,1.275)--(-7.164,2.28);
\draw[draw=black](-9.866,1.325)--(-7.165,2.332);
\draw[draw=black](-3.488,1.455)--(-7.165,2.332);
\draw[draw=black](-3.597,.677)--(-7.137,1.521);
\draw[draw=black](-9.731,.555)--(-7.137,1.521);
\draw[draw=black](-9.871,1.375)--(-7.165,2.383);
\draw[draw=black](-3.485,1.506)--(-7.165,2.383);
\draw[draw=black](-3.61,.623)--(-7.134,1.463);
\draw[draw=black](-9.715,.502)--(-7.134,1.463);
\draw[draw=black](-9.875,1.425)--(-7.166,2.435);
\draw[draw=black](-3.482,1.556)--(-7.166,2.435);
\draw[draw=black](-3.479,1.606)--(-7.167,2.486);
\draw[draw=black](-9.879,1.476)--(-7.167,2.486);
\draw[draw=black](-9.697,.448)--(-7.131,1.405);
\draw[draw=black](-3.624,.568)--(-7.131,1.405);
\draw[draw=black](-3.476,1.657)--(-7.168,2.537);
\draw[draw=black](-9.882,1.526)--(-7.168,2.537);
\draw[draw=black](-9.885,1.576)--(-7.168,2.588);
\draw[draw=black](-3.473,1.707)--(-7.168,2.588);
\draw[draw=black](-3.639,.513)--(-7.127,1.345);
\draw[draw=black](-9.678,.395)--(-7.127,1.345);
\draw[draw=black](-9.888,1.625)--(-7.169,2.639);
\draw[draw=black](-3.471,1.757)--(-7.169,2.639);
\draw[draw=black](-3.469,1.807)--(-7.169,2.69);
\draw[draw=black](-9.891,1.675)--(-7.169,2.69);
\draw[draw=black](-9.658,.341)--(-7.123,1.285);
\draw[draw=black](-3.655,.458)--(-7.123,1.285);
\draw[draw=black](-3.467,1.857)--(-7.17,2.74);
\draw[draw=black](-9.893,1.725)--(-7.17,2.74);
\draw[draw=black](-9.896,1.775)--(-7.17,2.791);
\draw[draw=black](-3.465,1.907)--(-7.17,2.791);
\draw[draw=black](-3.673,.402)--(-7.118,1.224);
\draw[draw=black](-9.636,.286)--(-7.118,1.224);
\draw[draw=black](-9.898,1.825)--(-7.171,2.841);
\draw[draw=black](-3.463,1.957)--(-7.171,2.841);
\draw[draw=black](-3.462,2.006)--(-7.171,2.891);
\draw[draw=black](-9.9,1.874)--(-7.171,2.891);
\draw[draw=black](-9.612,.231)--(-7.114,1.162);
\draw[draw=black](-3.692,.346)--(-7.114,1.162);
\draw[draw=black](-3.46,2.056)--(-7.172,2.941);
\draw[draw=black](-9.902,1.924)--(-7.172,2.941);
\draw[draw=black](-9.903,1.973)--(-7.172,2.991);
\draw[draw=black](-3.459,2.106)--(-7.172,2.991);
\draw[draw=black](-3.712,.289)--(-7.109,1.099);
\draw[draw=black](-9.587,.176)--(-7.109,1.099);
\draw[draw=black](-9.905,2.023)--(-7.172,3.041);
\draw[draw=black](-3.458,2.155)--(-7.172,3.041);
\draw[draw=black](-3.457,2.205)--(-7.172,3.091);
\draw[draw=black](-9.906,2.073)--(-7.172,3.091);
\draw[draw=black](-9.559,.12)--(-7.103,1.035);
\draw[draw=black](-3.734,.232)--(-7.103,1.035);
\draw[draw=black](-3.456,2.255)--(-7.173,3.141);
\draw[draw=black](-9.907,2.122)--(-7.173,3.141);
\draw[draw=black](-9.909,2.172)--(-7.173,3.191);
\draw[draw=black](-3.455,2.304)--(-7.173,3.191);
\draw[draw=black](-3.454,2.354)--(-7.173,3.241);
\draw[draw=black](-9.91,2.221)--(-7.173,3.241);
\draw[draw=black](-9.529,.064)--(-7.097,.97);
\draw[draw=black](-3.758,.174)--(-7.097,.97);
\draw[draw=black](-3.453,2.403)--(-7.173,3.291);
\draw[draw=black](-9.911,2.27)--(-7.173,3.291);
\draw[draw=black](-9.912,2.32)--(-7.174,3.34);
\draw[draw=black](-3.452,2.453)--(-7.174,3.34);
\draw[draw=black](-3.451,2.502)--(-7.174,3.39);
\draw[draw=black](-9.913,2.369)--(-7.174,3.39);
\draw[draw=black](-9.497,.007)--(-7.091,.904);
\draw[draw=black](-3.784,.115)--(-7.091,.904);
\draw[draw=black](-3.451,2.552)--(-7.174,3.44);
\draw[draw=black](-9.913,2.419)--(-7.174,3.44);
\draw[draw=black](-9.914,2.468)--(-7.174,3.489);
\draw[draw=black](-3.45,2.601)--(-7.174,3.489);
\draw[draw=black](-3.45,2.65)--(-7.174,3.539);
\draw[draw=black](-9.915,2.517)--(-7.174,3.539);
\draw[draw=black](-9.463,-.05)--(-7.084,.836);
\draw[draw=black](-3.811,.056)--(-7.084,.836);
\draw[draw=black](-3.449,2.7)--(-7.174,3.588);
\draw[draw=black](-9.915,2.567)--(-7.174,3.588);
\draw[draw=black](-9.916,2.616)--(-7.174,3.638);
\draw[draw=black](-3.449,2.749)--(-7.174,3.638);
\draw[draw=black](-3.448,2.798)--(-7.175,3.687);
\draw[draw=black](-9.916,2.665)--(-7.175,3.687);
\draw[draw=black](-9.426,-.108)--(-7.076,.767);
\draw[draw=black](-3.841,-.005)--(-7.076,.767);
\draw[draw=black](-3.873,-.066)--(-7.068,.696);
\draw[draw=black](-9.386,-.167)--(-7.068,.696);
\draw[draw=black](-9.343,-.227)--(-7.06,.624);
\draw[draw=black](-3.907,-.128)--(-7.06,.624);
\draw[draw=black](-3.944,-.191)--(-7.051,.551);
\draw[draw=black](-9.297,-.287)--(-7.051,.551);
\draw[draw=black](-9.248,-.348)--(-7.041,.475);
\draw[draw=black](-3.983,-.254)--(-7.041,.475);
\draw[draw=black](-4.025,-.319)--(-7.03,.398);
\draw[draw=black](-9.196,-.409)--(-7.03,.398);
\draw[draw=black](-9.139,-.472)--(-7.019,.318);
\draw[draw=black](-4.07,-.385)--(-7.019,.318);
\draw[draw=black](-4.118,-.452)--(-7.007,.237);
\draw[draw=black](-9.079,-.535)--(-7.007,.237);
\draw[draw=black](-9.015,-.6)--(-6.995,.153);
\draw[draw=black](-4.17,-.521)--(-6.995,.153);
\draw[draw=black](-4.224,-.59)--(-6.981,.067);
\draw[draw=black](-8.947,-.665)--(-6.981,.067);
\draw[draw=black](-8.874,-.732)--(-6.966,-.021);
\draw[draw=black](-4.282,-.661)--(-6.966,-.021);
\draw[draw=black](-4.344,-.733)--(-6.951,-.112);
\draw[draw=black](-8.797,-.8)--(-6.951,-.112);
\draw[draw=black](-8.715,-.868)--(-6.934,-.205);
\draw[draw=black](-4.41,-.807)--(-6.934,-.205);
\draw[draw=black](-4.479,-.882)--(-6.917,-.3);
\draw[draw=black](-8.628,-.938)--(-6.917,-.3);
\draw[draw=black](-8.536,-1.009)--(-6.898,-.399);
\draw[draw=black](-4.552,-.959)--(-6.898,-.399);
\draw[axe](-3.354,-1.25)--(-6.708,-2.5);
\draw[draw=black](-8.439,-1.082)--(-6.879,-.5);
\draw[draw=black](-4.63,-1.037)--(-6.879,-.5);
\draw[draw=black](-4.712,-1.117)--(-6.859,-.605);
\draw[draw=black](-8.337,-1.155)--(-6.859,-.605);
\draw[draw=black](-8.229,-1.23)--(-6.837,-.712);
\draw[draw=black](-4.798,-1.198)--(-6.837,-.712);
\draw[axe](-6.336,-2.361)--(-10.528,-1.361);
\draw[draw=black](-4.889,-1.281)--(-6.815,-.822);
\draw[draw=black](-8.116,-1.307)--(-6.815,-.822);
\draw[axe](-10.062,-1.472)--(-10.062,3.038);
\draw[draw=black](-7.996,-1.384)--(-6.79,-.935);
\draw[draw=black](-4.984,-1.366)--(-6.79,-.935);
\draw[draw=red,thick](-7.805,-1.922)--(-7.763,-1.886)--(-7.722,-1.851)--(-7.681,-1.816)--(-7.641,-1.781)--(-7.601,-1.746)--(-7.562,-1.712)--(-7.523,-1.677)--(-7.485,-1.643)--(-7.448,-1.609)--(-7.411,-1.575)--(-7.375,-1.541)--(-7.339,-1.508)--(-7.304,-1.474)--(-7.269,-1.441)--(-7.234,-1.408)--(-7.201,-1.375)--(-7.168,-1.343)--(-7.135,-1.31)--(-7.103,-1.278)--(-7.071,-1.246)--(-7.04,-1.214)--(-7.01,-1.183)--(-6.98,-1.151)--(-6.95,-1.12)--(-6.921,-1.089)--(-6.893,-1.058)--(-6.865,-1.027)--(-6.838,-.997)--(-6.811,-.966);
\draw[draw=black](-5.084,-1.452)--(-6.865,-1.027);
\draw[draw=black](-7.971,-1.439)--(-6.865,-1.027);
\draw[draw=black](-7.952,-1.493)--(-6.95,-1.12);
\draw[draw=black](-5.188,-1.54)--(-6.95,-1.12);
\draw[draw=black](-5.298,-1.63)--(-7.04,-1.214);
\draw[draw=black](-7.933,-1.547)--(-7.04,-1.214);
\draw[draw=black](-7.914,-1.601)--(-7.135,-1.31);
\draw[draw=black](-5.411,-1.721)--(-7.135,-1.31);
\draw[draw=black](-5.53,-1.815)--(-7.234,-1.408);
\draw[draw=black](-7.895,-1.654)--(-7.234,-1.408);
\draw[draw=black](-7.876,-1.708)--(-7.339,-1.508);
\draw[draw=black](-5.653,-1.91)--(-7.339,-1.508);
\draw[draw=black](-5.78,-2.007)--(-7.448,-1.609);
\draw[draw=black](-7.858,-1.762)--(-7.448,-1.609);
\draw[draw=black](-7.84,-1.815)--(-7.562,-1.712);
\draw[draw=black](-5.913,-2.105)--(-7.562,-1.712);
\draw[draw=black](-6.049,-2.205)--(-7.681,-1.816);
\draw[draw=black](-7.822,-1.868)--(-7.681,-1.816);
\draw[draw=black](-7.805,-1.922)--(-7.805,-1.922);
\draw[draw=black](-6.19,-2.307)--(-7.805,-1.922);
\path (-3.727,-1.389) node[below]{\small $1$};  
\path (-4.472,-1.667) node[below]{\small $1.2$};
\path (-5.217,-1.944) node[below]{\small $1.4$};
\path (-5.963,-2.222) node[below]{\small $1.6$};
\path (-10.062,-.652) node[left]{\small $1$};
\path (-10.062,.168) node[left]{\small $2$};
\path (-10.062,.988) node[left]{\small $3$};
\path (-10.062,1.808) node[left]{\small $4$};
\path (-10.062,2.628) node[left]{\small $5$};
\path (-7.267,-2.139) node[below]{\small $0.2$};
\path (-8.199,-1.917) node[below]{\small $0.4$};
\path (-9.131,-1.694) node[below]{\small $0.6$};
\path (-10.062,-1.472) node[below]{\small $0.8$};
\path 	(-6.708,-2.5) node[below]	{$R_A$}
					(-10.062,3.038) node[above]	{$R_C$}
					(-10.528,-1.361) node[left]	{$\Delta$};
\end{tikzpicture}
\caption{Achievable tuples in the quadratic Gaussian case ($\rho_C=0.8$, $\rho_E=0.6$, $D=0.1$).}
\label{fig:gaussian:inner_region}
\end{figure}

\subsubsection{Variable $W$--Rate at Charlie}

We first define $\rho_W\in[0,1)$ by $\rho_W^2 = 1 - 2^{-2R_C}$, and choose random variable $W$ as follows:
\[
W = \rho_W C + N_W	\ ,
\]
where $N_W\sim\cN(0,1-\rho_W^2)$ is an independent random noise.
With these definitions, 
\begin{IEEEeqnarray*}{rCl}
I(W;C)	&=&		\frac12\log\left(\frac1{\Var{C|W}}\right)	\\
		&=&		\frac12\log\left(\frac1{1-\rho_W^2}\right)	\\
		&=&		R_C	\ .
\end{IEEEeqnarray*}

\subsubsection{Variable $V$--Distortion at Bob and Rate at Alice}

We then define $\rho_V\in[0,1)$ by 
\begin{equation}
\label{eq:gaussian:defV}
\rho_V^2 = \left\{
\begin{array}{ll}
\frac{1-(\rho_W\rho_C)^2-D}{1-(\rho_W\rho_C)^2-D(\rho_W\rho_C)^2}
	& \text{ if } 	D < 1-(\rho_W\rho_C)^2	\ ,\\
0	& \text{ otherwise.}
\end{array}
\right.
\end{equation}
and choose random variable $V$ as follows:
\[
W = \rho_V A + N_V	\ ,
\]
where $N_V\sim\cN(0,1-\rho_V^2)$ is an independent random noise.
Note that if large distortion levels are allowed, then Alice will not transmit anything ($V=\emptyset$).

With these definitions, 
\begin{IEEEeqnarray*}{rCl}
\bE\big[d(A,\hat A(V,W))\big] 
	&=&		\Var{A|VW}													\\
	&=&		\frac{(1-\rho_V^2)(1-(\rho_W\rho_C)^2)}{1-(\rho_V\rho_W\rho_C)^2}	\\
	&\leq&	D															\ ,
\end{IEEEeqnarray*}
and
\begin{IEEEeqnarray*}{rCl}
I(V;A|W)	&=&	\frac12\log\left(\frac{\Var{A|W}}{\Var{A|VW}}\right)	\\
			&=&	\frac12\log\left(\frac	{1-(\rho_W\rho_C)^2}
											{\Var{A|VW}}		\right)	\\
			&=&	\frac12	\left[
					\log\left( \frac{1-\rho_C^2+\rho_C^2\,2^{-2R_C}} D \right)
						\right]_+										\ .
\end{IEEEeqnarray*}

\subsubsection{Variable $U$--Equivocation Rate at Eve}

The above rates and distortion level can be achieved with the following equivocation rate, depending on the choice of $U$:
\begin{itemize}
\item If $U=\emptyset$:
\begin{IEEEeqnarray*}{l}
h(A|VW) + I(A;W|U) - I(A;E|U)									\\
\hspace{2.3cm}=	h(A|E) - I(V;A|W) 								\\
\hspace{2.3cm}=	\frac12\log\left(2\pi e(1-\rho_E^2)\right) \\
\hspace{2.3cm}			- \frac12 \left[ \log\left( \frac{1-\rho_C^2+\rho_C^2\,2^{-2R_C}} D \right) \right]_+	\ .
\end{IEEEeqnarray*}

\item If $U=V$:
\begin{IEEEeqnarray*}{l}
h(A|VW) + I(A;W|U) - I(A;E|U)									\\
=	h(A|E) - I(V;A|E)									\\
=	\frac12\log\left(2\pi e(1-\rho_E^2)\right) 
			- \frac12\log\left(\frac{1-(\rho_V\rho_E)^2}{1-\rho_V^2}\right)	\\
=	\frac12\log\left(2\pi e(1-\rho_E^2)\right) \\
			- \frac12\log\left( 1 + (1-\rho_E^2)\left[ \frac1D - \frac1{1-\rho_C^2+\rho_C^2\,2^{-2R_C}} \right]_+ \right)	\ ,
\end{IEEEeqnarray*}
where the last equality follows from definition~\eqref{eq:gaussian:defV} after some straightforward derivations.
\end{itemize}

This proves Proposition~\ref{prop:gaussian:inner_region}.
\end{IEEEproof}

If Eve has no side information \emph{i.e.}, $\rho_E=0$, then the inner bound given by Proposition~\ref{prop:gaussian:inner_region}, and corresponding to Oohama coding~\cite{oohama1997gaussian}, is optimal.

\begin{proposition}
\label{prop:gaussian:optimal}
If $\rho_E=0$, then region $\cR_\text{Gaussian}^*$ reduces to the set of all tuples $(R_A,R_C,D,\Delta)\in\bR_+^2\times\bR_+^*\times\bR$ verifying the following inequalities:
\begin{IEEEeqnarray*}{rCl}
R_A		&\geq&	\frac12 \left[ \log\left( \frac{1-\rho_C^2+\rho_C^2\,2^{-2R_C}}{D} \right)\right]_+		\ ,\\
\Delta	&\leq&	\frac12\log(2\pi e) - \frac12 \left[ \log\left( \frac{1-\rho_C^2+\rho_C^2\,2^{-2R_C}}{D} \right)\right]_+		\ .
\end{IEEEeqnarray*}
\end{proposition}

\begin{IEEEproof}
\setcounter{subsubsection}{0}
The achievability follows from Proposition~\ref{prop:gaussian:inner_region}.
The proof of the converse part follows the argument of~\cite{oohama1997gaussian}.
Let $(R_A,R_C,D,\Delta)\in\cR_\text{Gaussian}^*$ and $\varepsilon>0$.
There exists an $(n,R_A+\varepsilon,R_C+\varepsilon)$-code $(f_A,f_C,g)$ s.t.:
\begin{IEEEeqnarray*}{rCl}
\bE\big[ d\big(A^n,g(f_A(A^n),f_C(C^n)) \big) \big]	
								&\leq& D+\varepsilon \ ,\label{eq:gaussian:distortion}\\
\dfrac1n\,h(A^n|f_A(A^n),E^n) = \dfrac1n\,h(A^n|f_A(A^n))	
								&\geq& \Delta-\varepsilon 	\ .
\end{IEEEeqnarray*}
Denote by $J=f_A(A^n)$ and $K=f_C(C^n)$ the messages transmitted by Alice and Charlie, respectively.

\subsubsection{Rate at Alice}

The rate at Alice verifies the following sequence of inequalities:
\begin{IEEEeqnarray*}{rCl}
n(R_A + \varepsilon)
	&\geq&	H(J)					\\
	&\geq&	I(J;A^n|K)					\\
	&\geq&	h(A^n|K) - h(A^n|JK)	\ .	
\end{IEEEeqnarray*}	

We now study each term of the r.h.s. of the above equation. 
First, note that from the Gaussian distribution of $(A,C)$ and $K=f_C(C^n)$, there exists random variables $N_{A,i}\sim\cN(0,1-\rho_C^2)$, independent of $C^n$ (and hence of $K$) such that $A_i = \rho_C C_i + N_{A,i}$, for each $i\in\{1,\dots,n\}$.
The conditional entropy power inequality (EPI)~\cite{cover2006elements,elgamal2010lecture} thus yields
\begin{IEEEeqnarray}{rCl}
2^{\frac2n h(A^n|K)}
	&\geq&	2^{\frac2n h(\rho_C C^n|K)} + 2^{\frac2n h(N_A^n|K)}\nonumber\\
	&=&		\rho_C^2\,2^{\frac2n h(C^n|K)} + 2\pi e (1-\rho_C^2)\label{eq:EPI}\ .
\end{IEEEeqnarray}

On the other hand, the rate at Charlie can be lower bounded as follows:
\begin{IEEEeqnarray*}{rCl}
n(R_C + \varepsilon)	
	&\geq&	H(K) 						\\
	&=&		I(K;C^n)					\\
	&=&		h(C^n) - h(C^n|K)			\ .
\end{IEEEeqnarray*}

Equation~\eqref{eq:EPI} thus yields
\begin{IEEEeqnarray*}{rCl}
2^{\frac2n h(A^n|K)}
	&\geq&	\rho_C^2\,2^{\frac2n (h(C^n)-n(R_C + \varepsilon))} + 2\pi e (1-\rho_C^2)	\\
	&=&		\rho_C^2\,2\pi e\,2^{-2(R_C + \varepsilon)} + 2\pi e (1-\rho_C^2) \ .
\end{IEEEeqnarray*}

Term $h(A^n|JK)$ can be easily upper bounded:
\begin{IEEEeqnarray*}{rCl}
h(A^n|JK)
	&\stackrel{(a)}{=}&		\sum_{i=1}^n h(A_i|JK A^{i-1})										\\
	&\stackrel{(b)}{\leq}&	\sum_{i=1}^n h(A_i|J,K)												\\
	&\leq&					\sum_{i=1}^n \frac12\log\left( 2\pi e\,\Var{A_i|J,K} \right)		\\
	&\stackrel{(c)}{\leq}&	\sum_{i=1}^n \frac12\log\left( 2\pi e\,\bE{(A_i-g_i(J,K))^2}\right)	\\
	&\stackrel{(d)}{\leq}&	\frac n2\log\left( \frac{2\pi e}n\sum_{i=1}^n\bE{(A_i-g_i(J,K))^2}\right)	\\
	&\stackrel{(e)}{\leq}&	\frac n2\log\left( 2\pi e(D+\varepsilon) \right)	\ ,
\end{IEEEeqnarray*}
where
\begin{itemize}
\item step~$(a)$ follows from the chain rule for conditional entropy,
\item step~$(b)$ from the fact that conditioning reduces the entropy,
\item step~$(c)$ from the fact $\Var{A_i|J,K}$ is the minimum mean square error (over all possible estimators of $A_i$), for each $i\in\{1,\dots,n\}$,
\item step~$(d)$ from the fact that function $\log(\cdot)$ is concave, and Jensen inequality,
\item step~$(e)$ from the distortion constraint~\eqref{eq:gaussian:distortion}. 
\end{itemize}

Putting everything together, we proved that
\begin{IEEEeqnarray*}{rCl}
n(R_A + \varepsilon)
	&\geq&	h(A^n|K) - h(A^n|JK)								\\
	&\geq&	\frac n2\log\left( \rho_C^2\,2\pi e\,2^{-2(R_C + \varepsilon)} + 2\pi e (1-\rho_C^2) \right) \\
			&- &\frac n2\log\left( 2\pi e(D+\varepsilon) \right)	\\
	&=&		\frac n2\log\left( \frac{ 1 - \rho_C^2 + \rho_C^2\,2^{-2(R_C + \varepsilon)}}{D+\varepsilon} \right) \ .
\end{IEEEeqnarray*}

\subsubsection{Equivocation Rate at Eve}

The above argument also provides an upper bound on the equivocation rate:
\begin{IEEEeqnarray*}{rCl}
n(\Delta-\varepsilon)
	&\leq&	h(A^n) - H(J)				  			\\
	&\leq&	\frac n2\log\left( 2\pi e\right)
			- \frac n2\log\left( \frac{ 1 - \rho_C^2 + \rho_C^2\,2^{-2(R_C + \varepsilon)}}{D+\varepsilon} \right) 	\ .
\end{IEEEeqnarray*}

This proves Proposition~\ref{prop:gaussian:optimal}.
\end{IEEEproof}

\begin{remark}
In case of \emph{uncoded} side information at Bob \emph{i.e.}, $R_C\to\infty$, the inner bound provided by Proposition~\ref{prop:gaussian:inner_region} is optimal if $\rho_C\geq\rho_E$ \emph{i.e.}, $C\lessnoisy{A}E$.
The authors conjecture that it also holds if $\rho_C<\rho_E$, while the proof seems more tricky.
\end{remark}

\begin{figure}[ht]
\centering
\begin{tikzpicture}
	\node	(A) 	at (0,.5) 	{$A$};
	\node	(A0) 	at (0,0) 	{$0$};
	\node	(A1)	at (0,-2)	{$1$};
	
	\node	(C) 	at (3,.5) 	{$C$};
	\node	(C0) 	at (3,0) 	{$0$};
	\node	(Ce)	at (3,-1)	{$e$};
	\node	(C1)	at (3,-2)	{$1$};
	
	\node	(E) 	at (-3,.5) 	{$E$};
	\node	(E0) 	at (-3,0) 	{$0$};
	\node	(E1)	at (-3,-2)	{$1$};
	
	\draw[->] (A0) to node[auto,swap]{$1-p$}		(E0);
	\draw[->] (A0) to node[auto,swap]{$p$}			(E1);
	\draw[->] (A1) to node[auto]{$p$}				(E0);
	\draw[->] (A1) to node[auto]{$1-p$}				(E1);
	\draw[->] (A0) to node[auto]{$1-\epsilon$}		(C0);
	\draw[->] (A0) to node[auto,swap]{$\epsilon$}	(Ce);
	\draw[->] (A1) to node[auto]{$\epsilon$}		(Ce);
	\draw[->] (A1) to node[auto,swap]{$1-\epsilon$}	(C1);
\end{tikzpicture}
\caption{Binary source with BEC/BSC side informations.}
\label{fig:binary}
\end{figure}
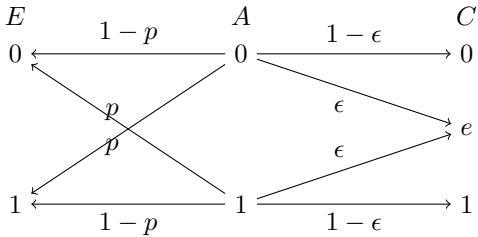
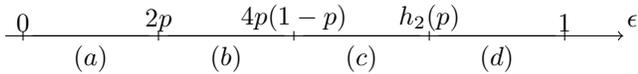
\begin{figure}[ht]
\centering
\begin{tikzpicture}[scale=.9]
	\node	(ci)	at (-.4,0)	{};
	\node	(c0) 	at (0,0) 	{};
	\node	(c1)	at (2,0)	{};
	\node	(c2)	at (4,0)	{};
	\node	(c3)	at (6,0)	{};
	\node	(c4)	at (8,0)	{};
	\node	(e) 	at (9,0) 	{};
	
	\draw[->] (ci) to (e);
	\draw (c0) -- (c1) node[midway,anchor=north] {$(a)$};
	\draw (c1) -- (c2) node[midway,anchor=north] {$(b)$};
	\draw (c2) -- (c3) node[midway,anchor=north] {$(c)$};
	\draw (c3) -- (c4) node[midway,anchor=north] {$(d)$};
	\draw (0,2pt) -- (0,-2pt)	node[anchor=south]	{$0$};
	\draw (2,2pt) -- (2,-2pt)	node[anchor=south]	{$2p$};
	\draw (4,2pt) -- (4,-2pt) 	node[anchor=south] 	{$4p(1-p)$};
	\draw (6,2pt) -- (6,-2pt) 	node[anchor=south] 	{$h_2(p)$};
	\draw (8,2pt) -- (8,-2pt) 	node[anchor=south] 	{$1$};
	\draw (e) node[anchor=south] 	{$\epsilon$};
\end{tikzpicture}
\caption{The different regions as a function of $\epsilon$.}
\label{fig:binary:cas}
\end{figure}

\subsection{Binary Source with (Uncoded) BEC/BSC Side Informations}

Consider the source model depicted in Fig.~\ref{fig:binary} where the source is binary and the side information at Bob, resp. Eve, is the output of a binary erasure channel (BEC) with erasure probability $\epsilon\in[0,1/2]$, resp. a binary symmetric channel (BSC) with crossover probability $p\in[0,1/2]$, with input $A$. 

This model is of interest since neither Bob nor Eve can always  be a lessnoisy decoder for all values of $(p,\epsilon)$.
Let $h_2$ denote the binary entropy function given by $h_2(x) = -x\log(x) -(1-x)\log(1-x)$. According to the values of the parameters $(p,\epsilon)$, it can be shown by means of standard manipulations~\cite{nair2010capacity} that the broadcast channel with input $A$ and outputs $(C,E)$ satisfies the following properties (see Fig.~\ref{fig:binary:cas}):
\begin{enumerate}
\item[(a)] $0\leq \epsilon\leq 2p$: The side information $E$ is a stochastically degraded version of $C$,
\item[(b)] $2p\leq \epsilon\leq 4p(1-p)$: The side information $C$ is less noisy than $E$ \emph{i.e.}, $C\lessnoisy{A}E$,
\item[(c)] $4p(1-p)\leq \epsilon\leq h_2(p)$: The side information $C$ is more capable than $E$, \emph{i.e.}, $I(A;C)\geq I(A;E)$,
\item[(d)] $h_2(p)< \epsilon\leq 1$: Any of the above relations hold between the side informations $C$ and $E$.
\end{enumerate}

Corollary~\ref{coro:uncoded:lessnoisy} thus provides an optimal characterization of the rate-distortion-equivocation region $\cR_\text{uncoded}^*$ when $\epsilon$ lies in region~$(a)$ or $(b)$.
Otherwise, only Theorem~\ref{th:uncoded} applies for the general case and variable $U$ is neither constant nor equal to $V$.

\begin{figure}
\centering
\begin{tikzpicture}
	\node	(A) 	at (0,.5) 	{$A$};
	\node	(A0) 	at (0,0) 	{$0$};
	\node	(A1)	at (0,-2)	{$1$};
		
	\node	(V) 	at (-3,.5) 	{$V$};
	\node	(V0) 	at (-3,0) 	{$0$};
	\node	(V1)	at (-3,-2)	{$1$};
	
	\node	(U) 	at (-6,.5) 	{$U$};
	\node	(U0) 	at (-6,0) 	{$0$};
	\node	(U1)	at (-6,-2)	{$1$};
	
	\draw[->] (A0) to node[auto,swap]	{$1-\alpha$}	(V0);
	\draw[->] (A0) to node[auto,swap]	{$\alpha$}		(V1);
	\draw[->] (A1) to node[auto]		{$\alpha$}		(V0);
	\draw[->] (A1) to node[auto]		{$1-\alpha$}	(V1);
	\draw[->] (V0) to node[auto,swap]	{$1-\beta$}		(U0);
	\draw[->] (V0) to node[auto,swap]	{$\beta$}		(U1);
	\draw[->] (V1) to node[auto]		{$\beta$}		(U0);
	\draw[->] (V1) to node[auto]		{$1-\beta$}		(U1);
\end{tikzpicture}
\caption{Binary auxiliary random variables.}
\label{fig:binary:aux}
\end{figure}
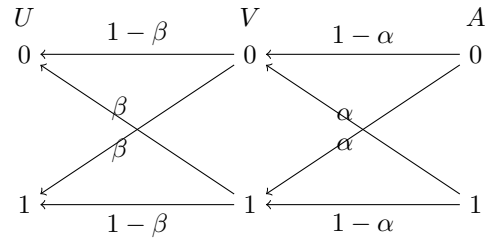

From now on, let the distortion function at Bob $d$ be the Hamming distance and assume for simplicity that the source is uniformly distributed, i.e., $\pr{A=0}=\pr{A=1}=1/2$. We know from the cardinality constraints given in Proposition~\ref{prop:uncoded:card} that it suffices to consider sets $\cU$ and $\cV$  such that $\norm{\cU}\leq 4$ and $\norm{\cV}\leq 12$. As a matter of fact, according to the following proposition, we can restrict our attention to the auxiliary variables $(U,V)$ obtained as the outputs of a degraded binary symmetric broadcast channel with input $A$, as depicted in Fig.~\ref{fig:binary:aux}. Notice that $V$ is identical to the auxiliary variable used  by Wyner and Ziv~\cite{wyner1976rate} for the rate-distortion function of a binary source in the case where there is no eavesdropper.

\begin{proposition}
\label{prop:binary}
In the case considered in this section, region $\cR_\text{uncoded}^*$ reduces to the set of all tuples $(R_A,D,\Delta)\in\bR_+^3$ such that there exist $\alpha,\beta\in[0,1/2]$ satisfying
\begin{IEEEeqnarray*}{rCl}
R_A		&\geq& \epsilon\,(1-h_2(\alpha)) 			\ ,\\
D 		&\geq& \epsilon\,\alpha 						\ ,\\
\Delta	&\leq& \epsilon\,h_2(\alpha) + (1-\epsilon)\,h_2(\alpha\star\beta) - h_2(p\star\alpha\star\beta) + h_2(p) \ .
\end{IEEEeqnarray*}
\end{proposition}

\begin{IEEEproof}
The achievability part of Proposition~\ref{prop:binary} is a direct application of Theorem~\ref{th:uncoded}: define auxiliary random variables $U$ and $V$ as depicted in Fig.~\ref{fig:binary:aux}, and function $\hat A$ on $\cV\times\cC=\{0,1\}\times\{0,e,1\}$ by 
\[
\hat A(v,c) = \left\{
\begin{array}{cl}
c	&	\text{ if } c\neq e	\ ,\\
v	&	\text{ otherwise}	\ .
\end{array}
\right.
\]
Expressions of Proposition~\ref{prop:binary} follow after some straightforward derivations.

The converse part needs more arguments. Let $(R_A,D,\Delta)$ be an achievable tuple. From Theorem~\ref{th:uncoded}, there exist finite sets $\cU$, $\cV$, random variables $U$ on $\cU$, $V$ on $\cV$ and  a function $\hat A : \cV \to \cA$, s.t. $U\mkv V\mkv A\mkv (C,E)$ form a Markov chain and
\begin{IEEEeqnarray*}{rCl}
R_A		&\geq& \epsilon\,I(V;A) 											\ ,\\
D 		&\geq& \epsilon\,\bE\big[d(A,\hat A(V))\big] 					\ ,\\
\Delta	&\leq& \epsilon H(A|V) +(1-\epsilon)H(A|U) - H(E|U) + h_2(p)	\ .
\end{IEEEeqnarray*}
The proof of the above expressions is straightforward, and hence it is omitted here. 
We now prove that there exist $\alpha,\beta \in [0,1/2]$ satisfying the inequalities of Proposition~\ref{prop:binary}:

\begin{table*}[!ht]
\centering
\caption{Some achievable tuples and corresponding parameters for auxiliary random variables ($p=0.1$, $\epsilon=h_2(p)\approx 0.469$).}
\label{tab:numeric}
\begin{IEEEeqnarraybox}[\IEEEeqnarraystrutmode]{x/s/V/t/t/v/t/t/x}
&				&& Secure source coding		& Slepian-Wolf	
				&& Secure source coding		& Wyner-Ziv 	&\\
\IEEEeqnarrayrulerow										\\
&Rate $R$					&& 0.469		& 0.469		
							&& 0.375		& 0.375			&\\
&Distortion $D$				&& 0			& 0			
							&& 0.015		& 0.015			&\\
&Equivocation Rate $\Delta$	&& 0.039		& 0			
							&& 0.133		& 0.126			&\\
\IEEEeqnarrayrulerow										\\
&$\alpha$					&& 0			& 0			
							&& 0.031		& 0.031			&\\
&$\beta$					&& 0.078		& 0			
							&& 0.050		& 0 			&
\end{IEEEeqnarraybox}
\end{table*}

\subsubsection{Rate}

Random variable $A$ is uniformly distributed on $\{0;1\}$, thus:
\begin{IEEEeqnarray*}{rCl}
I(V;A)
	&=& H(A) - H(A|V) \\
	&=& 1 - H(A|V) \ .
\end{IEEEeqnarray*}
Since $0\leq H(A|V)\leq H(A)=1$, and function $h_2$ is a continuous one-to-one mapping from $[0,1/2]$ to $[0,1]$, there exists $\alpha\in[0,1/2]$ such that $H(A|V) = h_2(\alpha)$, and
\[
I(V;A) = 1 - h_2(\alpha) \ .
\]

\subsubsection{Distortion at Bob}

Since distortion~$d$ is the Hamming distance, we can write:
\[
\bE\big[d(A,\hat A(V))\big] = \pr{\hat A(V) \neq A} \ ,
\]
and, from Fano's inequality~\cite{cover2006elements}:
\begin{IEEEeqnarray*}{rCl}
h_2\left(\pr{\hat A(V) \neq A}\right) &+& \pr{\hat A(V) \neq A} \log(\norm{\cA}-1)\\
	\ &\geq&\ H(A|V) \ ,
\end{IEEEeqnarray*}
\emph{i.e.},
\[
h_2\left(\pr{\hat A(V) \neq A}\right) \geq h_2(\alpha)\ .
\]
Function~$h_2$ is increasing on~$[0,1/2]$, and $\alpha\in[0,1/2]$. The last inequality thus implies
\[
\pr{\hat A(V) \neq A} \geq \alpha \ .
\]

\subsubsection{Equivocation Rate at Eve}

Define r.v. $\hat V$ on $\{0,1\}$ as the output of a BSC with crossover probability $\alpha$ and input $A$. Since $A$ is uniformly distributed on~$\{0,1\}$, $A$ is also the output of a BSC with crossover probability $\alpha$ and input $\hat V$. From Mrs. Gerber's lemma~\cite{wyner1973theorem}, we can write, for each $u\in\cU$:
\[
H(A|U=u) = h_2\left(\alpha\star h_2^{-1}\big(H(\hat V|U=u)\big)\right)\ ,
\] 
and hence,
\[
H(A|U) = \sum_{u\in\cU} h_2\left(\alpha\star h_2^{-1}\big(H(\hat V|U=u)\big)\right) p(u) \ .
\]
Following the same argument, since $E$ is the output of a BSC with crossover probability $p$ and input $A$, it is also the output of a BSC with crossover probability $p\star\alpha$ and input $\hat V$, and:
\[
H(E|U) = \sum_{u\in\cU} h_2\left( (p\star\alpha)\star h_2^{-1}\big(H(\hat V|U=u)\big)\right) p(u) \ .
\]
Now, for each $u\in\cU$, $0\leq H(\hat V|U=u)\leq H(\hat V)\leq 1$, and there exists $\beta_u\in[0,1/2]$ such that $H(\hat V|U=u) = h_2(\beta_u)$. Consequently,
\begin{IEEEeqnarray*}{ll}
(1-\epsilon)&H(A|U) - H(E|U)\\
	&= 	\sum_{u\in\cU} \Big[ (1-\epsilon)\,h_2(\alpha\star\beta_u) - h_2(p\star\alpha\star\beta_u) \Big] p(u) \\
	&\leq	(1-\epsilon)\,h_2(\alpha\star\beta) - h_2(p\star\alpha\star\beta) \ ,
\end{IEEEeqnarray*} 
where $\beta=\beta_{u^*}$ for some $u^*\in\cU$.

This proves Proposition~\ref{prop:binary}.
\end{IEEEproof}

\begin{remark}
In this binary case with Hamming distance as distortion measure, an achievable distortion level $D$ is an upper bound on the average bit error rate (BER) at Bob (while estimating $A$):
\[
\bE\big[ d(A^n,g(f(A^n),C^n)) \big] 
	=	\frac1n \sum_{i=1}^n \pr{\hat A_i\neq A_i}	\ ,
\]
where $\hat A_i \triangleq g_i(f_A(A^n),C^n)$ is the $i$-th coordinate of the estimate of $A^n$ at Bob.
At the same time, an achievable equivocation rate $\Delta$ provides a lower bound on the BER at Eve, as shown by the following sequence of inequalities:
\begin{IEEEeqnarray*}{rCl}
\frac1n H(A^n|J E^n)
	&\stackrel{(a)}{\leq}&	\frac1n H(A^n|\breve A^n)							\\
	&\stackrel{(b)}{\leq}&	\frac1n \sum_{i=1}^n H(A_i|\breve A_i)				\\
	&\stackrel{(c)}{=}&		\frac1n \sum_{i=1}^n H(W_i | \breve A_i) + H(A_i | W_i\breve A_i)	\\
	&\stackrel{(d)}{\leq}&	\frac1n \sum_{i=1}^n H(W_i)							\\
	&=&						\frac1n \sum_{i=1}^n h_2 \left(\pr{\breve A_i\neq A_i} \right)		\\
	&\stackrel{(e)}{\leq}&	h_2 \left(\frac1n \sum_{i=1}^n \pr{\breve A_i\neq A_i} \right)\ ,
\end{IEEEeqnarray*}
where 
\begin{itemize}
\item step~$(a)$ holds for any $\breve A^n\in\cA^n$ such that $\breve A^n\mkv (J,E^n)\mkv A^n$ form a Markov chain, 
\item step~$(b)$ follows from the chain rule for conditional entropy and the fact that conditioning reduces the entropy,
\item step~$(c)$ from $W_i \triangleq A_i\oplus\breve A_i$, for each $i\in\{1,\dots,n\}$,
\item step~$(d)$ from identity $A_i = W_i\oplus\breve A_i$ and the fact that conditioning reduces the entropy,
\item step~$(e)$ from the fact that function $h_2$ is concave, and Jensen inequality. 
\end{itemize}
\end{remark}

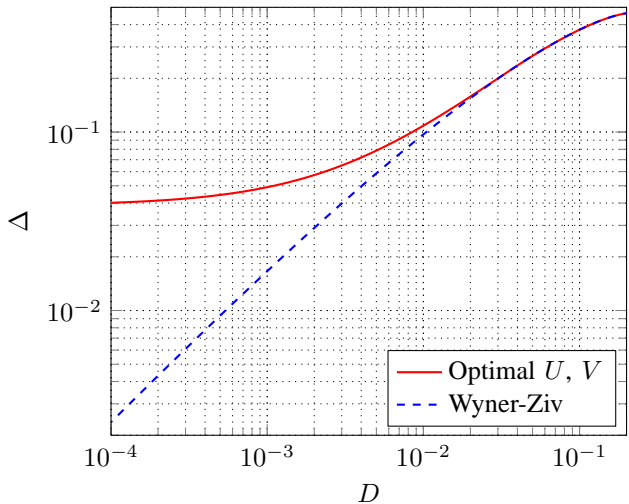
\begin{figure}[t]
\centering
\begin{tikzpicture}

\tikzstyle{optimal}=[color=red,	thick]
\tikzstyle{wyner}=[dashed, color=blue,	thick]

\newcommand{\xmin}{1e-4}
\newcommand{\xmax}{2e-1}
\newcommand{\ymin}{2e-3}
\newcommand{\ymax}{5e-1}

\newcommand{\xlabel}{$D$}
\newcommand{\ylabel}{$\Delta$}

\begin{loglogaxis}[xlabel={\xlabel}, ylabel={\ylabel}, xmin=\xmin, xmax=\xmax, ymin=\ymin, ymax=\ymax, grid=both]

\pgfplotsset{every axis grid/.style={dotted}} 
\pgfplotsset{every axis legend/.append style={
	cells={anchor=west},
	at={(.98,.02)},
	anchor=south east
}}


\addplot[optimal] coordinates{
(9.3799e-005,0.040059)(9.756e-005,0.040105)(0.00010147,0.040153)(0.00010554,0.040203)(0.00010977,0.040254)(0.00011418,0.040307)(0.00011875,0.040361)(0.00012352,0.040418)(0.00012847,0.040477)(0.00013362,0.040538)(0.00013898,0.040601)(0.00014455,0.040666)(0.00015035,0.040733)(0.00015638,0.040803)(0.00016265,0.040875)(0.00016917,0.04095)(0.00017595,0.041028)(0.00018301,0.041108)(0.00019035,0.041191)(0.00019798,0.041276)(0.00020592,0.041365)(0.00021418,0.041457)(0.00022277,0.041552)(0.0002317,0.04165)(0.00024099,0.041752)(0.00025065,0.041857)(0.00026071,0.041966)(0.00027116,0.042079)(0.00028203,0.042196)(0.00029334,0.042316)(0.00030511,0.042441)(0.00031734,0.04257)(0.00033007,0.042704)(0.0003433,0.042842)(0.00035707,0.042985)(0.00037139,0.043133)(0.00038628,0.043286)(0.00040177,0.043444)(0.00041788,0.043608)(0.00043464,0.043777)(0.00045207,0.043952)(0.00047019,0.044133)(0.00048905,0.044321)(0.00050866,0.044514)(0.00052906,0.044715)(0.00055027,0.044922)(0.00057234,0.045136)(0.00059529,0.045357)(0.00061916,0.045586)(0.00064399,0.045823)(0.00066981,0.046068)(0.00069667,0.046321)(0.00072461,0.046582)(0.00075366,0.046853)(0.00078389,0.047132)(0.00081532,0.047421)(0.00084801,0.04772)(0.00088202,0.048029)(0.00091739,0.048348)(0.00095418,0.048678)(0.00099244,0.049019)(0.0010322,0.049371)(0.0010736,0.049735)(0.0011167,0.050111)(0.0011615,0.0505)(0.001208,0.050902)(0.0012565,0.051317)(0.0013069,0.051746)(0.0013593,0.052189)(0.0014138,0.052647)(0.0014705,0.05312)(0.0015294,0.053608)(0.0015908,0.054113)(0.0016545,0.054634)(0.0017209,0.055172)(0.0017899,0.055728)(0.0018617,0.056302)(0.0019363,0.056895)(0.002014,0.057507)(0.0020947,0.05814)(0.0021787,0.058792)(0.0022661,0.059466)(0.002357,0.060162)(0.0024515,0.060881)(0.0025498,0.061622)(0.002652,0.062388)(0.0027584,0.063178)(0.002869,0.063993)(0.002984,0.064835)(0.0031037,0.065703)(0.0032282,0.0666)(0.0033576,0.067524)(0.0034923,0.068479)(0.0036323,0.069463)(0.0037779,0.070479)(0.0039294,0.071526)(0.004087,0.072607)(0.0042509,0.073721)(0.0044214,0.074871)(0.0045987,0.076056)(0.0047831,0.077279)(0.0049749,0.078539)(0.0051744,0.079839)(0.0053818,0.081179)(0.0055977,0.08256)(0.0058221,0.083984)(0.0060556,0.085451)(0.0062984,0.086963)(0.006551,0.088521)(0.0068137,0.090127)(0.0070869,0.091782)(0.0073711,0.093486)(0.0076667,0.095242)(0.0079741,0.09705)(0.0082939,0.098912)(0.0086265,0.10083)(0.0089724,0.1028)(0.0093322,0.10484)(0.0097064,0.10693)(0.010096,0.10909)(0.0105,0.1113)(0.010922,0.11358)(0.011359,0.11593)(0.011815,0.11835)(0.012289,0.12083)(0.012782,0.12339)(0.013294,0.12602)(0.013827,0.12872)(0.014382,0.1315)(0.014958,0.13436)(0.015558,0.13729)(0.016182,0.14031)(0.016831,0.14341)(0.017506,0.14659)(0.018208,0.14986)(0.018938,0.15322)(0.019697,0.15667)(0.020487,0.16021)(0.021309,0.16384)(0.022163,0.16757)(0.023052,0.1714)(0.023976,0.17532)(0.024938,0.17934)(0.025938,0.18347)(0.026978,0.18769)(0.02806,0.19202)(0.029185,0.19646)(0.030355,0.20101)(0.031573,0.20566)(0.032839,0.21042)(0.034155,0.21529)(0.035525,0.22028)(0.03695,0.22537)(0.038431,0.23055)(0.039972,0.23581)(0.041575,0.24113)(0.043242,0.24654)(0.044976,0.25201)(0.04678,0.25755)(0.048656,0.26316)(0.050607,0.26884)(0.052636,0.27458)(0.054747,0.28039)(0.056942,0.28625)(0.059226,0.29217)(0.061601,0.29815)(0.064071,0.30417)(0.06664,0.31024)(0.069312,0.31635)(0.072092,0.3225)(0.074983,0.32868)(0.07799,0.33488)(0.081117,0.34111)(0.08437,0.34735)(0.087753,0.3536)(0.091272,0.35984)(0.094932,0.36608)(0.098739,0.3723)(0.1027,0.37849)(0.10682,0.38464)(0.1111,0.39073)(0.11555,0.39677)(0.12019,0.40272)(0.12501,0.40858)(0.13002,0.41433)(0.13523,0.41996)(0.14066,0.42543)(0.1463,0.43073)(0.15216,0.43583)(0.15827,0.44071)(0.16461,0.44535)(0.17121,0.44969)(0.17808,0.45372)(0.18522,0.45739)(0.19265,0.46065)(0.20037,0.46346)(0.20841,0.46577)(0.21676,0.46751)(0.22546,0.46861)(0.2345,0.469)
};
\addlegendentry{Optimal $U$, $V$};

\addplot[wyner] coordinates{
(9.3799e-005,0.0022391)(9.756e-005,0.0023171)(0.00010147,0.0023977)(0.00010554,0.0024811)(0.00010977,0.0025673)(0.00011418,0.0026565)(0.00011875,0.0027487)(0.00012352,0.0028439)(0.00012847,0.0029425)(0.00013362,0.0030443)(0.00013898,0.0031496)(0.00014455,0.0032584)(0.00015035,0.0033709)(0.00015638,0.0034872)(0.00016265,0.0036074)(0.00016917,0.0037316)(0.00017595,0.00386)(0.00018301,0.0039927)(0.00019035,0.0041298)(0.00019798,0.0042715)(0.00020592,0.0044179)(0.00021418,0.0045692)(0.00022277,0.0047255)(0.0002317,0.004887)(0.00024099,0.0050539)(0.00025065,0.0052263)(0.00026071,0.0054044)(0.00027116,0.0055883)(0.00028203,0.0057784)(0.00029334,0.0059747)(0.00030511,0.0061774)(0.00031734,0.0063868)(0.00033007,0.0066031)(0.0003433,0.0068265)(0.00035707,0.0070571)(0.00037139,0.0072953)(0.00038628,0.0075412)(0.00040177,0.0077952)(0.00041788,0.0080574)(0.00043464,0.008328)(0.00045207,0.0086075)(0.00047019,0.0088959)(0.00048905,0.0091937)(0.00050866,0.009501)(0.00052906,0.0098183)(0.00055027,0.010146)(0.00057234,0.010484)(0.00059529,0.010832)(0.00061916,0.011192)(0.00064399,0.011563)(0.00066981,0.011946)(0.00069667,0.012341)(0.00072461,0.012749)(0.00075366,0.01317)(0.00078389,0.013603)(0.00081532,0.014051)(0.00084801,0.014512)(0.00088202,0.014988)(0.00091739,0.015479)(0.00095418,0.015985)(0.00099244,0.016506)(0.0010322,0.017044)(0.0010736,0.017599)(0.0011167,0.01817)(0.0011615,0.01876)(0.001208,0.019367)(0.0012565,0.019993)(0.0013069,0.020638)(0.0013593,0.021302)(0.0014138,0.021987)(0.0014705,0.022692)(0.0015294,0.023419)(0.0015908,0.024168)(0.0016545,0.024939)(0.0017209,0.025733)(0.0017899,0.026551)(0.0018617,0.027393)(0.0019363,0.02826)(0.002014,0.029153)(0.0020947,0.030072)(0.0021787,0.031018)(0.0022661,0.031991)(0.002357,0.032993)(0.0024515,0.034024)(0.0025498,0.035085)(0.002652,0.036177)(0.0027584,0.0373)(0.002869,0.038454)(0.002984,0.039642)(0.0031037,0.040864)(0.0032282,0.04212)(0.0033576,0.043411)(0.0034923,0.044738)(0.0036323,0.046103)(0.0037779,0.047505)(0.0039294,0.048946)(0.004087,0.050427)(0.0042509,0.051948)(0.0044214,0.053511)(0.0045987,0.055116)(0.0047831,0.056765)(0.0049749,0.058457)(0.0051744,0.060195)(0.0053818,0.061979)(0.0055977,0.06381)(0.0058221,0.065689)(0.0060556,0.067618)(0.0062984,0.069596)(0.006551,0.071625)(0.0068137,0.073706)(0.0070869,0.075841)(0.0073711,0.078029)(0.0076667,0.080273)(0.0079741,0.082572)(0.0082939,0.084928)(0.0086265,0.087343)(0.0089724,0.089817)(0.0093322,0.09235)(0.0097064,0.094945)(0.010096,0.097602)(0.0105,0.10032)(0.010922,0.10311)(0.011359,0.10595)(0.011815,0.10887)(0.012289,0.11185)(0.012782,0.1149)(0.013294,0.11802)(0.013827,0.1212)(0.014382,0.12446)(0.014958,0.12779)(0.015558,0.13119)(0.016182,0.13467)(0.016831,0.13821)(0.017506,0.14183)(0.018208,0.14553)(0.018938,0.1493)(0.019697,0.15315)(0.020487,0.15708)(0.021309,0.16108)(0.022163,0.16516)(0.023052,0.16932)(0.023976,0.17356)(0.024938,0.17787)(0.025938,0.18227)(0.026978,0.18674)(0.02806,0.1913)(0.029185,0.19593)(0.030355,0.20064)(0.031573,0.20543)(0.032839,0.2103)(0.034155,0.21525)(0.035525,0.22027)(0.03695,0.22537)(0.038431,0.23055)(0.039972,0.23581)(0.041575,0.24113)(0.043242,0.24654)(0.044976,0.25201)(0.04678,0.25755)(0.048656,0.26316)(0.050607,0.26884)(0.052636,0.27458)(0.054747,0.28039)(0.056942,0.28625)(0.059226,0.29217)(0.061601,0.29815)(0.064071,0.30417)(0.06664,0.31024)(0.069312,0.31635)(0.072092,0.3225)(0.074983,0.32868)(0.07799,0.33488)(0.081117,0.34111)(0.08437,0.34735)(0.087753,0.3536)(0.091272,0.35984)(0.094932,0.36608)(0.098739,0.3723)(0.1027,0.37849)(0.10682,0.38464)(0.1111,0.39073)(0.11555,0.39677)(0.12019,0.40272)(0.12501,0.40858)(0.13002,0.41433)(0.13523,0.41996)(0.14066,0.42543)(0.1463,0.43073)(0.15216,0.43583)(0.15827,0.44071)(0.16461,0.44535)(0.17121,0.44969)(0.17808,0.45372)(0.18522,0.45739)(0.19265,0.46065)(0.20037,0.46346)(0.20841,0.46577)(0.21676,0.46751)(0.22546,0.46861)(0.2345,0.469)
};
\addlegendentry{Wyner-Ziv};

\end{loglogaxis}
\end{tikzpicture}

\caption{Equivocation rate at Eve as a function of the distortion level at Bob ($p=0.1$, $\epsilon=h_2(p)\approx 0.469$).}
\label{fig:binary:Delta=f(D)}
\end{figure}

\subsubsection*{Numerical evaluation}
Using the inequalities of Proposition~\ref{prop:binary}, we now numerically compute some achievable tuples for $p=0.1$ and $\epsilon=h_2(p)\approx 0.469$ (see Fig.~\ref{fig:binary:Delta=f(D)}).
In case of lossless compression (columns \#1 and \#2 of Table~\ref{tab:numeric}), the auxiliary random variable $V$ is set to $A$ \emph{i.e.}, $\alpha=0$. 
Variable $U$ actually enables a non-zero equivocation level.
Now assume that the coding rate is limited to a maximum of $80\%$ of the required rate for perfect reconstruction of the source (column \#3). 
This induces a distortion of $0.015$ at Bob and an equivocation rate of $0.133\,$bits at Eve. 
Even a small increase in the distortion at Bob can be fully exploited by Alice to achieve very significant gains (more than third times in this case) in terms of equivocation rate at Eve. 
Moreover, for distortion levels higher than $0.036$, Wyner-Ziv coding actually achieves the optimal performance, as shown in Fig.~\ref{fig:binary:Delta=f(D)}.

\section{Summary and Discussions}
\label{sec:summary}

In this paper, we have addressed the general problem of secure lossy source coding with coded side information. Inner and outer bounds on the corresponding achievable region have been derived. 
This setting can be seen as the natural extension of the Berger \emph{et al.} problem~\cite{berger1979upper} by taking the security requirements into account. 
It should be mentioned here that the latter is a fundamental information-theoretic problem for which the best known inner bound is not optimal in general. 
In the same way, our proposed bounds do not match in general, but the achievable inner region turns to be optimal for two cases of particular interest. 
Namely, secure lossy source coding with uncoded side information, and secure distributed lossless compression. 
Interestingly enough, it is proved for both cases that there is no loss in coding to provide a \emph{common} description of the source to both receivers, the legitimate one and the eavesdropper. 
The remaining information is intended to the legitimate receiver and considered at the eavesdropper as ``raw'' bits. 
Furthermore, under certain conditions (\emph{e.g.}, \emph{less noisy}), the standalone Wyner-Ziv (or Slepian-Wolf) coding scheme can achieve the entire region and hence the highest security is guaranteed without additional efforts. 

Application examples to secure lossy source coding of Gaussian and binary sources have been considered. 
The binary model is of interest since neither Bob nor Eve can always be a lessnoisy decoder and thus the encoding strategy needed to achieve the whole region is rather novel. 
In the Gaussian quadratic case, the results by Oohama~\cite{oohama1997gaussian} suggest an inner bound which has been proved to be optimal in some cases. 
A deep analysis along with recent extremal inequalities~\cite{liu2007extremal,rioul2011information} may yield the expected converse. 
However, in the light of known results on Gaussian quadratic multiterminal compression~\cite{oohama1997gaussian,oohama2005rate-distortion,wagner2008rate,tavildar2010gaussian,rahman2010rate}, this might be a tricky problem.

Several possible extensions of this work can be identified. 
First of all, we can think about an extension of the CEO problem~\cite{flynn1987encoding,berger1996ceo} under some security constraints, where the purpose of the legitimate decoder is to estimate a common underlying random variable. 
Recent results~\cite{chen2008successive} indicate that Wyner-Ziv-like coding works well in this setup, and the quadratic Gaussian case has already been solved by Oohama~\cite{oohama1998rate-distortion}. 
Since the quantity of interest is the underlying variable, the secrecy of the system could be measured by the equivocation at the eavesdropper about this variable rather than the observation of one encoder. 

Further extensions could include the introduction of multiple eavesdroppers in order to consider the fact that the encoder cannot reliably know the statistics of the information at the eavesdropper. 
As a matter of fact, if the observations of these multiple eavesdroppers are degraded (or maybe less noisy), as it will be the case with scalar Gaussian variables, then a multi-layer superposition coding scheme may yield a characterization of the equivocation rate at each eavesdropper.

Through this work, error-free rate-limited links were assumed between the encoders and receivers, while noisy channels could provide additional security, as in the traditional wiretap setting. 
A result of optimality for the case of degraded channels and side informations has already been derived~\cite{merhav2006shannon}. 
A comprehensive study of the more general setup of \emph{secure joint source/channel coding} seems promising.

\appendices

\section{Useful Notions and Results}

The appendices below provide basic notions on some concepts used in this paper.

\subsection{Strongly Typical Sequences and Delta-Convention}
\label{sec:typical}

Following~\cite{csiszar1982information}, we use in this paper \emph{strongly typical sets} and the so-called \emph{Delta-Convention}. 
Some useful facts are recalled here.
Let $X$ and $Y$ be random variables on some finite sets $\cX$ and $\cY$, respectively. We denote by $P_{X,Y}$ (resp. $P_{Y|X}$, and $P_X$) the joint probability distribution of $(X,Y)$ (resp. conditional distribution of $Y$ given $X$, and marginal distribution of $X$). 

\begin{definition}
For any sequence $x^n\in\cX^n$ and any symbol $a\in\cX$, notation $N(a|x^n)$ stands for the number of occurrences of $a$ in $x^n$.
\end{definition}

\begin{definition}
A sequence $x^n\in\cX^n$ is called \emph{(strongly) $\delta$-typical} w.r.t. $X$ (or simply \emph{typical} if the context is clear) if
\[
\abs{\frac1n N(a|x^n) - P_X(a)} \leq \delta \ \text{ for each } a\in\cX \ ,
\]
and $N(a|x^n)=0$ for each $a\in\cX$ such that $P_X(a)=0$.
The set of all such sequences is denoted by $\typ{X}$.
\end{definition}

\begin{definition}
Let $x^n\in\cX^n$.
A sequence $y^n\in\cY^n$ is called \emph{(strongly) $\delta$-typical (w.r.t. $Y$) given $x^n$}  if
\[
\abs{\frac1n N(a,b|x^n,y^n) - \frac1n N(a|x^n)P_{Y|X}(b|a)} \leq \delta \  \ ,
\]
for each $a\in\cX, b\in\cY$ and, $N(a,b|x^n,y^n)=0$ for each $a\in\cX$, $b\in\cY$ such that $P_{Y|X}(b|a)=0$.
The set of all such sequences is denoted by $\typ{Y|x^n}$.
\end{definition}

\emph{Delta-Convention~\cite{csiszar1982information}:\ }
For any sets $\cX$, $\cY$, there exists a sequence $\{\delta_n\}_{n\in\bN^*}$ such that lemmas below hold.\footnote{As a matter of fact, $\delta_n\to0$ and $\sqrt{n}\,\delta_n\to\infty$ as $n\to\infty$.}
From now on, typical sequences are understood with $\delta=\delta_n$. 
Typical sets are still denoted by $\typ{\cdot}$.

\begin{lemma}[{\cite[Lemma~1.2.12]{csiszar1982information}}]
There exists a sequence $\eta_n\toas{n\to\infty}0$ such that
\[
P_X(\typ{X}) \geq 1 - \eta_n \ .
\]
\end{lemma}

\begin{lemma}[{\cite[Lemma~1.2.13]{csiszar1982information}}]
\label{lem:cardTyp}
There exists a sequence $\eta_n\toas{n\to\infty}0$ such that, for each $x^n\in\typ{X}$,
\begin{IEEEeqnarray*}{c}
\abs{\frac1n \log \norm{\typ{X}} - H(X)} \leq \eta_n 		\ ,\\
\abs{\frac1n \log \norm{\typ{Y|x^n}} - H(Y|X)} \leq \eta_n	\ .
\end{IEEEeqnarray*}
\end{lemma}

\begin{lemma}[Asymptotic equipartition property]
\label{lem:AEP}
There exists a sequence $\eta_n\toas{n\to\infty}0$ such that, for each $x^n\in\typ{X}$ and each $y^n\in\typ{Y|x^n}$,
\begin{IEEEeqnarray*}{c}
\abs{-\frac1n \log P_X(x^n) - H(X)} \leq \eta_n 				\ ,\\
\abs{-\frac1n \log P_{Y|X}(y^n|x^n) - H(Y|X)} \leq \eta_n		\ .
\end{IEEEeqnarray*}
\end{lemma}

\begin{lemma}[Joint typicality lemma~\cite{elgamal2010lecture}]
\label{lem:jointTypicality}
There exists a sequence $\eta_n\toas{n\to\infty}0$ such that
\[
\abs{-\frac1n \log P_Y(\typ{Y|x^n}) - I(X;Y)} \leq \eta_n \
\ 
\]
for each $x^n\in\typ{X}$.
\end{lemma}

\begin{IEEEproof}
\begin{IEEEeqnarray*}{rCl}
P_Y(\typ{Y|x^n})
	&=&						\sum_{y^n\in\typ{Y|x^n}} P_Y(y^n)				\\
	&\stackrel{(a)}{\leq}&	\norm{\typ{Y|x^n}}\,2^{-n[H(Y)-\alpha_n]}		\\
	&\stackrel{(b)}{\leq}&	2^{n[H(Y|X)+\beta_n]}\,2^{-n[H(Y)-\alpha_n]}	\\
	&=&						2^{-n[I(X;Y)-\beta_n-\alpha_n]}					\ ,
\end{IEEEeqnarray*}
where
\begin{itemize}
\item step~$(a)$ follows from the fact that $\typ{Y|x^n}\subset\typ{Y}$ and Lemma~\ref{lem:AEP}, for some sequence $\alpha_n\toas{n\to\infty}0$,
\item step~$(b)$ from Lemma~\ref{lem:cardTyp}, for some sequence $\beta_n\toas{n\to\infty}0$.
\end{itemize}
The reverse inequality $P_Y(\typ{Y|x^n})\geq 2^{-n[I(X;Y)+\beta_n+\alpha_n]}$ can be proved following similar argument. 
\end{IEEEproof}

\subsection{Graphical Representation of Probability Distributions}
\label{sec:graphical}

Following~\cite[Section~II]{permuter2010two-way}, we use in this paper a technique based on undirected graphs, that provides a sufficient condition for establishing Markov chains from a joint distribution. 
Such a technique for establishing conditional independence was introduced in~\cite{pearl1986fusion} for Bayesian networks, and further generalized to various types of graphs~\cite{kramer2003capacity}.
This paragraph recalls the main points of this technique.

Assume that a sequence of random variables $X^n$ has joint distribution with the following form:
\[
p(x^n) = f_1(x_{\cS_1}) f_2(x_{\cS_2}) \cdots f_k(x_{\cS_k})	\ ,
\]
where, for each $i\in\{1,\dots,k\}$, $\cS_i$ is a subset of $\{1,\dots,n\}$,  notation $x_{\cS_i}$ stands for collection $(x_j)_{j\in\cS_i}$, and $f_i$ is some nonnegative function.

\subsubsection{Drawing the graph}
Draw an undirected graph where all involved random variables \emph{e.g.}, $(X_j)_{j\in\{1,\dots,n\}}$, are nodes.
For each $i\in\{1,\dots,k\}$, draw edges between all the nodes in $X_{\cS_i}$.

\subsubsection{Checking Markov relations}
Let $\cG_1$, $\cG_2$, and $\cG_3$ be three disjoint subsets of $\{1,\dots,n\}$.
If all paths in the graph from a node in $X_{\cG_1}$ to a node in $X_{\cG_3}$ pass through a node in $X_{\cG_2}$, then $X_{\cG_1}\mkv X_{\cG_2}\mkv X_{\cG_3}$ form a Markov chain. 
The proof of this result can be found in~\cite{permuter2010two-way} and is omitted here.

\subsection{Csisz\'ar and K\"orner's Equality}
\label{sec:csiszarkorner}

\begin{lemma}[{Csisz\'ar and K\"orner's equality~\cite[Lemma~7]{csiszar1978broadcast}}]
Consider two i.i.d. sequences $X^n$ and $Y^n$, and a constant $C$.
The following identity holds true:
\[
\sum_{i=1}^n I(Y_{i+1}^n ; X_i | C X^{i-1})
	=	\sum_{j=1}^n I(X^{j-1} ; Y_j | C Y_{j+1}^n)	\ .
\]
\end{lemma}

\begin{IEEEproof}
From the chain rule for conditional mutual information, we can write:
\begin{IEEEeqnarray*}{rCl}
\sum_{i=1}^n I(Y_{i+1}^n ; X_i | C X^{i-1})
	&=& \sum_{i=1}^n \sum_{j=i+1}^n I(Y_j ; X_i | C X^{i-1} Y_{j+1}^n)		\\
	&=&	\sum_{i,j:\ i<j} I(Y_j ; X_i | C X^{i-1} Y_{j+1}^n)					\\
	&=&	\sum_{j=1}^n \sum_{i=1}^{j-1} I(X_i ; Y_j | C X^{i-1} Y_{j+1}^n)	\\
	&=&	\sum_{j=1}^n I(X^{j-1} ; Y_j | C Y_{j+1}^n)							\ .
\end{IEEEeqnarray*}
\end{IEEEproof}

\section{Proof of Theorem~\ref{th:inner_region} (Inner Bound)}
\label{sec:inner_region}

Let $U$, $V$, $W$ be three random variables on finite sets $\cU$, $\cV$, $\cW$, respectively, such that $p(uvwace)=p(u|v)p(v|a)p(w|c)p(ace)$, a function $\hat A : \cV\times\cW \to \cA$, and a tuple $(R_A,R_C,D,\Delta)\in\bR_+^4$.
In this section, we describe a scheme that achieves (under some sufficient conditions) tuple $(R_A,R_C,D,\Delta)$ \emph{i.e.}, for any $\varepsilon>0$, we construct an $(n,R_A+\varepsilon,R_C+\varepsilon)$-code $(f_A,f_C,g)$ such that:
\begin{IEEEeqnarray*}{rCl}
\bE\big[ d\big(A^n,g(f_A(A^n),f_C(C^n))\big) \big]	&\leq& D+\varepsilon 		\ ,\\
\dfrac1n\,H(A^n|f_A(A^n),E^n) 						&\geq& \Delta-\varepsilon 	\ .
\end{IEEEeqnarray*}

In this scheme, Alice (resp. Charlie) transmits to Bob a compressed version $(U,V)$, with $V$ on the top of $U$, (resp. $W$) of $A$ (resp. $C$) using random binning. From the three bin indices, Bob \emph{jointly} decodes variables $U$, $V$ and $W$.

Let $\varepsilon>0$, $R_1,R_2\in\bR_+^*$ such that $R_1+R_2=R_A+\varepsilon$, and $S_1\geq R_1$, $S_2\geq R_2$, $S_C\geq R_C+\varepsilon$.
Define $\gamma = \frac\varepsilon{8\,d_\text{max}}$.

\subsection{Codebook generation at Alice}

Randomly pick $2^{nS_1}$ sequences $u^n(s_1)$ from $\typ{U}$ and divide them into $2^{nR_1}$ equal size bins $\{B_1(r_1)\}_{r_1\in\left\{1,\dots,2^{nR_1}\right\}}$.
Then, for each codeword  $u^n(s_1)$, randomly pick $2^{nS_2}$ sequences $v^n(s_1,s_2)$ from $\typ{V|u^n(s_1)}$ and divide them into $2^{nR_2}$ equal size bins $\{B_2(s_1,r_2)\}_{r_2\in\left\{1,\dots,2^{nR_2}\right\}}$.

\subsection{Codebook generation at Charlie}

Randomly pick $2^{nS_C}$ sequences $w^n(s)$ from $\typ{W}$ and divide them into $2^{n(R_C+\varepsilon)}$ equal size bins $\{B_C(r)\}_{r\in\left\{1,\dots,2^{n(R_C+\varepsilon)}\right\}}$.

\subsection{Encoding at Alice}
\label{sec:encoding}

Assume that sequence $A^n$ is produced at Alice.
Look for the first codeword $u^n(s_1)$ such that $(u^n(s_1),A^n)\in\typ{U,A}$.
Then look for a codeword $v^n(s_1,s_2)$ such that $(v^n(s_1,s_2),A^n)\in\typ{V,A|u^n(s_1)}$.
Let $B_1(r_1)$ and $B_2(s_1,r_2)$ be the bins of $u^n(s_1)$ and $v^n(s_1,s_2)$, respectively.
Alice sends the message $J = f_A(A^n) \triangleq (r_1,r_2)$ on her error-free link.

\subsection{Encoding at Charlie}

Assume that sequence $C^n$ is produced at Charlie.
Look for a codeword $w^n(s)$ such that $(w^n(s),C^n)\in\typ{W,C}$.
Let $B_C(r)$ be the bin of $w^n(s)$.
Charlie sends the message $K = f_C(C^n) \triangleq r$ on his error-free link.

\subsection{Decoding at Bob}

Assume that Bob receives $J=(r_1,r_2)$ from Alice and $K=r$ from Charlie.
Look for the unique \emph{jointly typical} codewords $(u^n,v^n,w^n)$ with bin indices $(r_1,r_2,r)$ \emph{i.e.}, look for the unique indices $(s_1,s_2,s)$ such that $(u^n(s_1),v^n(s_1,s_2),w^n(s)) \in (B_1(r_1)\times B_2(s_1,r_2)\times B_C(r)) \cap \typ{U,V,W}$.
Then compute the estimate $g(J,K)\in\cA^n$ using the component-wise
relation $g_i(J,K) \triangleq \hat A(v_i(s_1,s_2),w_i(s))$ for each $i=\{1,\dots,n\}$.

\subsection{Errors and constraints}
\label{sec:errors}

Denoting by $\err$ the event ``An error occurred during the encoding or decoding steps,'' we expand its probability (averaged over the set of all possible codebooks) as follows: $\pr{\err}\leq P_0+P_{e,1}+P_{e,2}+P_{e,3}+P_d$, where each term corresponds to a particular error event, as detailed below.
We derive sufficient conditions on the parameters that make each of these probabilities small.

\subsubsection{}
From standard properties of typical sequences (see Appendix~\ref{sec:typical}), there exists a sequence $\eta_n\toas{n\to\infty}0$ such that
$P_0 \triangleq \pr{(A^n,C^n,E^n)\not\in\typ{A,C,E}} \leq \eta_n$.
Consequently, $P_0 \leq \gamma$ for some sufficiently large $n$.

\subsubsection{}
In the first encoding step, Alice needs to find (at least) one codeword $u^n(s_1)$ such that $(u^n(s_1),A^n)\in\typ{U,A}$.
The corresponding error probability $P_{e,1}$ writes:
\begin{IEEEeqnarray*}{rCl}
P_{e,1}	&\triangleq& \pr{ \nexists s_1 \text{ s.t. } (u^n(s_1),A^n)\in\typ{U,A} }	\\
	&=& \left( \Pr\Big\{ (U^n,A^n)\notin\typ{U,A}\ \Big|\ U^n\in\typ{U}, \right. \\ 
	&& \hspace{4.4cm}\left.  A^n\in\typ{A} \Big\} \right)^{2^{nS_1}}	\\
	&=& \left( 1- \Pr\Big\{ (U^n,A^n)\in\typ{U,A}\ \Big|\ U^n\in\typ{U}, \right.\\
	&&\hspace{4.4cm} \left. A^n\in\typ{A} \Big\} \right)^{2^{nS_1}}	\\
	&\leq& 2^{ - 2^{nS_1} \pr{ (U^n,A^n)\in\typ{U,A}\ \middle|\ U^n\in\typ{U}, A^n\in\typ{A} } }		\\
	&\leq& 2^{ - 2^{nS_1} 2^{-n(I(U;A)+\eta_n)} } \ ,
\end{IEEEeqnarray*}
for some sequence $\eta_n\toas{n\to\infty}0$ (see Lemma~\ref{lem:jointTypicality} in Appendix~\ref{sec:typical}).
If $S_1>I(U;A)$, then probability $P_{e,1}$ vanishes as $n$ tends to infinity, and hence can be upper bounded by $\gamma$ for some sufficiently large $n$.

Similarly, the second encoding step requires condition $S_2>I(V;A|U)$ to succeed with probability $1-P_{e,2}\geq1-\gamma$.

\subsubsection{}
In his encoding step, Charlie needs to find (at least) one codeword $w^n(s)$ such that $(w^n(s),C^n)\in\typ{W,C}$. Following the above argument, this requires condition $S_C>I(W;C)$ to succeed with probability $1-P_{e,3}\geq1-\gamma$.

\subsubsection{}
The decoding error probability $P_d$ must be carefully handled. An error occurs when the decoded tuple differ from the original one $(s_1,s_2,s)$. There are three meaningful possible events so that $P_d$ writes:\footnote{We denote by $\check s$ the event ``Index $s$ has been correctly decoded'', and $\cancel{\,s\,}$ its complement. Same notation holds for indices $s_1$, $s_2$, and any tuple of indices.}
\begin{IEEEeqnarray*}{rCl}
P_d &\triangleq& 	\pr{ \cancel{(s_1,s_2,s)} }											\\
	&=& 	\pr{	
				\{	\cancel{\,s\,}	\} \cup
				\{	\cancel{s_1},					\check s		\} \cup
				\{	\check s_1,		\cancel{s_2},	\check s		\}  }	\\
	&\leq&	\pr{	\cancel{\,s\,}									}	+
			\pr{	\cancel{s_1},					\check s		} 	+
			\pr{	\check s_1,		\cancel{s_2},	\check s		}	\ .
\end{IEEEeqnarray*}

We now study each term of the r.h.s. of the above equation.

\begin{IEEEeqnarray*}{rCl}
\pr{ \cancel{\,s\,}	}
	&=&		\Pr	\Big\{ \exists\ s_1', s_2', s'\neq s \text{ s.t. } (u^n(s_1'),v^n(s_1',s_2'),w^n(s')) 	\\
	&&\hspace{3mm}	\in (B_1(r_1)\times B_2(s_1',r_2)\times B_C(r)) \cap \typ{U,V,W} \Big\}				\\
	&\leq&	2^{n(S_1-R_1+S_2-R_2+S_C-R_C-\varepsilon)} \times\Pr\Big\{ (U^n,V^n,W^n)\in			\\
	&& \typ{U,V,W}\ \Big|	\ (U^n,V^n)\in\typ{U,V}, W^n\in\typ{W} \Big\}					\\
	&\leq&	2^{n(S_1-R_1+S_2-R_2+S_C-R_C-\varepsilon)}\ 2^{-n(I(UV;W)-\eta_n)}	\\
	&=& 	2^{n(S_1-R_1+S_2-R_2+S_C-R_C-\varepsilon - I(V;W)+\eta_n)}			\ ,
\end{IEEEeqnarray*}
for some sequence $\eta_n\toas{n\to\infty}0$ (see Lemma~\ref{lem:jointTypicality} in Appendix~\ref{sec:typical}).
If $S_1-R_1+S_2-R_2+S_C-R_C-\varepsilon < I(V;W)$, then the above probability vanishes as $n$ tends to infinity, and hence can be upper bounded by $\gamma$ for some sufficiently large $n$.

\begin{IEEEeqnarray*}{lr}
&\pr{ \cancel{s_1},\check s }=	\Pr	\Big\{ \exists\ s_1'\neq s_1, s_2' \text{ s.t. } (u^n(s_1'),v^n(s_1',s_2'),w^n(s)) 	\\
	&\in  (B_1(r_1)\times B_2(s_1',r_2)\times B_C(r)) \cap \typ{U,V,W} \Big\}			\\
	&\leq 2^{n(S_1-R_1+S_2-R_2)} \times\Pr\Big\{ (U^n,V^n,W^n)	\in \typ{U,V,W}\ 				\\
	& \Big|	(U^n,V^n)\in\typ{U,V},  W^n\in\typ{W} \Big\}					\ ,
\end{IEEEeqnarray*}
Following the above argument, if $S_1-R_1+S_2-R_2 < I(V;W)$, then the above probability can be upper bounded by $\gamma$ for some sufficiently large $n$.

\begin{IEEEeqnarray*}{ll}
\pr{ \check s_1,\cancel{s_2},\check s } &=		\Pr	\Big\{ \exists\ s_2'\neq s_2 \text{ s.t. } (u^n(s_1),v^n(s_1,s_2'),	w^n(s))\\
	&	\hspace{-5mm} \in (B_1(r_1)\times B_2(s_1,r_2)\times B_C(r)) \cap \typ{U,V,W} \Big\}				\\
	&\hspace{-5mm}\leq	2^{n(S_2-R_2)} \times\Pr\Big\{ (U^n,V^n,W^n)\in\typ{U,V,W}\\
	& \hspace{-3mm}\Big|\ (U^n,V^n)\in\typ{U,V}, (U^n,W^n)\in\typ{U,W} \Big\}					\\
	&\hspace{-5mm}\leq	2^{n(S_2-R_2)}\ 2^{-n(I(V;W|U)-\eta_n)}		\ ,
\end{IEEEeqnarray*}
for some sequence $\eta_n\toas{n\to\infty}0$.
If $S_2-R_2 < I(V;W|U)$, then the above probability vanishes as $n$ tends to infinity, and hence $\Pr\{\check s_1,\cancel{s_2},\check s\} \leq \gamma$, for some sufficiently large $n$.

\subsubsection{Summary}
In this paragraph, we proved that under some sufficient conditions, $\pr{\err} \leq 7\gamma$.

\subsection{Distortion at Bob}

We now check that our code achieves the required distortion level at Bob (averaged over the set of all possible codebooks):
\begin{IEEEeqnarray*}{ll}
\bE\Big[ d\big(A^n,&g(f_A(A^n),f_C(C^n))\big) \Big]\leq\pr{\err}d_\text{max}+	\\
	&	 (1-\pr{\err}) \bE\left[ d\big(A^n,\hat A\big(v^n(s_1,s_2),w^n(s)\big)\big) \middle| \Nerr \right]  \\
	&\leq		\bE\big[ d(A,\hat A(V,W)) \big] + \frac\varepsilon8 + \frac{7\varepsilon}8 \ ,
\end{IEEEeqnarray*}
where the last inequality holds for some sufficiently large $n$, and follows from $\pr{\err}\leq 7\gamma$, the definition of $\gamma$, and the argument below:
For each $(a^n,v^n,w^n)\in\typ{A,V,W}$,
\begin{IEEEeqnarray*}{ll}
d\big(a^n,&\hat A\big(v^n,w^n\big)\big)=		\frac1n \sum_{i=1}^n d\big(a_i,\hat A\big(v_i,w_i\big)\big)	\\
	&= 	\frac1n \sum_{(a,v,w)\in\cA\times\cV\times\cW} d\big(a,\hat A(v,w)\big) N(a,v,w|a^n,v^n,w^n)	\\
	&= 	\bE\big[ d(A,\hat A(V,W)) \big]	+	\sum_{(a,v,w)\in\cA \times \cV\times\cW} d\big(a,\hat A(v,w)\big) \\
	&	\times \left( \frac1n N(a,v,w|a^n,v^n,w^n) - p(a,v,w) \right)	\\
	&\leq	\bE\big[ d(A,\hat A(V,W)) \big]	+ d_\text{max} \norm{\cA}\norm{\cV}\norm{\cW} \delta_n	\ ,
\end{IEEEeqnarray*}
where the last inequality holds since $(a^n,v^n,w^n)\in\typ{A,V,W}$.
The result follows from the fact that $(A^n,v^n(s_1,s_2),w^n(s))\in\typ{A,V,W}$ when no error occurred, and $\delta_n\toas{n\to\infty}0$ (see the Delta-Convention in Appendix~\ref{sec:typical}). 
For some sufficiently large $n$, $d_\text{max} \norm{\cA}\norm{\cV}\norm{\cW} \delta_n\leq \frac\varepsilon8$.

Condition $D \geq \bE\big[ d(A,\hat A(V,W)) \big]$ is thus sufficient to achieve distortion $D + \varepsilon$ at Bob.

\subsection{Equivocation rate at Eve}
\label{sec:inner_equivocation}

The equivocation rate at Eve (averaged over the set of all possible codebooks) can be lower bounded as follows:
\begin{IEEEeqnarray*}{rCl}
\frac1n\,H(A^n|f_A(A^n),E^n)
	&=& 					\frac1n\,H(A^n|r_1 r_2 E^n)						\\
	&=&						\frac1n\,\Big[ H(A^n|r_1 E^n) - I(A^n;r_2|r_1 E^n)	\Big]	\\
	&\stackrel{(a)}{\geq}&	\frac1n\,\Big[ H(A^n|s_1 E^n) - H(r_2) \Big] 	\\
	&\stackrel{(b)}{\geq}&	H(A|UE) - R_2 - \varepsilon						\ ,
\end{IEEEeqnarray*}
where
\begin{itemize}
\item step~$(a)$ follows from the facts that the bin index $r_1$ is a deterministic function of the codeword index $s_1$, the bin index $r_2$ is a deterministic function of $A^n$, and conditioning reduces the entropy,
\item step~$(b)$ for some sufficiently large $n$, from the fact that the codewords $u^n(s_1)$ are drawn i.i.d. (see Lemma~\ref{lem:HAUE} below), and $r_2\in\{1,\dots,2^{nR_2}\}$.
\end{itemize}

Condition $\Delta \leq H(A|UE) - R_2$ is thus sufficient to achieve equivocation rate  $\Delta - \varepsilon$ at Eve.

\begin{lemma}
\label{lem:HAUE}
The following inequality holds for some sequence $\eta_n\toas{n\to\infty}0$:
\[
H(A^n|s_1 E^n) \geq n(H(A|UE)-\eta_n) \ .
\]
\end{lemma}

\begin{IEEEproof}
Since the codeword index $s_1$ is a deterministic function of $A^n$, term $H(A^n|s_1 E^n)$ writes
\begin{IEEEeqnarray}{rCl}
H(A^n|s_1 E^n) 
	&=& H(A^n E^n|s_1) - H(E^n|s_1) 		\nonumber\\
	&=& H(A^n E^n) - H(s_1) - H(E^n|s_1)	\label{eq:HAsE}\ .
\end{IEEEeqnarray}
We now study each term of the r.h.s. of the above equation.
 
Variables $A_i$, $E_i$ are i.i.d., hence $H(A^n E^n) = nH(AE)$.
 
The second term is studied through the distribution of index $s_1$, using classical argument of typical sequences and random coding. 
From the encoding procedure described in Section~\ref{sec:encoding}, the distribution of $s_1$ writes, for each $j\in\{1,\dots,2^{nS_1}\}$:
\begin{IEEEeqnarray*}{ll}
\pr{s_1=j}	=&\\
	\pr{ (u^n(j),A^n)\in\typ{U,A} \cap  \bigcap_{i=1}^{j-1} \left( (u^n(i),A^n)\notin\typ{U,A} \right)}&	\\
			=	 t_n (1-t_n)^{j-1}		\ ,&
\end{IEEEeqnarray*}
where $$t_n\triangleq \pr{ (U^n,A^n)\in\typ{U,A}\ \Big|\ U^n\in\typ{U}, A^n\in\typ{A} }.$$
The entropy of index $s_1$ thus writes
\[
H(s_1) = - \sum_{j=1}^{2^{nS_1}} t_n(1-t_n)^{j-1} \log(t_n(1-t_n)^{j-1}) - P_{e,1} \log(P_{e,1}) \ ,
\]
where $P_{e,1}$ is the error probability of this encoding step.
From Section~\ref{sec:errors}, if $S_1>I(U;A)$, then probability $P_{e,1}$ vanishes as $n$ tends to infinity.
Since each term $t_n(1-t_n)^{j-1}$ is non-negative and $x\log x\toas{x\to0^+}0^-$, the above entropy can be upper bounded as follows:
\begin{equation}
\label{eq:Hs1}
H(s_1)	\leq	- \sum_{j=1}^{\infty} t_n(1-t_n)^{j-1} \log(t_n(1-t_n)^{j-1}) + \eta^{(1)}_n \ ,
\end{equation}
for some sequence $\eta^{(1)}_n\toas{n\to\infty}0$.
The above series writes 
\begin{multline*}
\sum_{j=1}^{\infty} t_n(1-t_n)^{j-1} \log(t_n(1-t_n)^{j-1}) = t_n\log(t_n) \times	\\
	\sum_{j=1}^{\infty} (1-t_n)^{j-1} + t_n\log(1-t_n) \sum_{j=1}^{\infty} (j-1)(1-t_n)^{j-1}	\ .
\end{multline*}
Equation~\eqref{eq:Hs1} thus yields the following upper bound:
\[
H(s_1)	\leq	- \log(t_n) - \log(1-t_n) \frac{1-t_n}{t_n} + \eta^{(1)}_n	\ .
\]
Now, from standard results on typical sequences (see Appendix~\ref{sec:typical}), $2^{-n(I(U;A)+\eta^{(2)}_n)} \leq t_n \leq 2^{-n(I(U;A)-\eta^{(2)}_n)}$ for some sequence $\eta^{(2)}_n\toas{n\to\infty}0$. 
Since $\frac{\log(1-x)}{x}\toas{x\to0}-1$, this yields
\begin{IEEEeqnarray*}{rCl}
H(s_1)	&\leq&	n( I(U;A)+\eta^{(2)}_n) + 1+\eta^{(3)}_n  + \eta^{(1)}_n	\ ,
\end{IEEEeqnarray*}
for some sequence $\eta^{(3)}_n\toas{n\to\infty}0$. 

The third term can be studied following the argument of~\cite[Section~2.3]{liang2009information}  for the wiretap channel:
\begin{itemize}
\item First, we define the following random variable:
\begin{equation}
\label{eq:hatE}
\hat E^n = \left\{
\begin{array}{ll}
E^n			& \text{if } (E^n, u^n(s_1))\in\typ{E,U} \\
\emptyset	& \text{otherwise}
\end{array}
\right.\ ,
\end{equation}
and write:
\begin{IEEEeqnarray*}{rCl}
\label{eq:HEs1}
\!\!\!\frac1n H(E^n|s_1)	
	&=& 	\frac1n \sum_{j=1}^{2^{nS_1}} H(E^n|s_1=j) \pr{s_1=j} 			\\
	&=&		\frac1n \sum_{j=1}^{2^{nS_1}} H(E^n \hat E^n|s_1=j) \pr{s_1=j} 	\\
	&=&		\frac1n \sum_{j=1}^{2^{nS_1}} \Big( H(\hat E^n|s_1=j) \\
\IEEEeqnarraymulticol{3}{R}{+ H(E^n|\hat E^n, s_1=j) \Big) \pr{s_1=j}\ ,\ 	\yesnumber }
\end{IEEEeqnarray*}
where the second equality follows from the fact that random variable $\hat E^n$ is a deterministic function of $E^n$ and $s_1$.
We now study each term of the r.h.s. of~\eqref{eq:HEs1}.

\item The first term can be upper bounded as follows:
\begin{IEEEeqnarray*}{rCl}
\IEEEeqnarraymulticol{3}{l}{
\frac1n \sum_{j=1}^{2^{nS_1}} H(\hat E^n|s_1=j) \pr{s_1=j} } \\
\qquad	&\stackrel{(a)}{\leq}& 	\frac1n \sum_{j=1}^{2^{nS_1}} \log\left( \norm{\typ{E|u^n(j)}} + 1 \right) \pr{s_1=j} \\
		&\stackrel{(b)}{\leq}& 	\sum_{j=1}^{2^{nS_1}} \left( H(E|U) + \eta^{(4)}_n \right) \pr{s_1=j} \\
		&=& 					H(E|U) + \eta^{(4)}_n \ ,
\end{IEEEeqnarray*}
where 
step~$(a)$ follows from definition~\eqref{eq:hatE}, 
step~$(b)$ from Lemma~\ref{lem:cardTyp} for some sequence $\eta^{(4)}_n\toas{n\to\infty}0$.

\item Fano's inequality~\cite{cover2006elements} yields the following upper bound on the second term of the r.h.s. of Equation~\eqref{eq:HEs1}:
\begin{IEEEeqnarray*}{rCl}
\IEEEeqnarraymulticol{3}{l}{
\frac1n \sum_{j=1}^{2^{nS_1}} H(E^n|\hat E^n, s_1=j) \pr{s_1=j} } \\
\quad	&\leq& 	\frac1n \sum_{j=1}^{2^{nS_1}} \left( 1 + \pr{E^n \neq \hat E^n \middle| s_1=j} \log\norm{\cE^n} \right) \\
		\IEEEeqnarraymulticol{3}{R}{\times \pr{s_1=j} } \\
		&\leq& 	\frac1n + \sum_{j=1}^{2^{nS_1}} \pr{(E^n, u^n(s_1))\notin\typ{E,U} \middle| s_1=j} \\
		\IEEEeqnarraymulticol{3}{R}{\times \log\norm{\cE} \pr{s_1=j} } \\
		&\leq& 	\frac1n + P_0 P_{e,1} \log\norm{\cE} \ .
\end{IEEEeqnarray*}
From Section~\ref{sec:errors}, if $S_1>I(U;A)$, then quantity $P_0 P_{e,1}$ vanishes as $n$ tends to infinity, and the above equation yields
\begin{IEEEeqnarray*}{rCl}
\frac1n \sum_{j=1}^{2^{nS_1}} H(E^n|\hat E^n, s_1=j) \pr{s_1=j}
	&\leq& 	\eta^{(5)}_n \ ,
\end{IEEEeqnarray*}
for some sequence $\eta^{(5)}_n\toas{n\to\infty}0$.

\item Gathering the above inequalities, we proved the following equivalent of Equation~(2.54) of~\cite{liang2009information}:
\[
H(E^n|s_1) \leq n \left( H(E|U) + \eta^{(4)}_n + \eta^{(5)}_n \right) \ .
\]
\end{itemize}

Equation~\eqref{eq:HAsE} along with the above results yields
\begin{multline}
\frac1n H(A^n|s_1 E^n) \geq H(AE) - I(U;A) - \eta^{(2)}_n \\
	- \frac{1+\eta^{(3)}_n + \eta^{(1)}_n}n - H(E|U) - \eta^{(4)}_n - \eta^{(5)}_n	\ .
\end{multline}
Using the Markov chain $U\mkv A\mkv E$, this proves Lemma~\ref{lem:HAUE}.
\end{IEEEproof}

\subsection{End of Proof}

In this section, we proved that sufficient conditions for the achievability of a tuple $(R_A,R_C,D,\Delta)$ are given by the following system of inequalities, for each $\varepsilon>0$:
\[
\left\{
\begin{array}{rcl}
R_1										&>&		0								\\
R_2										&>&		0								\\
R_A+\varepsilon							&=&		R_1 + R_2						\\
R_C										&\geq&	0								\\
S_1										&\geq&	R_1								\\
S_2										&\geq&	R_2								\\
S_C										&\geq&	R_C+\varepsilon					\\
S_1										&>&		I(U;A)							\\
S_2										&>&		I(V;A|U)						\\
S_C										&>&		I(W;C)							\\
S_1-R_1 + S_2-R_2 + S_C-R_C-\varepsilon	&<&		I(V;W)							\\
S_1-R_1 + S_2-R_2 						&<&		I(V,W)							\\
		  S_2-R_2						&<&		I(V;W|U)						\\
D										&\geq&	\bE\big[ d(A,\hat A(V,W)) \big]	\\
\Delta									&\leq&	H(A|UE) - R_2
\end{array}
\right.
\]

Fourier-Motzkin elimination then yields:
\[
\!\left\{
\begin{array}{rcl}
R_A+\varepsilon			&>&		I(V;A|W) 						\\
R_C+\varepsilon			&>&		I(W;C|V) 						\\
R_A+R_C + 2\varepsilon	&>&		I(VW;AC)						\\
D 						&\geq&	\bE\big[d(A,\hat A(V,W))\big] 	\\
\Delta					&<&		H(A|VW) + I(A;W|U) 				\\
						& &			\hspace{3.5cm} - I(A;E|U)	\\	
\Delta-R_C-\varepsilon	&<&		H(A|V)-I(A;E|U)-I(W;C|V)		
\end{array}
\right.
\]
This proves Theorem~\ref{th:inner_region}.
\endproof

\section{Proof of Proposition~\ref{prop:card} (Bounds on the Cardinalities)}
\label{sec:card}

\subsection{Bound on $\norm{\cW}$}

First, note that the single-letter inequalities of Theorem~\ref{th:inner_region} can be written as follows:
\begin{IEEEeqnarray*}{rCl}
R_A			&\geq&	I(V;A|W) 							\ ,\\
R_C			&\geq&	H(C|V) - H(C|VW) 					\ ,\\
R_A+R_C		&\geq&	I(V;A) + H(C|V) - H(C|VW)			\ ,\\
D 			&\geq&	\bE\big[d(A,\hat A(V,W))\big] 		\ ,\\
\Delta		&\leq&	H(A|UE) - I(V;A|W) + I(U;A|W) 		\ ,\\	
\Delta-R_C	&\leq&	H(A|V)-I(A;E|U) - H(C|V) + H(C|VW)	\ .
\end{IEEEeqnarray*}

We then use Fenchel-Eggleston-Carath\'eodory's theorem and follow standard arguments (see~\cite[Appendix C]{elgamal2010lecture}). Consider the following~$\norm{\cC}+3$ continuous functions of~$p(c|w)$:
\begin{IEEEeqnarray*}{l}
p(c|w) 											\ ,\\
I(V;A|W=w) 										\ ,\\
H(C|V,W=w) = H(CV|W=w) - H(V|W=w)				\ ,\\
\bE\big[d(A,\hat A(V,W)) \big| W=w \big]		\ ,\\
I(U;A|W=w)
\end{IEEEeqnarray*}
From Fenchel-Eggleston-Carath\'eodory's theorem, there exists a random variable~$W'$ on~$\cW'$ with $\norm{\cW'}\leq\norm{\cC}+3$ such that $p(c)$, $I(V;A|W)$, $H(C|VW)$, $\bE\big[d(A,\hat A(V,W))\big]$ and $I(U;A|W)$ are preserved.

\subsection{Bounds on $\norm{\cU}$ and $\norm{\cV}$}

We now rewrite the inequalities of Theorem~\ref{th:inner_region} as follows:
\begin{IEEEeqnarray*}{rCl}
R_A			&\geq&	H(A|W) - H(A|VW)				\ ,\\
R_C			&\geq&	I(W;C|V) 						\ ,\\
R_A+R_C		&\geq&	I(W;C) + H(A|W) - H(A|VW)		\ ,\\
D 			&\geq&	\bE\big[d(A,\hat A(V,W))\big] 	\ ,\\
\Delta		&\leq&	H(A|VW) + I(A;W|U) - I(A;E|U) 	\ ,\\	
\Delta-R_C	&\leq&	H(A|V)-I(A;E|U)-I(W;C|V)		\ .
\end{IEEEeqnarray*}

Consider the following~$\norm{\cA}+5$ continuous functions of~$p(v|u)$:
\begin{IEEEeqnarray*}{l}
p(a|u) = \bE\big[ p(a|V) \big| U=u \big]	\ ,\\
H(A|VW,U=u) = H(AVW|U=u) - H(VW|U=u)		\ ,\\
I(W;C|V,U=u) = I(W;C|U=u) - I(W;V|U=u)		\ ,\\
\bE\big[d(A,\hat A(V,W)) \big| U=u \big]	\ ,\\
I(A;W|U=u) 									\ ,\\
I(A;E|U=u) 									\ ,\\
H(A|V,U=u) = H(AV|U=u) - H(V|u=u)			\ .
\end{IEEEeqnarray*}
From Fenchel-Eggleston-Carath\'eodory's theorem, there exists a random variable~$U'$ on~$\cU'$ with $\norm{\cU'}\leq\norm{\cA}+5$ such that $p(a)$, $H(A|VW)$, $I(W;C|V)$, $\bE\big[d(A,\hat A(V,W))\big]$, $I(A;W|U)$, $I(A;E|U)$, and $H(A|V)$ are preserved.

Now, for each $u'\in\cU'$, consider the following~$\norm{\cA}+3$ continuous functions of~$p(a|u',v)$:
\begin{IEEEeqnarray*}{ll}
&p(a|u',v) 											\ ,\\
&H(A|W,U'=u',V=v) = H(AW|U'=u',V=v) \\
&- H(W|U'=u',V=v)	\ ,\\
&I(W;C|U'=u',V=v) 									\ ,\\
&\bE\big[d(A,\hat A(V,W)) \big| U'=u', V=v \big]		\ ,\\
&H(A|U'=u',V=v)										\ .
\end{IEEEeqnarray*}
From Fenchel-Eggleston-Carath\'eodory's theorem, there exists a set $\cV'$ with $\norm{\cV'}\leq\norm{\cA}+3$ and, for each $u'\in\cU'$, a random variable~$V'|\{U'=u'\}$ on~$\cV'$  and a function $\hat A'_{u'}:\cV'\times\cW\to\cA$, such that $p(a|u')$, $H(A|VW,U'=u')$, $I(W;C|V,U'=u')$, $\bE\big[d(A,\hat A(V,W)) \big| U'=u'\big]$, and $H(A|V,U'=u')$ are preserved.

Then define set~$\cV''=\cU'\times\cV'$, random variable $V'' = (U',V')$
and function $\hat A'' : \cV''\times\cW\to\cA$ by 
$\hat A''(v'',w) = \hat A''(u',v',w)\triangleq \hat A'_{u'}(v',w)$.
From the above cardinality bounds, $\norm{\cV''} \leq (\norm{\cA}+5)(\norm{\cA}+3)$.
Note that $U'\mkv V''\mkv A\mkv (C,E)$ form a Markov chain.
From these new definitions and previous constructions, we check that quantities involving variable $V$ are preserved:
\begin{IEEEeqnarray*}{rCl}
H(A|V''W)
	&=& H(A|U'V'W)	\\
	&=& H(A|U'VW)	\\
	&=& H(A|VW)		\ ,
\end{IEEEeqnarray*}
\begin{IEEEeqnarray*}{rCl}
I(W;C|V'') 
	&=& I(W;C|U'V') \\
	&=& I(W;C|U'V) 	\\
	&=& I(W;C|V) 	\ ,
\end{IEEEeqnarray*}
\begin{IEEEeqnarray*}{rCl}
\bE\big[d(A,\hat A''(V'',W))\big]
	&=& \bE\big[d(A,\hat A'_{U'}(V',W))\big] \\
	&=& \bE\Big[\bE\big[d(A,\hat A'_{U'}(V',W)) \big| U'\big] \Big] \\	
	&=& \bE\Big[\bE\big[d(A,\hat A(V,W)) \big| U'\big] \Big] \\
	&=& \bE\big[d(A,\hat A(V,W)) \big] \ ,
\end{IEEEeqnarray*}
and
\begin{IEEEeqnarray*}{rCl}
H(A|V'')	
	&=& H(A|U'V')	\\
	&=& H(A|U'V)	\\
	&=& H(A|V) 		\ .
\end{IEEEeqnarray*}

This proves Proposition~\ref{prop:card}.
\endproof

\section{Proof of Theorem~\ref{th:outer_region} (Outer Bound)}
\label{sec:outer_region}

In this section, we prove Theorem~\ref{th:outer_region}. Let $(R_A,R_C,D,\Delta)$ be an achievable tuple and $\varepsilon>0$. There exists an $(n,R_A+\varepsilon,R_C+\varepsilon)$-code $(f_A,f_C,g)$ s.t.:
\begin{IEEEeqnarray*}{rCl}
\bE\big[ d(A^n,g(f_A(A^n),f_C(C^n))) \big]	&\leq& D+\varepsilon \ ,\\
\dfrac1n\,H(A^n|f_A(A^n),E^n) 				&\geq& \Delta-\varepsilon \ .
\end{IEEEeqnarray*}

\begin{figure}
\centering

\subfloat[]{ 
\begin{tikzpicture}[scale=1.4]	
	\tikzstyle{point}=[circle,fill,inner sep=0pt,minimum size=1.5mm]
	
	\node[point,label=left:$J$]	(J)	at (0,0)	{};
	
	\node[point,label=above:$A_{i+1}^n$]	(A+)at (1,1)	{};
	\node[point,label=above right:$A_i$]	(Ai)at (1,0)	{};
	\node[point,label=below:$A^{i-1}$]		(A-)at (1,-1) 	{};
	
	\node[point,label=above:$(C_{i+1}^n\,E_{i+1}^n)$]	(CE+)at (3,1)	{};
	\node[point,label=above:$(C_i\,E_i)$]				(CEi)at (3,0)	{};
	\node[point,label=below:$(C^{i-1}\,E^{i-1})$]		(CE-)at (3,-1) 	{};
		
	\draw	(J)	to (A+);
	\draw	(J)	to (Ai);
	\draw	(J)	to (A-);
	
	\draw	(A+)to (Ai);
	\draw	(Ai)to (A-);
	\draw	(A+)to [out=-110,in=110] (A-);
	
	\draw	(A+) to (CE+);
	\draw	(Ai) to (CEi);
	\draw	(A-) to (CE-);
\end{tikzpicture}
\label{fig:prob_vect_J}
}
\qquad
\subfloat[]{ 
\begin{tikzpicture}[scale=1.4]	
	\tikzstyle{point}=[circle,fill,inner sep=0pt,minimum size=1.5mm]
		
	\node[point,label=above:$(A_{i+1}^n\,E_{i+1}^n)$]	(AE+)at (0,1)	{};
	\node[point,label=above:$(A_i\,E_i)$]				(AEi)at (0,0)	{};
	\node[point,label=below:$(A^{i-1}\,E^{i-1})$]		(AE-)at (0,-1) 	{};
	
	\node[point,label=above:$C_{i+1}^n$]	(C+)at (2,1)	{};
	\node[point,label=above left:$C_i$]		(Ci)at (2,0)	{};
	\node[point,label=below:$C^{i-1}$]		(C-)at (2,-1) 	{};
	
	\node[point,label=right:$K$]	(K)	at (3,0)	{};
	
	\draw	(AE+) to (C+);
	\draw	(AEi) to (Ci);
	\draw	(AE-) to (C-);
	
	\draw	(C+)to (Ci);
	\draw	(Ci)to (C-);
	\draw	(C+)to [out=-70,in=70] (C-);
	
	\draw	(K)	to (C+);
	\draw	(K)	to (Ci);
	\draw	(K)	to (C-);
\end{tikzpicture}
\label{fig:prob_vect_K}
}
\\
\subfloat[]{ 
\begin{tikzpicture}[scale=1.4]	
	\tikzstyle{point}=[circle,fill,inner sep=0pt,minimum size=1.5mm]
				
	\node[point,label=left:$J$]					(J)	at (0,0)	{};
	
	\node[point,label=below right:$A_{i+1}^n$]	(A+)at (1,1)	{};
	\node[point,label=below right:$A_i$]		(Ai)at (1,0)	{};
	\node[point,label=below right:$A^{i-1}$]	(A-)at (1,-1) 	{};
	
	\node[point,label=below:$E_{i+1}^n$]		(E+)at (2.5,1)	{};
	\node[point,label=below:$E_i$]				(Ei)at (2.5,0)	{};
	\node[point,label=below:$E^{i-1}$]			(E-)at (2.5,-1)	{};
	
	\node[point,label=below left:$C_{i+1}^n$]	(C+)at (4,1)	{};
	\node[point,label=below left:$C_i$]			(Ci)at (4,0)	{};
	\node[point,label=below left:$C^{i-1}$]		(C-)at (4,-1) 	{};
		
	\node[point,label=right:$K$]				(K)	at (5,0)	{};
		
	\draw	(J)	to (A+);
	\draw	(J)	to (Ai);
	\draw	(J)	to (A-);
	
	\draw	(A+)to (Ai);
	\draw	(Ai)to (A-);
	\draw	(A+)to [out=-110,in=110] (A-);
	
	\draw	(A+) to (E+);
	\draw	(Ai) to (Ei);
	\draw	(A-) to (E-);
	
	\draw	(E+) to (C+);
	\draw	(Ei) to (Ci);
	\draw	(E-) to (C-);
		
	\draw	(A+)to [out=20,in=160] (C+);
	\draw	(Ai)to [out=20,in=160] (Ci);
	\draw	(A-)to [out=20,in=160] (C-);
			
	\draw	(C+)to (Ci);
	\draw	(Ci)to (C-);
	\draw	(C+)to [out=-70,in=70] (C-);
	
	\draw	(K)	to (C+);
	\draw	(K)	to (Ci);
	\draw	(K)	to (C-);
\end{tikzpicture}
\label{fig:prob_vect_JK}
}

\caption{
Outer bound--Graphical representation of probability distributions 
(a) $p(j,a^n,c^n,e^n)$, (b) $p(k,a^n,c^n,e^n)$ and (c) $p(j,k,a^n,c^n,e^n)$.
}
\label{fig:prob_vect}
\end{figure}
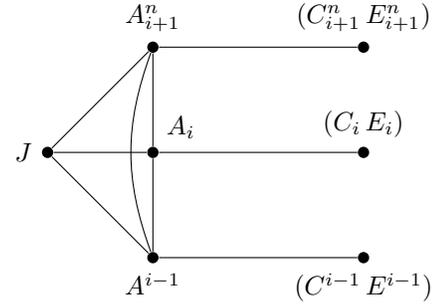
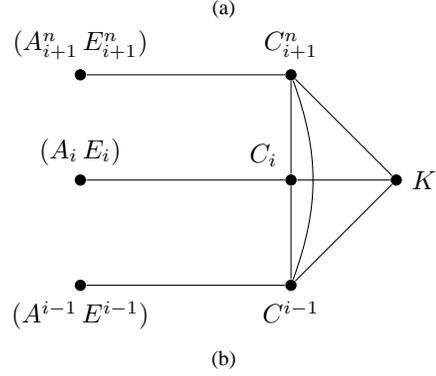
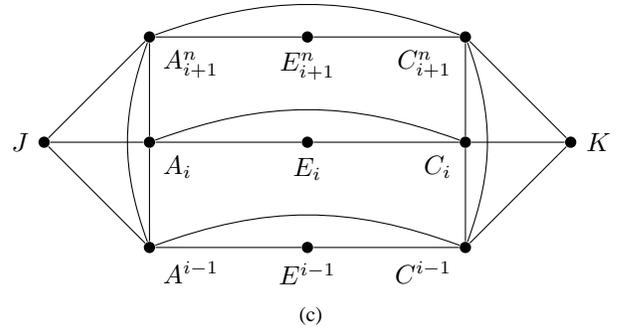

Denote by $J=f_A(A^n)$ and $K=f_C(C^n)$ the messages transmitted by Alice and Charlie, respectively.
From these definitions and the fact that random variables $A_i$, $C_i$, $E_i$ are independent across time, the joint distribution of $(J,K,A^n,C^n,E^n)$ can be written as follows:
\begin{multline*}
p(j,k,a^n,c^n,e^n)=	\ind{f_A(a^n)}(j)\,\ind{f_C(c^n)}(k) \\
\times p(a^{i-1},c^{i-1},e^{i-1})\,p(a_i,c_i,e_i)\,p(a_{i+1}^n,c_{i+1}^n,e_{i+1}^n)	\ .
\end{multline*}
Following the technique described in Appendix~\ref{sec:graphical} and using the above expansion, we can obtain the graphs of Fig.~\ref{fig:prob_vect}.

For each $i\in\{1,\dots,n\}$, define random variables $U_i$, $V_i$ and $W_i$ as follows:
\begin{IEEEeqnarray}{rCl}
U_i &=& (J,A^{i-1}, 			\phantom{C^{i-1},}	E^{i-1})	\label{eq:defU} \ ,\\
V_i &=& (J,A^{i-1}, 			C^{i-1},			E^{i-1})	\label{eq:defV} \ ,\\
W_i &=& (K,\phantom{A^{i-1},}	C^{i-1}\phantom{,	E^{i-1}})	\label{eq:defW} \ .
\end{IEEEeqnarray}
From Fig.~\ref{fig:prob_vect}, $U_i\mkv V_i\mkv A_i\mkv (C_i,E_i)$ and $W_i \mkv C_i \mkv (A_i,E_i)$ form Markov chains (see Appendix~\ref{sec:graphical} for details on this graphical technique for checking Markov relations).

Following the usual technique, we also define an independent random variable $Q$ uniformly distributed over the set $\{1,\dots,n\}$, and $A=A_Q$, $C=C_Q$, $E=E_Q$, $U=(Q,U_Q)$, $V=(Q,V_Q)$, and $W=(Q,W_Q)$.
Note that $U\mkv V\mkv A\mkv (C,E)$ and $W\mkv C\mkv (A,E)$ still form Markov chains, and that $(A,C,E)$ is distributed according to the joint distribution $p(a,c,e)$ \emph{i.e.}, the original distribution of $(A_i,C_i,E_i)$.

\subsection{Rate at Alice}

\begin{IEEEeqnarray*}{rCl}
n(R_A+\varepsilon)
	&\geq& 					H(J) \\
	&\stackrel{(a)}{=}&		I(J ; K A^n C^n E^n) \\
	&\stackrel{(b)}{\geq}&	I(J ; A^n C^n E^n | K) \\
	&\stackrel{(c)}{=}&		\sum_{i=1}^n I(J ; A_i C_i E_i | K A^{i-1} C^{i-1} E^{i-1}) \\
	&=&						\sum_{i=1}^n \Big[ I(J A^{i-1} E^{i-1} ; A_i C_i E_i | K C^{i-1})\\
	& &\hspace{5mm}- I(A^{i-1} E^{i-1} ; A_i C_i E_i | K C^{i-1}) \Big] \\
	&\stackrel{(d)}{=}&		\sum_{i=1}^n I(J A^{i-1} C^{i-1} E^{i-1} ; A_i C_i E_i | K C^{i-1}) \\
	&\stackrel{(e)}{\geq}&	\sum_{i=1}^n I(V_i ; A_i | W_i) \ ,
\end{IEEEeqnarray*}
where
\begin{itemize}
\item step~$(a)$ follows from $J=f_A(A^n)$,
\item step~$(b)$ from the non-negativity of mutual information,
\item step~$(c)$ from the chain rule for conditional mutual information,
\item step~$(d)$ from the Markov chain $(A_i,C_i,E_i)\mkv (K,C^{i-1})\mkv (A^{i-1},E^{i-1})$ (see Fig.~\ref{fig:prob_vect_K}),
\item step~$(e)$ from the non-negativity of mutual information and definitions~\eqref{eq:defV},~\eqref{eq:defW}.
\end{itemize}

Using random variable $Q$, this yields
\begin{IEEEeqnarray*}{rCl}
R_A + \varepsilon
 	&\geq&	\frac1n \sum_{i=1}^n I(V_Q ; A_Q | W_Q, Q=i)	\\
	&=&		I(V_Q ; A_Q | W_Q Q)							\\
	&=&		I(V;A|W) 										\ .
\end{IEEEeqnarray*}

\subsection{Rate at Charlie}

Using similar arguments with $K=f_C(C^n)$, we can obtain:
\begin{IEEEeqnarray*}{rCl}
n(R_C+\varepsilon)
	&\geq& 					H(K) \\
	&\stackrel{(a)}{=}&		I(K ; J A^n C^n E^n) \\
	&\stackrel{(b)}{\geq}&	I(K ; A^n C^n E^n | J) \\
	&\stackrel{(c)}{=}&		\sum_{i=1}^n I(K ; A_i C_i E_i | J A^{i-1} C^{i-1} E^{i-1}) \\
	&=&						\sum_{i=1}^n I(K C^{i-1} ; A_i C_i E_i | J A^{i-1} C^{i-1} E^{i-1}) \\
	&\stackrel{(d)}{\geq}&	\sum_{i=1}^n I(W_i ; C_i | V_i) \ ,
\end{IEEEeqnarray*}
where
\begin{itemize}
\item step~$(a)$ follows from $K=f_C(C^n)$,
\item step~$(b)$ from the non-negativity of mutual information,
\item step~$(c)$ from the chain rule for conditional mutual information,
\item step~$(d)$ from the non-negativity of mutual information and definitions~\eqref{eq:defV},~\eqref{eq:defW}.
\end{itemize}

Then, using auxiliary random variable $Q$,
\begin{IEEEeqnarray*}{rCl}
R_C + \varepsilon
 	&\geq&	\frac1n \sum_{i=1}^n I(W_Q ; C_Q | V_Q, Q=i)	=		I(W ; C | V) 									\ .
\end{IEEEeqnarray*}

\subsection{Sum-rate}

\begin{IEEEeqnarray*}{rCl}
n(R_A+R_C+2\varepsilon)
	&\geq& 					H(JK) 								\\
	&\stackrel{(a)}{=}&		I(JK ; A^n C^n E^n) 				\\
	&\stackrel{(b)}{=}&		\sum_{i=1}^n I(JK ; A_i C_i E_i | A^{i-1} C^{i-1} E^{i-1}) \\
	&=&						\sum_{i=1}^n \Big[ I(JK A^{i-1} C^{i-1} E^{i-1} ; A_i C_i E_i) \\
	&&\hspace{5mm}- I(A^{i-1} C^{i-1} E^{i-1} ; A_i C_i E_i) \Big]	\\
	&\stackrel{(c)}{=}&		\sum_{i=1}^n I(JK A^{i-1} C^{i-1} E^{i-1} ; A_i C_i E_i) \\
	&\stackrel{(d)}{\geq}&	\sum_{i=1}^n I(V_i W_i ; A_i C_i)	\ ,
\end{IEEEeqnarray*}
where
\begin{itemize}
\item step~$(a)$ follows from $J=f_A(A^n)$ and $K=f_C(C^n)$,
\item step~$(b)$ from the chain rule for mutual information,
\item step~$(c)$ from the fact that random variables $A_i$, $C_i$ and $E_i$ are independent across time,
\item step~$(d)$ from the non-negativity of mutual information and definitions~\eqref{eq:defV},~\eqref{eq:defW}.
\end{itemize}

Using random variable $Q$, this yields
\begin{IEEEeqnarray*}{rCl}
R_A+R_C+2\varepsilon
 	&\geq&	\frac1n \sum_{i=1}^n I(V_Q W_Q ; A_Q C_Q | Q=i)	\\
	&=&		I(VW ; AC)	 									\ .
\end{IEEEeqnarray*}

\subsection{Distortion at Bob}

Bob reconstructs $g(J,K)$. For each $i\in\{1,\dots,n\}$, define function $\hat A_i$ as the $i$-th coordinate of this estimate:
\[
\hat A_i(V_i,W_i) \triangleq g_i(J,K) \ .
\]

The component-wise mean distortion at Bob thus verifies
\begin{IEEEeqnarray*}{rCl}
D + \varepsilon
	&\geq&	\bE\big[ d(A^n,g(J,K)) \big] \\
	&=& 	\frac1n \sum_{i=1}^n \bE\left[ d(A_i,\hat A_i(V_i,W_i)) \right] \\
	&=& 	\frac1n \sum_{i=1}^n \bE\left[ d(A_Q,\hat A_Q(V_Q,W_Q))\ \middle|\ Q=i \right] \\
	&=& 	\bE\left[ d(A_Q,\hat A_Q(V_Q,W_Q)) \right] \\
	&=& 	\bE\left[ d(A,\hat A(V,W)) \right] \ ,
\end{IEEEeqnarray*}
where we defined function $\hat A$ by 
\[
\hat A(V,W) = \hat A(Q,V_Q,W_Q) \triangleq \hat A_Q(V_Q,W_Q) \ .
\]

\subsection{Equivocation rate at Eve}

\begin{IEEEeqnarray*}{rCl}
n(\Delta-\varepsilon)
	&\leq&					H(A^n|J E^n) \\
	&=& 					H(A^n|J) - I(A^n ; E^n | J) \\
	&\stackrel{(a)}{=}& 	H(A^n|J) - I(A^n ; E^n) + I(J ; E^n) \\
	&\stackrel{(b)}{=}&		\sum_{i=1}^n \Big[ H(A_i|J A^{i-1}) - I(A_i ; E_i) + I(JE^{i-1} ; E_i) \Big] \\
	&=&						\sum_{i=1}^n \Big[ H(A_i|J K A^{i-1} C^{i-1} E^{i-1}) - I(A_i ; E_i)\\
	&&+ I(A_i;K C^{i-1} E^{i-1} | J A^{i-1})  + I(JE^{i-1} ; E_i) \Big] \\
	&\stackrel{(c)}{\leq}&	\sum_{i=1}^n \Big[ H(A_i|J K A^{i-1} C^{i-1} E^{i-1}) \\
	&&+ I(A_i;K C^{i-1} | J A^{i-1} E^{i-1}) - I(A_i ; E_i) \\
	&&+I(A_i; E^{i-1} |J A^{i-1}) + I(J A^{i-1} E^{i-1} ; E_i) \Big]	\\
\end{IEEEeqnarray*}	
\begin{IEEEeqnarray*}{rCl}	
	&\stackrel{(d)}{=}&		\sum_{i=1}^n \Big[ H(A_i|J K A^{i-1} C^{i-1} E^{i-1})\\
	&& + I(A_i;K C^{i-1} | J A^{i-1} E^{i-1}) - I(A_i ; E_i | J A^{i-1} E^{i-1}) \Big] \\
	&\stackrel{(e)}{=}&		\sum_{i=1}^n \Big[ H(A_i|V_i W_i) + I(A_i;W_i|U_i) - I(A_i;E_i|U_i) \Big] \ ,
\end{IEEEeqnarray*}
where 
\begin{itemize}
\item step~$(a)$ follows from the Markov chain $J\mkv A^n\mkv E^n$ (see Fig.~\ref{fig:prob_vect_J}),
\item step~$(b)$ from the chain rules for conditional entropy and mutual information,
	and the fact that random variables $A_i$ and $E_i$ are independent across time,
\item step~$(c)$ from standard identities and the non-negativity of conditional mutual information,
\item step~$(d)$ from the Markov chain $E_i\mkv A_i\mkv (J A^{i-1})\mkv E^{i-1}$ (see Fig.~\ref{fig:prob_vect_J}),
\item step~$(e)$ from definitions~\eqref{eq:defU},~\eqref{eq:defV} and~\eqref{eq:defW}.
\end{itemize}

Now, using auxiliary random variable $Q$,
\begin{IEEEeqnarray*}{rCl}
\Delta-\varepsilon
	&\leq&	\frac1n \sum_{i=1}^n \Big[ H(A_Q|V_Q W_Q, Q=i) 					\\
	&&+\,I(A_Q;W_Q|U_Q, Q=i) - I(A_Q;E_Q|U_Q, Q=i) \Big]	\\
	&=&	H(A|VW) + I(A;W|U) - I(A;E|U) \ .
\end{IEEEeqnarray*}

\subsection{Public-link secrecy rate}

\begin{IEEEeqnarray*}{rCl}
n(\Delta-R_C-2\varepsilon)
	&\leq&					H(A^n|J E^n) - H(K)		\\
	&\stackrel{(a)}{\leq}&	H(A^n|J E^n) - H(K|J)	\\
	&\stackrel{(b)}{=}&		H(A^n|J E^n) - I(K ; A^n C^n|J)	\\
	&\stackrel{(c)}{=}&		\sum_{i=1}^n \Big[ H(A_i|J A^{i-1} E^n) \\
	&-& I(K ; A_i C_i|J A^{i-1} C^{i-1}) \Big]	\\
	&\stackrel{(d)}{\leq}&	\sum_{i=1}^n \Big[ H(A_i|J A^{i-1} E^i)\\
	&& - I(K ; C_i|J A^{i-1} C^{i-1})	\Big]	\\
	&\stackrel{(e)}{=}&		\sum_{i=1}^n \Big[ H(A_i|J A^{i-1} C^{i-1} E^{i-1}) \\
	&&- I(A_i;E_i|J A^{i-1} E^{i-1}) \\
	&&-I(K C^{i-1}; C_i|J A^{i-1} C^{i-1} E^{i-1}) \Big]	\\
	&\stackrel{(f)}{=}&		\sum_{i=1}^n \Big[ H(A_i|V_i) - I(A_i;E_i|U_i) \\
	&&- I(W_i;C_i|V_i) \Big] \ ,
\end{IEEEeqnarray*}
where 
\begin{itemize}
\item step~$(a)$ follows from the fact that conditioning reduces the entropy,
\item step~$(b)$ from $K=f_C(C^n)$,
\item step~$(c)$ from the chain rules for conditional entropy and conditional mutual information,
\item step~$(d)$ from the non-negativity of conditional mutual information,
\item step~$(e)$ from the Markov chains $A_i\mkv (J,A^{i-1})\mkv (C^{i-1},E^{i-1})$ (see Fig.~\ref{fig:prob_vect_J}) and $(K,C_i)\mkv (J,A^{i-1},C^{i-1})\mkv E^{i-1}$ (see Fig.~\ref{fig:prob_vect_JK}),
\item step~$(f)$ from definitions~\eqref{eq:defU},~\eqref{eq:defV} and~\eqref{eq:defW}.
\end{itemize}

Using auxiliary random variable $Q$,
\begin{IEEEeqnarray*}{ll}
\Delta-R_C&-2\varepsilon\leq	\frac1n \sum_{i=1}^n \Big[ H(A_Q|V_Q,Q=i)\\ 
	&- I(A_Q;E_Q|U_Q,Q=i) - I(W_Q;C_Q|V_Q,Q=i) \Big] \\
	&=	H(A|V) - I(A;E|U) - I(W;C|V) \ .
\end{IEEEeqnarray*}

\subsection{End of Proof}

We proved that, for each achievable tuple $(R_A,R_C,D,\Delta)$ and each $\varepsilon>0$, there exist random variables $U$, $V$ and $W$ such that $U\mkv V\mkv A\mkv (C,E)$ and $W \mkv C \mkv (A,E)$ form Markov chains, and a function $\hat A$ such that
\begin{IEEEeqnarray*}{rCl}
R_A	+ \varepsilon			&\geq&	I(V;A|W) 						\ ,\\
R_C	+ \varepsilon			&\geq&	I(W;C|V) 						\ ,\\
R_A+R_C	+ 2\varepsilon		&\geq&	I(VW;AC)						\ ,\\
D + \varepsilon				&\geq&	\bE\big[d(A,\hat A(V,W))\big] 	\ ,\\
\Delta - \varepsilon		&\leq&	H(A|VW) + I(A;W|U) - I(A;E|U)	\ ,\\	
\Delta - R_C - 2\varepsilon	&\leq&	H(A|V) - I(A;E|U) - I(W;C|V) 	\ ,
\end{IEEEeqnarray*}
\emph{i.e.}, $(R_A+\varepsilon,R_C+\varepsilon,D+\varepsilon,\Delta-\varepsilon)\in\cR_\text{out}$.
Recalling that region $\cR_\text{out}$ is closed, and letting $\varepsilon$ tend to zero prove Theorem~\ref{th:outer_region}.
\endproof

\section{Proof of the Converse Part of Theorem~\ref{th:uncoded}}
\label{sec:uncoded:converse}

Let $(R_A,D,\Delta)$ be an achievable tuple and $\varepsilon>0$.
There exists an $(n,R_A+\varepsilon)$-code $(f,g)$ s.t.:
\begin{IEEEeqnarray*}{rCl}
\bE\left[ d(A^n,g(f(A^n),C^n)) \right]	&\leq& D+\varepsilon \ ,\\
\dfrac1n\,H(A^n|f(A^n),E^n) 			&\geq& \Delta-\varepsilon \ .
\end{IEEEeqnarray*}

Denote by $J=f(A^n)$ the transmitted message, and define variables $U_i$ and $V_i$ as follows, for each $i\in\{1,\dots,n\}$:
\begin{IEEEeqnarray}{rCl}
U_i &=& (J,\phantom{A^{i-1},C^{i-1},}	C_{i+1}^n,E^{i-1})	
										\label{eq:uncoded:defU} \ ,\\
V_i &=& (J,			A^{i-1},C^{i-1},	C_{i+1}^n,E^{i-1})
										\label{eq:uncoded:defV} \ .
\end{IEEEeqnarray}
From Fig.~\ref{fig:prob_vect_J}, $U_i\mkv V_i\mkv A_i\mkv (C_i,E_i)$ form a Markov chain.

We also define an independent random variable $Q$ uniformly distributed over the set $\{1,\dots,n\}$, and $A=A_Q$, $C=C_Q$, $E=E_Q$, $U=(Q,U_Q)$, and $V=(Q,V_Q)$.
$U\mkv V\mkv A\mkv (C,E)$ still form a Markov chain and $(A,C,E)$ is distributed according to the joint distribution $p(a,c,e)$ \emph{i.e.}, the original distribution of $(A_i,C_i,E_i)$.

\subsection{Rate}

\begin{IEEEeqnarray*}{rCl}
n(R_A+\varepsilon)
	&\geq& 					H(J) \\
	&\stackrel{(a)}{=}&		I(J ; A^n C^n E^n) \\
	&\stackrel{(b)}{\geq}&	I(J ; A^n E^n | C^n) \\
	&\stackrel{(c)}{=}&		\sum_{i=1}^n I(J ; A_i E_i | A^{i-1} C^n E^{i-1}) \\
	&=&						\sum_{i=1}^n \Big[ I(J A^{i-1} C^{i-1} C_{i+1}^n E^{i-1} ; A_i E_i | C_i)\\ 
	&&- I(A^{i-1} C^{i-1} C_{i+1}^n E^{i-1} ; A_i E_i | C_i) \Big] \\
	&\stackrel{(d)}{=}&		\sum_{i=1}^n I(J A^{i-1} C^{i-1} C_{i+1}^n E^{i-1} ; A_i E_i | C_i) \\
	&\stackrel{(e)}{\geq}&	\sum_{i=1}^n I(V_i ; A_i | C_i) \ ,
\end{IEEEeqnarray*}
where
\begin{itemize}
\item step~$(a)$ follows from $J=f(A^n)$,
\item step~$(b)$ from the non-negativity of mutual information,
\item step~$(c)$ from the chain rule for conditional mutual information,
\item step~$(d)$ from the fact that random variables $A_i$, $C_i$ and $E_i$ are independent across time,
\item step~$(e)$ from the non-negativity of mutual information and definition~\eqref{eq:uncoded:defV}.
\end{itemize}

Then, using random variable $Q$, 
\begin{IEEEeqnarray*}{rCl}
R_A + \varepsilon
 	&\geq&	\frac1n \sum_{i=1}^n I(V_Q ; A_Q | C_Q, Q=i)	\\
	&=&		I(V_Q ; A_Q | C_Q Q) 							\\
	&=&		I(V;A|C) 										\ .
\end{IEEEeqnarray*}

\subsection{Distortion at Bob}

Bob reconstructs $g(J,C^n)$. 
For each $i\in\{1,\dots,n\}$, define function $\hat A_i$ as the $i$-th coordinate of this estimate:
\[
\hat A_i(V_i,C_i) \triangleq g_i(J,C^{i-1},C_i,C_{i+1}^n)	\ .
\]

The component-wise mean distortion at Bob thus verifies
\begin{IEEEeqnarray*}{rCl}
D + \varepsilon
	&\geq&	\bE\left[ d(A^n,g(J,C^n)) \right] 							\\
	&=& 	\frac1n \sum_{i=1}^n \bE\left[ d(A_i,\hat A_i(V_i,C_i)) \right] \\
	&=& 	\frac1n \sum_{i=1}^n \bE\left[ d(A_Q,\hat A_Q(V_Q,C_Q))\ \middle|\ Q=i \right] \\
	&=& 	\bE\left[ d(A_Q,\hat A_Q(V_Q,C_Q)) \right] \\
	&=& 	\bE\left[ d(A,\hat A(V,C)) \right] \ ,
\end{IEEEeqnarray*}
where we defined function $\hat A$ on $\cV\times\cC$ by 
\[
\hat A(V,C) = \hat A(Q,V_Q,C_Q) \triangleq \hat A_Q(V_Q,C_Q) \ .
\]

\subsection{Equivocation Rate at Eve}

\begin{IEEEeqnarray*}{rCl}
n(\Delta - \varepsilon)
	&\leq&				H(A^n|J,E^n)				\\
	&=& 				H(A^n|J) - I(A^n ; E^n | J) \\
	&=& 				H(A^n|J C^n) + I(A^n;C^n | J) - I(A^n ; E^n | J) \\
	&\stackrel{(a)}{=}& H(A^n|J C^n) + I(A^n ; C^n) - I(J ; C^n) \\
	&&- I(A^n ; E^n) + I(J ; E^n) \\
	&\stackrel{(b)}{=}&	\sum_{i=1}^n \Big[ H(A_i|J A^{i-1} C^n) + I(A_i;C_i) \\
	&&- I(J C_{i+1}^n ; C_i) - I(A_i ; E_i) + I(J E^{i-1}; E_i) \Big] \\
	&\stackrel{(c)}{=}&	\sum_{i=1}^n \Big[ H(A_i|J A^{i-1} C^n E^{i-1}) + I(A_i;C_i) \\
	&&- I(J C_{i+1}^n ; C_i) - I(A_i ; E_i) +I(J E^{i-1}; E_i)\\
	&& + I(E_i ; C_{i+1}^n | J E^{i-1}) - I(C_i ; E^{i-1} | J C_{i+1}^n) \Big]	\\
	&=&					\sum_{i=1}^n \Big[ H(A_i|J A^{i-1} C^n E^{i-1}) + I(A_i;C_i)  \\
	&&- I(A_i ; E_i)+I(E_i ; J C_{i+1}^n E^{i-1}) \\
	&&- I(C_i ; J C_{i+1}^n E^{i-1}) \Big]	\\
	&\stackrel{(d)}{=}&	\sum_{i=1}^n \Big[ H(A_i|V_i C_i) + I(A_i;C_i)\\
	&& - I(A_i ; E_i) + I(E_i ; U_i) - I(C_i ; U_i) \Big] \\
	&\stackrel{(e)}{=}&	\sum_{i=1}^n \Big[ H(A_i|V_i C_i) + I(A_i;C_i|U_i)\\ 
	&&- I(A_i;E_i|U_i) \Big] \ ,
\end{IEEEeqnarray*}
where 
\begin{itemize}
\item step~$(a)$ follows from the Markov chain $J\mkv A^n\mkv (C^n,E^n)$,
\item step~$(b)$ from the chain rules for conditional entropy and mutual information, and the fact that random variables $A_i$, $C_i$ and $E_i$ are independent across time,
\item step~$(c)$ from the Markov chain $(A_i,C^i)\mkv (J A^{i-1})\mkv(C^{i-1},E^{i-1})$ (see Fig.~\ref{fig:prob_vect_J}) and Csisz\'ar and K\"orner's equality~\cite{csiszar1978broadcast} (see Appendix~\ref{sec:csiszarkorner}),
\item step~$(d)$ from definitions~\eqref{eq:uncoded:defU} and~\eqref{eq:uncoded:defV},
\item step~$(e)$ from the Markov chain $U_i\mkv A_i\mkv (C_i,E_i)$.
\end{itemize}

Using auxiliary random variable $Q$, this yields
\begin{IEEEeqnarray*}{rCl}
\Delta - \varepsilon
	&\leq&	\frac1n \sum_{i=1}^n \Big[ H(A_Q|V_Q C_Q, Q=i) 					\\
	&&\IEEEeqnarraymulticol{1}{r}{	+\,I(A_Q;C_Q|U_Q, Q=i) - I(A_Q;E_Q|U_Q, Q=i) \Big]	}\\
	&=&	H(A|V C) + I(A;C|U) - I(A;E|U) \ .
\end{IEEEeqnarray*}

\subsection{End of Proof}

We proved that, for each achievable tuple $(R_A,D,\Delta)$ and each $\varepsilon>0$, there exist random variables $U$, $V$ such that $U\mkv V\mkv A\mkv (C,E)$ forms a Markov chain, and
\begin{IEEEeqnarray*}{rCl}
R_A	+ \varepsilon		&\geq& I(V;A|C)							\ ,\\
D + \varepsilon			&\geq& \bE\big[d(A,\hat A(V,C))\big]	\ ,\\
\Delta - \varepsilon	&\leq& H(A|VC) + I(A;C|U) - I(A;E|U)	\ .
\end{IEEEeqnarray*}
Recalling that region $\cR^*_\text{uncoded}$ is closed, and letting $\varepsilon$ tend to zero prove the converse part of Theorem~\ref{th:uncoded}.
\endproof

\section{Proof of the Converse Part of Theorem~\ref{th:lossless}}
\label{sec:lossless:converse}

Let $(R_A,R_C,\Delta)$ be an achievable tuple and $\varepsilon>0$. There exists an $(n,R_A+\varepsilon,R_C+\varepsilon)$-code $(f_A,f_C,g)$ s.t.:
\begin{IEEEeqnarray*}{rCl}
\pr{ g(f_A(A^n),f_C(C^n))\neq(A^n,C^n) }	&\leq& \varepsilon	 \ ,\\
\dfrac1n\,H(A^n|f_A(A^n),E^n)				&\geq& \Delta-\varepsilon		\ .
\end{IEEEeqnarray*}

Denote by $J=f_A(A^n)$ and $K=f_C(C^n)$ the messages transmitted by Alice and Charlie, respectively.
For each $i\in\{1,\dots,n\}$, define random variable $U_i$ by
\begin{equation}
\label{eq:lossless:defU}
U_i = (J,C_{i+1}^n,E^{i-1}) \ .
\end{equation}
From Fig.~\ref{fig:prob_vect_J}, $U_i\mkv A_i\mkv (C_i,E_i)$ form a Markov chain.

We also define an independent random variable $Q$ uniformly distributed over the set $\{1,\dots,n\}$, and $A=A_Q$, $C=C_Q$, $E=E_Q$, $U=(Q,U_Q)$. Note that $U\mkv A\mkv (C,E)$ still form a Markov chain, and that $(A,C,E)$ is distributed according to the joint distribution $p(a,c,e)$ \emph{i.e.}, the original distribution of $(A_i,C_i,E_i)$.

\subsection{Rate at Alice}

Following the argument of the converse for the Slepian-Wolf theorem~\cite[Section~15.4.2]{cover2006elements}, we prove lower bounds on the rates:
\begin{IEEEeqnarray*}{rCl}
n(R_A+\varepsilon)
	&\geq& 					H(J) 							\\
	&\stackrel{(a)}{\geq}& 	H(J|C^n) 						\\
	&\stackrel{(b)}{=}&		I(A^n;J|C^n) 					\\
	&\stackrel{(c)}{=}&		H(A^n|C^n) - H(A^n|J K C^n)	 	\\
	&\stackrel{(d)}{\geq}& 	nH(A|C) - n O(\varepsilon)		\ ,
\end{IEEEeqnarray*}
where
\begin{itemize}
\item step~$(a)$ follows from the fact that conditioning reduces the entropy,
\item step~$(b)$ from $J=f_A(A^n)$,
\item step~$(c)$ from $K=f_C(C^n)$,
\item step~$(d)$ from the fact that random variables $A_i$ and $C_i$ are i.i.d., and Fano's inequality\footnote{Landau-like notation $O(\varepsilon)$ stands for a term $X$ such that $0\leq X\leq k\varepsilon$ for some constant $k>0$.}.
\end{itemize}

\subsection{Rate at Charlie}

Using similar arguments with $K=f_C(C^n)$, we can obtain:
\begin{IEEEeqnarray*}{rCl}
n(R_C+\varepsilon)
	&\geq& 					H(K) 												\\
	&\stackrel{(a)}{\geq}& 	H(K|J) 												\\
	&\stackrel{(b)}{=}&		I(K;C^n|J) 											\\
	&=&						H(C^n|J) - H(C^n|J K)	 							\\
	&\stackrel{(c)}{\geq}& 	\sum_{i=1}^n H(C_i|J C_{i+1}^n) - nO(\varepsilon)	\\
	&\stackrel{(d)}{\geq}& 	\sum_{i=1}^n H(C_i|U_i) - nO(\varepsilon)		\ ,
\end{IEEEeqnarray*}
where
\begin{itemize}
\item step~$(a)$ follows from the fact that conditioning reduces the entropy,
\item step~$(b)$ from $K=f_C(C^n)$,
\item step~$(c)$ from the chain rule for conditional entropy and Fano's inequality,
\item step~$(d)$ from the fact that conditioning reduces the entropy, and definition~\eqref{eq:lossless:defU}.
\end{itemize}

Now, using auxiliary random variable $Q$,
\begin{IEEEeqnarray}{rCl}
R_C+\varepsilon
	&\geq&	\frac1n\sum_{i=1}^n H(C_Q|U_Q,Q=i) - O(\varepsilon)	\nonumber\\
	&=&		H(C|U) - O(\varepsilon)								\label{eq:lossless:RA}\ .
\end{IEEEeqnarray}

\subsection{Sum-rate}

A lower bound on the sum-rate can be derived as well:
\begin{IEEEeqnarray*}{rCl}
n(R_A+R_C+2\varepsilon)
	&\geq& 					H(JK) 							\\
	&\stackrel{(a)}{=}&		I(A^nC^n;JK) 					\\
	&\stackrel{(b)}{\geq}& 	nH(AC) - n O(\varepsilon)	 	\ ,
\end{IEEEeqnarray*}
where
\begin{itemize}
\item step~$(a)$ follows from $J=f_A(A^n)$ and $K=f_C(C^n)$,
\item step~$(b)$ from the fact that random variables $A_i$ and $C_i$ are i.i.d., and Fano's inequality.
\end{itemize}

\subsection{Equivocation rate at Eve}

\begin{IEEEeqnarray*}{rCl}
n(\Delta-\varepsilon)
	&\leq&					H(A^n|J E^n) 									\\
	&=& 					H(A^n|J) - I(A^n ; E^n | J) 					\\
	&=& 					H(A^n|JK) + I(A^n;K|J) - I(A^n;E^n|J) 		\\
	&\stackrel{(a)}{\leq}&	n O(\varepsilon) + I(A^n;C^n|J) - I(A^n;E^n|J) 	\\
	&\stackrel{(b)}{=}& 	n O(\varepsilon) + I(A^n;C^n) - I(J;C^n) - I(A^n;E^n)\\
	&& + I(J;E^n) 	\\
	&\stackrel{(c)}{=}& 	n O(\varepsilon) + \sum_{i=1}^n \Big[ I(A_i;C_i) - I(J C_{i+1}^n;C_i)\\
	&& - I(A_i;E_i) + I(J E^{i-1};E_i) \Big]	
\end{IEEEeqnarray*}	

\begin{IEEEeqnarray*}{rCl}	
	&\stackrel{(d)}{=}& 	n O(\varepsilon) + \sum_{i=1}^n \Big[ I(A_i;C_i) - I(J C_{i+1}^n;C_i) \\
	&&- I(A_i;E_i) + I(J E^{i-1};E_i) 	\\
	&& +I(E_i;C_{i+1}^n|JE^{i-1}) -  I(C_i;E^{i-1}|JC_{i+1}^n) \Big]	\\
	&=&						n O(\varepsilon) + \sum_{i=1}^n \Big[ I(A_i;C_i) - I(JC_{i+1}^nE^{i-1};C_i) \\
	&&- I(A_i;E_i) + I(JC_{i+1}^nE^{i-1};E_i)  \Big]	\\
	&\stackrel{(e)}{=}&		n O(\varepsilon) + \sum_{i=1}^n \Big[ I(A_i;C_i|U_i) - I(A_i;E_i|U_i) \Big]	\ ,
\end{IEEEeqnarray*}
where 
\begin{itemize}
\item step~$(a)$ follows from Fano's inequality, and $K=f_C(C^n)$,
\item step~$(b)$ from the Markov chain $J\mkv A^n\mkv (C^n,E^n)$,
\item step~$(c)$ from the chain rule for mutual information, and the fact that random variables $A_i$, $C_i$, and $E_i$ are independent across time,
\item step~$(d)$ from Csisz\'ar and K\"orner's equality~\cite{csiszar1978broadcast} (see Appendix~\ref{sec:csiszarkorner}),
\item step~$(e)$ from definition~\eqref{eq:lossless:defU}, and the Markov chain $U_i\mkv A_i\mkv (C_i,E_i)$.
\end{itemize}

Now, using auxiliary random variable $Q$,
\begin{IEEEeqnarray*}{rCl}
\Delta-\varepsilon
	&\leq&	\frac1n\sum_{i=1}^n \Big[ I(A_Q;C_Q|U_Q,Q=i)\\
	&& \hspace{25mm}- I(A_Q;E_Q|U_Q,Q=i) \Big] + O(\varepsilon) \\
	&=&	I(A;C|U) - I(A;E|U) + O(\varepsilon)\ .
\end{IEEEeqnarray*}

\subsection{End of Proof}

We proved that, for each achievable tuple $(R_A,R_C,\Delta)$ and each $\varepsilon>0$, there exists a random variable $U$ such that $U\mkv A\mkv (C,E)$ form a Markov chain, and
\begin{IEEEeqnarray*}{rCl}
R_A	+ O(\varepsilon)			&\geq&	H(A|C)					\ ,\\
R_C	+ O(\varepsilon)			&\geq&	H(C|U) 					\ ,\\
R_A+R_C	+ O(\varepsilon)		&\geq&	H(AC)					\ ,\\
\Delta - O(\varepsilon)			&\leq&	I(A;C|U) - I(A;E|U)		\ .
\end{IEEEeqnarray*}
Recalling that region $\cR^*_\text{lossless}$ is closed, and letting $\varepsilon$ tend to zero prove the converse part of Theorem~\ref{th:lossless}.
\endproof

\section{Proof of the Converse Part of Proposition~\ref{prop:lossless_bis}}
\label{sec:lossless_bis:proof}

The proof of the converse part of Proposition~\ref{prop:lossless_bis} follows the same argument that Appendix~\ref{sec:lossless:converse}. In particular, definition~\eqref{eq:lossless:defU} remains the same.
The only difference lies in the lower bound for the rate at Alice:
\begin{IEEEeqnarray*}{rCl}
n(R_A+\varepsilon)	
	&\geq& 				H(J)											\\
	&\stackrel{(a)}{=}&	I(J ; A^n | C^n) + I(J ; C^n) 					\\
	&\stackrel{(b)}{=}&	H(A^n | C^n) - H(A^n | J K C^n) + I(J ; C^n) 	\\
	&\stackrel{(c)}{\geq}&	-nO(\varepsilon) + \sum_{i=1}^n \Big[ H(A_i|C_i) + I(J C_{i+1}^n ; C_i) \Big] \\
	&\stackrel{(d)}{=}&	-nO(\varepsilon) + \sum_{i=1}^n \Big[ H(A_i|C_i) + I(J C_{i+1}^n ; C_i) \\
	&&+I(E^{i-1} ; C_i | J C_{i+1}^n) - I(C_{i+1}^n ; E_i | J E^{i-1}) \Big]	\\
	&\stackrel{(e)}{\geq}&	-nO(\varepsilon) +  \sum_{i=1}^n \Big[H(A_i|C_i) \\
	&&+ I(J C_{i+1}^n E^{i-1} ; C_i) - I(J C_{i+1}^n E^{i-1} ; E_i) \Big] \\
	&\stackrel{(f)}{=}&	-nO(\varepsilon) + \sum_{i=1}^n \Big[ H(A_i|C_i) + I(U_i ; C_i) \\
	&&- I(U_i ; E_i) \Big] \ ,
\end{IEEEeqnarray*}
where
\begin{itemize}
\item step~$(a)$ follows from $J=f_A(A^n)$,
\item step~$(b)$ from $K=f_C(C^n)$,
\item step~$(c)$ from Fano's inequality, the chain rule for conditional mutual information and the fact that random variables $A_i$, $C_i$ are independent across time,
\item step~$(d)$ from Csisz\'ar and K\"orner's equality~\cite{csiszar1978broadcast},
\item step~$(e)$ from the fact that random variables $A_i$, $C_i$ and $E_i$ are independent across time, and the non-negativity of mutual information,
\item step~$(f)$ from definition~\eqref{eq:lossless:defU}.
\end{itemize}
Using random variable $Q$ and following the argument of Appendix \ref{sec:lossless:converse}, we proved the following lower bound:
\[
R_A+\varepsilon \geq H(A|C) + I(U;C) - I(U;E) - O(\varepsilon) \ .
\]
Since Equation~\eqref{eq:lossless:RA} still holds, we proved the bound on $R_A$ given by Proposition~\ref{prop:lossless_bis}.
Other steps of the proof remain unchanged.
\endproof

\section*{Acknowledgment}
The authors would like to thank Prof. Shlomo Shamai (Shitz) for many helpful discussions, the Associate Editor Prof. Yasutada Oohama and the anonymous reviewers for their valuable comments and suggestions  that contributed significantly to improve the quality of this paper.

\nocite{villard2010secure,villard2011secure}
\bibliographystyle{IEEEtran}
\bibliography{slsc_trans_IT}

\begin{IEEEbiographynophoto}{Joffrey Villard} 
(S'09-M'12) was born in Saint-\'{E}tienne, France, in 1985. He received the Dipl.Ing. degree in digital communication and electronics in 2008, the M.Sc. degree in wireless communication systems in 2008, and the Ph.D. degree in 2011, all from SUPELEC, Gif-sur-Yvette, France. From 2008 to 2011, he was with the Department of Telecommunications of SUPELEC. He is currently a Platform R\&D Engineer at WITHINGS, Issy-les-Moulineaux, France. His research interests include information theory, source coding, statistical inference, and signal processing for wireless sensor networks.
\end{IEEEbiographynophoto}

\begin{IEEEbiographynophoto}{Pablo Piantanida}
Pablo Piantanida received the B.Sc. and M.Sc degrees (with honors) in Electrical Engineering from the University of Buenos Aires (Argentina), in 2003, and the Ph.D. from the Paris-Sud University (France) in 2007. In 2006, he has been with the Department of Communications and Radio-Frequency Engineering at Vienna University of Technology (Austria). Since October 2007 he has joined in 2007 the Department of Telecommunications, SUPELEC, as an Assistant Professor in network information theory. His research interests include multi-terminal information theory, Shannon theory, cooperative communications, physical-layer security and coding theory for wireless applications. 
\end{IEEEbiographynophoto}


\end{document}